\RequirePackage{rotating}
\documentclass[10pt]{iopart}
\usepackage{iopams}
\expandafter\let\csname equation*\endcsname\relax
\expandafter\let\csname endequation*\endcsname\relax
\usepackage{amsmath,amssymb,amsthm,mathtools}
\usepackage{commath,dsfont,empheq,mathrsfs,slashed,bm,graphicx,wrapfig,chngcntr,color,xcolor,multicol,multirow,relsize,wrapfig,varioref,enumitem,mdframed,colortbl,needspace,setspace,letterspace,footmisc,hyperref,url,caption,subcaption,comment,soul,verbatim,pdfpages,booktabs,soul,dashbox}
\usepackage{etoolbox}
\makeatletter
\def\@mkboth#1#2{}
\newlength\appendixwidth
\preto\appendix{\addtocontents{toc}{\protect\patchl@section}}
\newcommand{\patchl@section}{%
  \settowidth{\appendixwidth}{\textbf{Appendix }}%
  \addtolength{\appendixwidth}{1.5em}%
  \patchcmd{\l@section}{1.5em}{\appendixwidth}{}{\ddt}%
}
\makeatother
\usepackage[outercaption]{sidecap}\sidecaptionvpos{figure}{c}
\usepackage[square,sort&compress,numbers]{natbib}
\usepackage{soul}
\usepackage[toc,indexonlyfirst]{glossaries}\makeglossaries\newglossaryentry{evolution}{%
  name={evolution},
  description={In biology, the change in inherited traits over successive generations in populations of organisms}
}

\newglossaryentry{heredity}{%
  name={heredity},
  description={(also \emph{inheritance}). Transmission of traits from one generation to another. The study of heredity in biology is genetics}
}

\newglossaryentry{central dogma}{%
  name={central dogma},
  description={Originally stated by F. Crick, it says that the genetic information flow progresses from DNA to RNA (transcription) to proteins (translation). Exceptions are well-known}
}

\newglossaryentry{DNA}{%
  name={DNA},
  description={ \emph{Deoxyribonucleic Acid}. A macro-molecule usually consisting of nucleotide polymers comprising antiparallel chains in which the sugar residues are deoxyribose and which are held together by hydrogen bonds between base pairs}
}

\newglossaryentry{nucleobases}{%
  name={nucleobases},
  description={(also \emph{nitrogenous bases}). Nitrogen-containing biological compounds, the most common being adenine (\sml{A}), cytosine (\sml{C}), guanine (\sml{G}), thymine (\sml{T}), and uracil (\sml{U}). The bases  \sml{A,T,C,G} are found in the DNA, in the RNA \sml{T} is replaced by \sml{U}}
}

\newglossaryentry{transcription}{%
  name={transcription},
  description={Transfer of genetic information from DNA by the synthesis of a complementary RNA molecule using a DNA template}
}

\newglossaryentry{translation}{%
  name={translation},
  description={The derivation of the amino acid sequence of a polypeptide from the base sequence of an mRNA molecule in association with a ribosome and tRNAs}
}

\newglossaryentry{chromosome}{%
  name={chromosome},
  description={In prokaryotes, DNA molecule containing the organism’s genome. In eukaryotes, DNA molecule complexed with RNA and proteins to form a threadlike structure that contains genetic information arranged in a linear sequence}
}

\newglossaryentry{diploid}{%
  name={diploid},
  description={ ($2n$) A cell (by extension, an organism) that contains two copies of each chromosome}
}

\newglossaryentry{haploid}{%
  name={haploid},
  description={($n$) A cell (by extension, an organism) having one member of each pair of homologous chromosomes}
}

\newglossaryentry{population}{%
  name={population},
  description={A group of organisms of a species that interbreed and live in the same place at a same time. They are capable of reproduction}
}

\newglossaryentry{population genetics}{%
  name={population genetics},
  description={Study of the genetic composition of populations, including distributions and changes in genotype and phenotype frequency in response to the processes of natural selection, genetic drift, mutation and gene flow}
}

\newglossaryentry{natural selection}{%
  name={natural selection},
  description={Differential reproduction among members of a species owing to variable fitness due to genotypic and phenotypic differences}
}

\newglossaryentry{mutation}{%
  name={mutation},
  description={Any process that produces an alteration in DNA or chromosome structure; in genes, the source of new alleles. Among the most common: insertion, duplication, deletion, translocation, inversion, point mutation. They are \emph{silent} if they do not alter the polypeptide chain, \emph{missense} if they cause a substitution of a different amino acid in the resulting protein, \emph{nonsense} if they result in a premature stop codon}
}

\newglossaryentry{recombination}{%
  name={recombination},
  description={A process that leads to the formation of new allele combinations on chromosomes. In eukaryotes, genetic recombination during \emph{meiosis} can lead to a novel set of genetic information that can be passed on from the parents to the offspring. In bacteria recombination happens by \emph{transformation} (ability to take up DNA from the surroundings),  \emph{transduction}  (transfer  of  genetic material by the intermediary of viruses), and \emph{conjugation} (direct transfer of DNA from a donor to a recipient)}
}

\newglossaryentry{phenotype}{%
  name={phenotype},
  description={Ensamble of the observable characteristics or traits of an organism}
}

\newglossaryentry{genotype}{%
  name={genotype},
  description={The allelic or genetic constitution of an organism; often, the allelic composition of one or a limited number of genes under investigation}
}

\newglossaryentry{fitness}{%
  name={fitness},
  description={ Expected reproductive success of an organism. For modelling purposes, this is often equated to the average number of offspring in the subsequent generation (absolute fitness)}
}

\newglossaryentry{epistasis}{%
  name={epistasis},
  description={Effects on fitness that depend on 
  variations at two or more loci. 
  Sometimes a distinction is made between epistasis
  where two variation reinforce each other, and epistasis 
  where they act against each other. The second type is
  referred to as \textit{sign epistasis}}
}

\newglossaryentry{crossing-over}{%
  name={crossing-over},
  description={The exchange of genetic material during sexual reproduction between two homologous chromosomes. It happens during meiosis}
}

\newglossaryentry{meiosis}{%
  name={meiosis},
  description={The process of cell division in sexually-reproducing organisms during which the diploid number of chromosomes is reduced to the haploid number}
}

\newglossaryentry{allele}{%
  name={allele},
  description={One of the possible alternative forms of a genomic locus. Two different alleles may or may not be distinguishable based on the induced phenotypic effects}
}

\newglossaryentry{bottleneck}{%
  name={bottleneck},
  description={A drastic reduction in population size and consequent loss of genetic diversity, followed by an increase in population size. It causes a loss of diversity in the rebuilt population}
}

\newglossaryentry{genetic drift}{%
  name={genetic drift},
  description={Random sampling of the individuals that survive from one generation to another or fluctuations of genotypes or allele frequencies. Typically observed in small populations}
}

\newglossaryentry{hybridization}{%
  name={hybridization},
  description={A scenario where two strains of one species that have been evolving in isolation for some time come in contact again}
}

\newglossaryentry{quantitative genetics}{%
  name={quantitative genetics},
  description={Study of the genetic basis underlying phenotypic variation among individuals, with a focus primarily on traits that take a continuous range of values \emph{e.g.} height, weight, longevity}
}

\newglossaryentry{fds}{%
  name={frequency-dependent selection},
  description={The fitness of a genotype depends of the frequency of that genotype in the population. It is \emph{positive} if fitness increases with frequency, \emph{negative} in the opposite case}
}

\newglossaryentry{RNA}{%
  name={RNA},
      description={Similar to DNA but characterized by the pentose sugar ribose, the pyrimidine uracil (instead of thymine), and the single-stranded nature of the polynucleotide chain. RNA molecules exist in different types in the cells: messenger RNA (mRNA), ribosomal RNA (rRNA), and transfer RNA (tRNA)}
}

\newglossaryentry{germ line cells}{%
  name={germ line cell},
  description={In mammals, haploid cell able to unite with one from the opposite sex to form a new individual}
}

\newglossaryentry{soma line cells}{%
  name={soma line cell},
  description={In mammals, any cell of the body except germ cells. Somatic cells are diploid. Mutations in somatic cells are not passed on to offspring}
}

\newglossaryentry{transduction}{
    name = {transduction},
    description={Genetic material is transferred from one bacterium to another one by a virus}
}

\newglossaryentry{transformation}{
    name = {transformation},
    description={One bacterium releases \gls{DNA} into the environment where it is picked up by another one}
}

\newglossaryentry{conjugation}{
    name = {conjugation},
    description={Similarly to animal sex, a donor cell injects genetic material into a recipient cell through a dedicated conduit (a pilus). In contrast to animal sex, the transfer is only one way and does not give rise to a new organism; the effect is to transform the recipient cell}
}

\definecolor{Gray}{gray}{0.9}
\usepackage{empheq,mathrsfs}
\renewcommand{\vec}{\boldsymbol}
\newcommand{\de}{\partial}

\newcommand{\av}[1]{\langle #1 \rangle}

\newcommand{\pf}{\mathcal{Z}}
\newcommand{\sml}[1]{\texttt{#1}}
\DeclareMathOperator*{\argmax}{arg\,max}

\DeclareMathOperator*{\arctanh}{arc\,tanh}
\newtheorem*{definition-non}{Definition}

\definecolor{mLightBrown}{HTML}{EB811B}
\definecolor{mLightGreen}{HTML}{14B03D}
\definecolor{mDarkRed}{HTML}{CF0A0A}

\theoremstyle{definition}

\begin{document}
    \title[Statistical genetics in and out of quasi-linkage equilibrium]{Statistical genetics in and out of quasi-linkage equilibrium}

\author{Vito Dichio$^{1}$, Hong-Li Zeng$^{2}$, Erik Aurell$^{3}$}
\address{$^1$ Sorbonne Université, Paris Brain Institute - ICM, CNRS, Inria, Inserm, AP-HP, Hôpital de la Pitié Salpêtrière, F-75013, Paris, France}
\address{$^2$ School of Science, Nanjing University of Posts and Telecommunications, New Energy Technology Engineering Laboratory of Jiangsu Province, Nanjing, 210023, China}
\address{$^3$ Department of Computational Science and Technology,
AlbaNova University Center, SE-106 91 Stockholm, Sweden;}
\ead{eaurell@kth.se}

\vspace{10pt}
\begin{indented}
\item[]\today
\end{indented}

\begin{abstract}
This review is about statistical genetics, an interdisciplinary topic between statistical physics and population biology. The focus is on the phase of \emph{quasi-linkage equilibrium} (QLE). 
Our goals here are to clarify under which conditions the QLE phase can be expected to hold in population biology and how the stability of the QLE phase is lost. The QLE state, which has many similarities to 
a thermal equilibrium state in statistical mechanics, was discovered by M Kimura for a two-locus two-allele model, and was extended and generalized to the global genome scale by \emph{Neher} \& \emph{Shraiman (2011)}. What we will refer to as the Kimura-Neher-Shraiman (KNS) theory describes a population evolving due to the  mutations, recombination, natural selection and possibly genetic drift.
A QLE phase exists at sufficiently high recombination rate ($r$) and/or mutation rates $\mu$ with respect to selection strength. We show how in QLE it is possible to infer the epistatic parameters of the fitness function from the knowledge of the (dynamical) distribution of genotypes in a population.
We further consider the breakdown of the QLE regime for high enough selection strength. We review recent results for the selection-mutation and selection-recombination dynamics. Finally, we identify and characterize a new phase which we call the non-random coexistence (NRC) where variability persists in the population without either fixating or disappearing. 
\end{abstract}
\noindent{\it Keywords\/}: statistical genetics,  quasi-linkage equilibrium, direct coupling analysis, inference.

\newpage
\tableofcontents
\newpage
\printglossaries

\newpage

%%%%%%%%%%

\section{Introduction}
\label{sec:introduction}

This review is in the field of statistical genetics. In theoretical biology, this is the area concerned with the development of statistical methods 
to describe the distribution of genotypes in a population.
In this Introduction we will state what the review is about,
and what are its goals. For both tasks we need to use technical
terms concisely defined in the Glossary, where the page of their first occurrence in the manuscript is also indicated.
We will explain these terms and other 
biological concepts in more detail as we will need them 
further into the review.

The beginning of statistical genetics can be taken to be the discovery by Weinberg and Hardy more than a century ago of the \textit{Hardy-Weinberg equilibrium}. 
This is the result that the proportion of major to minor alleles at one locus in a population
evolving only under recombination will stay constant. 
The next important concept is \textit{linkage equilibrium} (LE) 
which can be thought of as a two-locus version of Hardy-Weinberg, as a property of haplotypes,
where there is also selection based on the alleles at two loci separately, and mutations.
That is, under recombination, mutations and a restricted type of selection,
the two-locus distributions of alleles can evolve to become 
independent. In difference to Hardy-Weinberg, in LE the frequencies at both
loci will not in general stay constant, but will each tend to a one-locus balance 
between selection and mutation.

All other distributions of genotypes in a population, where allele distributions between loci are not independent, 
can be defined as instances of a phase of \textit{linkage disequilibrium} (LD).

Statistical dependency between two loci implies that the allele distributions are correlated in the sense of having non-zero pairwise correlation function. In the population genetics literature the term LD is therefore also used in the sense of a norm of the correlation function. A positive value of LD (as a norm) then implies LD (as a state of the population). While for
bi-allelic loci this use of LD is unambiguous, for multi-allele distributions different norms have been used \cite{SLD}. In this review LD means only a phase of statistical dependency.

The first general mechanism behind LD is inheritance (phylogeny).
When a beneficial mutation arises in some individuals
and there is no recombination, the individual's descendants
inherit the ancestral allele also at other loci.
Those alleles will thus be present or absent together,
and hence correlated. This effect occurs also when recombination 
acts at finite speed, but then only between loci which are close enough.
The second general mechanism behind LD is selection dependent 
on variations at more than one locus. This mechanism then competes with recombination
so that even without mutations the joint distributions at two
loci will tend to be correlated \textit{i.e.} statistically dependent.
Both mechanisms can act in the same population at the same time.
Depending of their relative strengths, 
the wide phase of LD includes different sub-phases, with different properties.

This review is about the sub-class \textit{quasi-linkage equilibrium} (QLE)
where the main driving mechanism is the second
one above.
Historically, the QLE state was discovered by Kimura in 1965 in a bi-allelic 2-loci model~\cite{Kimura}, and developed extensively by Neher and Shraiman in a genome-wide setting (multi-loci models) \cite{NS-1}. 
In these presentations recombination was the fastest process. 
As a result, allele variations at different loci were statistically dependent,
though from the high recombination rate,
only weakly so. From this smallness of locus-locus correlations
one can say that
out of the different possibilities in LD, a QLE phase 
by the Kimura-Neher-Shraiman mechanism
is close to LE.
More generally QLE 
and LE are similar in that there is only
one distribution of genomes in the population.
This is in contrast to LD due to phylogeny,
where there can be two or more clones, sets of individuals with similar genomes
related by common descent, competing for dominance.

We have found it convenient to introduce a formal definition of QLE,
which we state in sec.(\ref{sINST}). One reason for doing so is that for there to be a QLE phase,
satisfying all reasonable requirements, the assumption of fast recombination is
sufficient but not necessary. As we will show in 
sec.(\ref{sec:Gaussian})
one can have a QLE phase also when the fastest process is mutation, provided
there is also some recombination.
The first part of our formal definition is that
the full distribution of genotypes in a population is a Boltzmann distribution. In previous discussions this property was a consequence of specific model assumptions.
LE is in this view the special case of QLE when there are no
interactions in the energy function.
The parameters of this Boltzmann distribution are related to evolutionary parameters by definite (different) relations.
Those relations will be a main
tool of this review. 
The second part of our formal definition is that multi-genome 
distributions factorize. This has been either assumed or derived from 
specific model assumptions in all previous discussions. We will postpone
a discussion of this point to sec.(\ref{sINST}).

There are many conceptual similarities between population genetics and statistical physics, reviewed from the side of physics multiple times, \emph{e.g.} \cite{Peliti,BMK,ZA-2}. More recently, the genotype-phenotype map was reviewed in \cite{manrubia2021} and the possible predictability and control of evolution in \cite{Lassig}. 
The Boltzmann distribution of QLE makes it an obvious additional intersection.
There are also differences. An important one is that the Boltzmann distribution characterizing QLE is not a consequence of underlying detailed balance, but 
arises for other reasons.

The review has three main objectives. The first is to put the spotlight on QLE as an important topic for statistical physicists interested in fundamental questions of population genetics. The second is to show that the presence (or not) of QLE in a simulated population with known parameters can be assessed with techniques borrowed from statistical inference, and collectively known as direct coupling analysis (DCA). The third is to leverage these assessments to clarify when QLE holds in population biology, and how the stability of the QLE phase can be lost. We will review, supplement and extend earlier theoretical investigations in the literature, and add new numerical tests.

The review is structured as follows. Section~\ref{sec:statistical-genetics} introduces the biological concepts of interest, and introduces Kimura-Neher-Shraiman (KNS) theory of statistical genetics.
Section \ref{cISF} is about the retrieval from samples of parameters of Gibbs-Boltzmann distributions with Ising/Potts Hamiltonian. Techniques to achieve this task in a computationally efficient yet accurate way are collectively known as ``direct coupling analysis" (DCA). DCA has been reviewed multiple times, and we will therefore only describe the most common variants. Section~\ref{sINST} introduces a formal definition of QLE and relates that to small variations in growth rate due to fitness, strong recombination and/or mutations. We derive in two different limits inference formulae whereby fitness parameters are related to statistics of the data analyzed by DCA.
These inference formulae are then compared and used to quantitatively map out when the KNS theory holds. Section~\ref{cVI} starts by reviewing previous theoretical approaches to population genetics out of the QLE regime. We then describe a new phase of \textit{non-random coexistence} (NRC) where variability persist in the population without either fixating or disappearing.  We identify an intermediate region in the parameter space where a finite population jumps stochastically between a QLE-like state and NRC-like behaviour. Finally, section~\ref{sec:discussion} summarizes the results and gives an outlook for the future.

An extended version of this manuscript - including derivations of the main results and supplementary information - can be found in \cite{dichio2021}.

%%%%%%%%%%

\section{Statistical genetics}
\label{sec:statistical-genetics}

\subsection{Subject matter: population genetics in a nutshell}\label{sSMPGiaN}

This section contains a brief introduction to the biology relevant to this review. It is primarily aimed to physicists not conversant with these matters; biological physicists and biologists may skip to the next section. As noted above, technical terms are defined in the Glossary.

The crucial difference between living and non-living forms of matter is Darwin's evolution, that selects the most apt individuals to the environment. 

\begin{figure}
    \centering
        \includegraphics[width=0.6\linewidth]{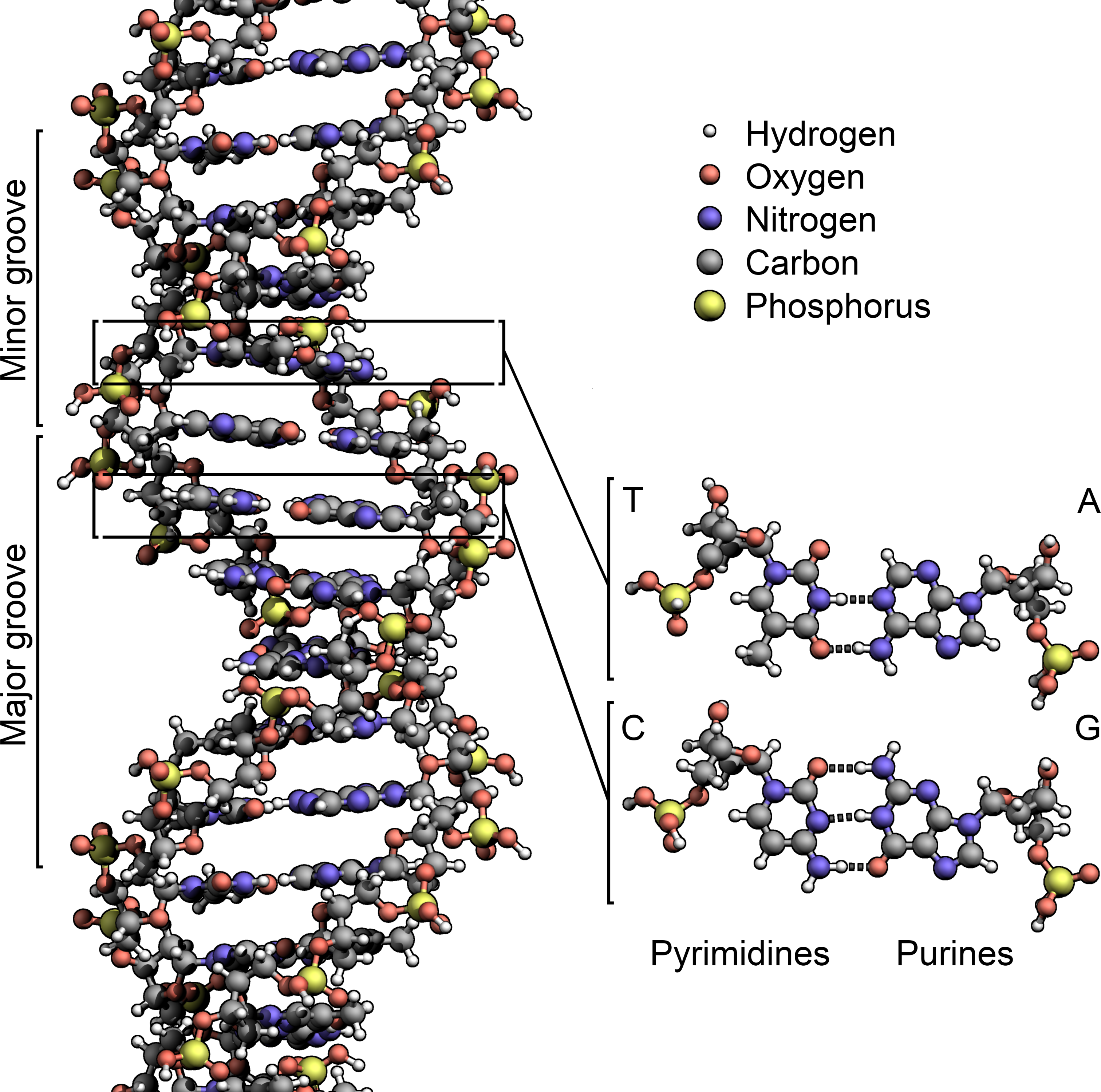}
        \caption{The DNA double helix as proposed by J. Watson and F. Crick. The interwining strands are sugar-phosphate backbones, while the horizontal rungs depict the nitrogenous base pairs (\sml{A-T,G-C}). Credits to Richard Wheeler, via \href{https://en.wikipedia.org/wiki/File:DNA_Structure\%2BKey\%2BLabelled.pn_NoBB.png}{Wikimedia Commons}.}\label{fdna}
    \end{figure}

% evolution
The key element of \gls{evolution} is \gls{heredity}, \emph{i.e.}, the possibility of inheriting information across generations. According to the \gls{central dogma} of molecular biology, biological information is encoded in the \gls{DNA}, a macro-molecule present in each cell that consists in two sugar-phosphate ribbon-like strands that coil around to form a double helix and whose horizontal rungs are pairs of complementary \gls{nucleobases}: \sml{A-T,G-C}, see fig.(\ref{fdna}). The information is encoded in the precise sequence of nucleobases of each strand. By means of \gls{transcription} \gls{DNA} is converted into the closely related molecule \gls{RNA}, and by means of \gls{translation}, a stretch of \gls{RNA} is translated into a polypeptide chain. The latter will eventually result in a protein, performing one of the many different functions needed to sustain the life of the cells.
   
The word `gene' refers to a stretch of DNA which is transcribed together, and a `gene product' is the protein produced from the corresponding part of the \gls{RNA}. In microscopic organisms (bacteria, viruses) as well as in higher organisms a number of biological mechanisms allow for the possibility that from one single gene more than one protein are generated. In this way it is possible for \textit{e.g.} humans to have about $20,000$ genes but more than $100,000$ proteins. Some types of heredity (epigenetics) exist that do not involve the sequence of nucleotides in \gls{DNA}, the most well-known being chemical modifications of \gls{DNA} (methylation and other) which is important in \textit{e.g.} heritable gene silencing. Even if there are exceptions to the central dogma, it describes the overwhelming majority of biological information processing as pertaining to information stored as chemical molecules.

% Regulation of gene expression
Complex regulatory mechanisms of the gene expression weave an intricate and largely unknown network of interactions within genes. Some such gene expression patterns can be inherited over many generations and comprise another type of epigenetics, even if often enhanced by methylation and similar processes in higher organisms. Many other gene expression patterns on the other hand change on fairly rapid time scale in response to changes in the environment. In bacteria this is in fact the main form of cellular information processing and it is vital in higher organisms as well, even if often overlayed by other and faster pathways. 

% Individual & reproduction
In eukaryotes the DNA is often condensed in the form of \glspl{chromosome}. A population in which each cell has one complete set of chromosomes is named \gls{haploid}, if there are two such sets, \gls{diploid}. In mammals the \glspl{germ line cells} (egg cell and sperm) are haploid and the \glspl{soma line cells} (the rest) are diploid. This form of life is hence diploid-dominated, and the organisms reproduce by going through an obligatory haploid phase where two germ line cells mix in sex. In other organisms very many different forms of reproduction and diploid/haploid division of labour are possible. Asexual reproduction has been found to occur naturally everywhere except among mammals  (among birds in domesticated turkeys and chicken), but is generally less frequent the more complex the organism. 

% Population - variability
We now broaden the perspective and consider an entire \gls{population}. Typically a gene can be found in one or several variants in a population. Such variants are called \glspl{allele} and can also be found in different proportions in different sub-populations. Alleles can differ either at one genomic position (single nucleotide polymorphisms, or SNPs), or in ways that involve changes at more than one genomic position. The latter can be through multiple SNPs in a single gene or by insertions and deletions. 

% Population genetics
The goal of \gls{population genetics} is to study the genetic composition and dynamics of evolving biological populations. 
% selection
A major driver of the evolutionary process is \gls{natural selection}. At the \gls{phenotype} level, advantageous features enhance the probability for an individual to survive and reproduce (high \gls{fitness}). This has consequences at the \gls{genotype} level, even though the exact map between these two layers may be complex, \emph{i.e.} it is not clear which characteristics of genotype elements lead to which phenotypic traits. 
% mutations
The variability on which selection acts can be fuelled by \glspl{mutation}, which can arise by chance in a genomic sequence \textit{e.g.} because of transcription errors. Mutations can have no consequences at the level of protein (\emph{synonymous}) or cause alterations in the polypeptide chain they code for (\emph{non-synonymous}). They represent a major source of variability for the evolutionary process.
% recombinations
Interactions between individuals by the exchange of genetic material (\gls{recombination}) can lead to the emergence of new genotypes, too. In a general sense they can all be called forms of sex, even if acting quite differently than sex in mammals. In prokaryotes, the main forms of recombination are \gls{transduction}, \gls{transformation}, \gls{conjugation}. In eukaryotes, recombination happens during \gls{meiosis} where the mixing between two chromosomes from each of the parents is enhanced by the \gls{crossing-over} mechanism.
% randomness
Finally, random events can additionally alter the genetic pool of a population:  \glspl{bottleneck}, \gls{genetic drift}, \gls{hybridization} (...)

% population genetics vs quantitative genetics
In the biological literature, a distinction is made between population genetics and \gls{quantitative genetics}. This review is almost exclusively about the first. The second deals with the genetics of continuously varying characteristics, such as height and skin color in human. Historically this was referred to as \textit{quantitative} (measured by a number), in opposition to characteristics that appear in only a few different types, as do \textit{qualities} in classical philosophy. Inherited qualities are due to differences in genotypes on one or a few positions. On the contrary, most quantitative characteristics of higher organisms are due both inheritance ("nature") and environment ("nurture"). In addition, if one could isolate the genetic component of such quantitative traits, they would be typically due to variations in many positions. 

% data availability
In the pre-sequencing era population genetics was the realm of theory and explanations, while quantitative genetics was the realm of what could be measured and of direct interest to biology. In modern times  this relationship is partly upended: whole genome sequences of many organisms can (and have been) obtained and the predictions of population genetics can be compared to such data. This is the approach we have followed in this work.
Measuring quantitative traits remains however time-consuming and difficult, and the relationship between genotype and phenotype is one of the most complex and least known (though most studied) in all of the science. In the spirit of statistical physics it is therefore natural to focus on the genotype scale (microstate), once that is measurable.

\subsection{A brief historical overview}
As noted in Introduction, the mathematical theory of population genetics started in 1908 when Hardy and Weinberg showed that in a population with diploid genomes evolving only due to recombination (sex) 
the frequencies of genotypes $AA$, $Aa$ and $aa$ at a bi-allelic locus tend to $p^2$, $2p(1-p)$ and $(1-p)^2$. The parameter $p$ is the total frequency of allele $A$ in the population which does not change under only recombination.
The publication date of Hardy's paper \cite{Hardy1908} is some months earlier than Weinberg's \cite{Weinberg1908}, but the latter was based on a public lecture Weinberg had given at the beginning of the year. In the English-language literature, the attribution of the result to both Hardy and Weinberg was first made in~\cite{Stern1943}.
The relation between Hardy-Weinberg equilibrium and 
linkage equilibrium (LE) was summarily discussed in Introduction.

Dynamic evolutionary models describe the changes of distribution of genotypes in a population in time.
When simulated numerically on a computer they produce evolutionary trajectories from which both one-time and multi-time characteristics can be computed. Many levels of detail can be included in such models. The first and simplest such models are the Wright-Fisher model and the Moran model which describe the evolution of populations under the influence of only mutations and random genetic drift. Mathematically these are discrete-time discrete-variable stochastic processes. For a single bi-allelic locus, the evolution of one population can hence be pictured as a jump process in a lattice of sites labeled by $n$ which can take values $0,1,\ldots,N$, $N$ being the total number of individuals. The Wright-Fisher and Moran models can be straight-forwardly extended to mechanisms of selection and migration (island models) \cite{BMK}.
The software used for numerical tests in sec.(\ref{sINST}-\ref{cVI})
can be said to simulate an extension of the Moran model where also selection and recombination
are taken into account, for precise description, see below.

The evolution of already the Wright-Fisher and Moran models
is more complex over many loci than at one locus. As a mathematical simplification, it is interesting to first consider genetic drift acting independently on each locus. The evolution of a population is then analogous to a jump process in an $L$-dimensional lattice with sites labeled $n_1, n_2,\ldots, n_L$, $L$ being the number of loci, with an independent source of randomness in each direction. Such evolution laws are non-degenerate stochastic processes, and the evolution of an ensemble of genomes is described by the associated Fokker-Planck equation (forward Kolmogorov equation). The physical flavour of this change of perspective from a distribution over genomes to a distribution over allele frequencies was succinctly stated by R Fisher in the 1953 Croonian Lecture to the Royal Society

\begin{quote}
"the frequencies with which the different genotypes occur define the gene ratios characteristic of the population, so that it is often convenient to consider a natural population not so much as an aggregate of living individuals as an aggregate of gene ratios. Such a change of viewpoint is similar to that familiar in the theory of gases, where the specification of the population of velocities is often more useful than that of a population of particles"
\end{quote}
\begin{flushright}  Ronald A. Fisher \cite{Fisher1953}
\end{flushright} 

Again similarly to physics, in the proper limit the evolution laws of the distribution are parabolic partial differential equations. The first model of such a law was proposed by Fisher in 1922 in the form of a standard diffusion \cite{Fisher1923,Fisher}. This model overestimated the amount of genetic drift when one allele is close to fixation. At the time the mathematical theory of state-dependent diffusions had not yet been developed, and it was Kolmogorov who in 1935 first wrote down the correct expression, where the strength of the random drift vanishes as one allele tends towards fixation \cite{Kolmogorov1935}. This expression was independently re-derived by Wright \cite{Wright1945} and Kimura \cite{Kimura-2,Kimura1964}, and is usually referred to as the \textit{diffusion approximation} or \textit{Kimura's diffusion approximation}. The rigorous mathematical aspects of this diffusion limit have been addressed by many authors from different communities \textit{cf.} \cite{shimakura1981formulas,Hofbauer1985,Huillet2017}. The resulting diffusion process (as well as the underlying discrete process) can be or not be in detailed balance. The condition for detailed balance here translates to that mutations satisfy an integrability condition relative to a measure induced by the random drift on genotype space, known as the Shahshahani-Svirezhev condition \cite{Svirezhev-Passekov2011,Shahshahani1979}. If this condition holds one can include both additive and epistatic terms of the fitness function to the model and still deduce a simple form for the stationary state (analogous to thermal equilibrium in a potential) \cite{AurellEkebergKoski2019}. Properties of reversible evolutionary dynamics were considered in \cite{Manhart2012}, and papers cited therein.

Genetic drift is in Wright-Fisher and Moran models implemented on top of mutations and selection by each individual in a population replaced by another randomly picked individual, to which one has applied random changes (mutations), with different probabilities (selection). Genetic drift hence does not actually act independently at each locus. 

The way biology nevertheless approaches Fisher's proposition is by the process of recombination (or sex) which mixes up the genotypes at different loci. Recombination plays in population genetics the role of collisions in gas theory, and the assumption of genetic drift acting independently at each locus is formally similar to Boltzmann's molecular chaos.

\subsection{Kimura-Neher-Shraiman Theory (KNS)}\label{sMNST}
The theory of evolution of a population under recombination as well as other forces was pioneered by Kimura~\cite{Kimura} and developed further by R. Neher and B. Shraiman in \cite{NS-1}. We will call it the Kimura-Neher-Shraiman (KNS) theory.

We will describe KNS by adopting the following simplification: by ``genotype'' we will always mean one genome out of all possible genomes of the same length. Although processes that change the length of genomes are important in biology, the restriction to genomes of the same length brings out clearly the analogies to equilibrium and non-equilibrium spin systems.

Additionally, we make the following simplifying
assumptions:
\begin{enumerate}
    \item \textbf{Genomic structure}. An haploid genome is a vector $g=(s_1,\dots,s_L)$ of $L$ loci $s_i$ where $i=1,\dots,L$. The number $L$ of loci is fixed and equal for all the individual genomes. A population is a collection $\{g^{\alpha}\}_{\alpha\in A}$, where $A$ is a set of indices. Each genome $g$ appears in the population with probability $P(g)$. 
    \item \textbf{Ising loci}. Loci are bi-allelic \textit{i.e.}     there are two alleles at each locus. They can then be coded by spin-like variables
    $s_i=\pm 1 \ \forall i$.     The genotype space is then represented by the $2^L$ vertices of the hypercube $\{-1,1\}^L$. 
    \item \textbf{Constant population}. The average number of individuals is fixed. This hypothesis can model \textit{e.g.} the struggle for survival in an environment with limited resources. Except when explicitly stated, the population is further assumed to be infinite ($|A|=\infty$).
    \item \textbf{One-genome evolution}. The distribution of one genome in a population is given by a genome distribution $P(g,t)$. This distribution evolves in time driven by three operators representing natural selection, mutations and recombination. For recombination, which fundamentally is a process acting on more than one genome at a time, this is a substantial assumption. This point will be discussed below. The action of the three evolutionary forces is then encoded in a master equation \textit{i.e.} a phenomenological first-order differential equation 
    \begin{equation}\label{eMEI} 
        \frac{d}{dt} P(g,t) = \frac{d}{dt}\Big|_{\substack{fit}} P(g,t) + \frac{d}{dt}\Big|_{\substack{mut}} P(g,t) + \frac{d}{dt}\Big|_{\substack{rec}} P(g,t)\ .
    \end{equation}
\end{enumerate}

We now turn to analyse each single terms in eq.(\ref{eMEI}) separately.

\subsubsection{Selection}\label{ssF}
The model for natural selection is based on a \emph{fitness function} $F$ defined to be proportional to the average number of offspring of an individual of genotype $g$. In other words, $F(g)$ expresses the propensity of a genotype to transfer its genomic material to the next generations. The explicit form of $F(g)$ defines the fitness landscape of the population.
A fitness function $F(g)$ does not capture all forms of natural selection. In particular, it implies
\begin{enumerate}
    \item[$\circ$] that fitness depends only on the genotype.
    In general, the reproductive rate of a given genome (or genomic trait) may depend on its frequency in the population \emph{e.g.} because of some feedback regulation system.
    \item[$\circ$]
    that effects of cooperation and strategic behaviour (games) are ignored.
    \item[$\circ$]
    that issues related to a possible fluctuating environment (\emph{fitness seascapes}) and related  time-dependence of selection are ignored. 
\end{enumerate}

With the above limitations the first term in eq.(\ref{eMEI}) can be written as
\begin{equation}\label{eFT}
    \frac{d}{dt}\Big|_{\substack{fit}} P(g,t) = [F(g)-\av{F}_t] P(g,t)\ ;
\end{equation}
where $\av{F}_t = \sum_g F(g) P(g,t)$ is the population-average fitness that ensuring the normalisation of $P(g,t)$. Fit individuals ($[F(g)-\av{F}_t]>0$) will grow in proportion, and an unfit ones ($[F(g)-\av{F}_t]<0$) will decrease; therefore, also in this simplified model, whether an individual is fit or not depends on which other individuals are present in the population.

We will here only consider fitness functions with linear and pairwise interactions:
\begin{equation}\label{eFL} 
    F(g) = \bar F + \sum_{i} f_i s_i + \sum_{i<j} f_{ij} s_i s_j\ .
\end{equation}
Other possibilities have been explored in the literature, see \cite{Peliti} and references therein. In above, $\bar F$ is a constant, irrelevant for eq.(\ref{eFT}). The first order contribution $f_i$ represents \emph{additive fitness terms} at locus $i$. This influences fitness independently of all other loci in the genome. Higher terms such as $f_{ij}$ (and  $f_{ijk}, f_{ijkl,\dots}$ if they were present) represent genetic interactions between loci, also called \gls{epistasis}. The total fitness can be characterized as a functional of the \textit{a priori} fitness function as 
\begin{equation}\label{eCFT}
\sigma(f)= \sqrt{\sum_i f_i^2 + \sum_{i<j}f_{ij}^2}\ ,
\end{equation}
Additive and epistatic terms of the fitness are quantified by $\sigma_a$ and $\sigma_e$, which have the same definition as eq.(\ref{eCFT}) except that respectively only the additive and epistatic contributions appear. $F(g)$ has dimension $[t^{-1}]$; the same is true for all the coefficients $f_i,f_{ij},\dots$ and for $\sigma,\sigma_e,\sigma_a$.
 
Fitness in statistical genetics plays a similar role as energy (modulo a minus sign) in statistical mechanics. 
The evolutionary process of a population can be pictured as an erratic motion of a point \emph{on} the fitness landscape.
In contrast to statistical mechanics, 
a point particle here does not slide down towards energy minima, but climbs fitness hills.
  
\subsubsection{Mutations}\label{ssM}
The model for mutations is single-locus swaps $s_i\rightarrow-s_i$. In mathematical terms an operator $M_i$ acts on a genomic sequence by swapping the $i$-th bi-allelic gene \textit{i.e.} $M_i(g)=(-s_i,\bm{s}_{\backslash i})\ . $ Let $\mu$ be the tunable \emph{mutation rate}, constant in time and the same for all loci; same as $\sigma$, it has dimensions $[t^{-1}]$. The mutation term in the master equation then takes the simple form 
\begin{equation}\label{eMT} 
    \frac{d}{dt}\Big|_{\substack{mut}} P(g,t) = \mu \sum_{i=1}^L [P(M_i g, t)- P(g,t)]\ .
\end{equation}
Same as for selection, the above simplification excludes potentially important mechanisms. Those are
\begin{enumerate} 
    \item[$\circ$] that mutations do not have to be only single nucleotide changes; insertions and deletions are in many settings at least as important.
    \item[$\circ$] that even single nucleotide changes do not have to proceed with the same rate at all positions; \textit{mutation hot-spots} are well-documented.
    \item[$\circ$] that the mutation rate does not have to be the same in both directions. In the more general setting of multi-allele loci mutations have in general to be specified by mutation matrices. 
\end{enumerate}

\subsubsection{Recombination}\label{ssR}
The model for recombination is that two parents $g^{(1)},g^{(2)}$ mix their genomic sequences and give birth to two new individuals $g,g'$ where eventually the second ($g'$) is ignored.  Several biological mechanisms give rise to recombination thus defined. First, sex in diploid organisms means that two gametes from two parents merge to form one new individual. These gametes are haploid; $g'$ then comprises the remaining genomic material of the parents. The formation of gametes includes the process of \textit{crossover} by which the (one-chromosome) gamete inherits parts of the two chromosomes of the parent. Second, this type of recombination models bacterial sex by \textit{transformation} or \textit{transduction} (where material goes in both ways), as well as recombination in several RNA viruses including HIV and coronaviruses. On the other hand, this type of recombination does not model bacterial sex by \textit{conjugation} (where material goes only one way).

Following \cite{NS-1} it is convenient to introduce a  set of random variables $\{\xi_i\}$  to describe recombination by defining a crossover pattern. Consider the allele at locus $i$ of the new individual $g$, if it has been inherited from $g^{(1)}$ then $\xi_i=1$ while if it comes from $g^{(2)}$ then $\xi_i=0$. The sequence $g'$ is simply complementary to $g$. In symbols $g, g'$ can be written as
\begin{equation}\label{eRPC} 
    \begin{split}
    \mathbf{g} \ :\  s_i&= \xi_i s_i^{(1)} + (1-\xi_i) s_i^{(2)}\ , \\ 
    \mathbf{g'} :\  s_i' &= (1-\xi_i) s_i^{(1)} + \xi_i s_i^{(2)}\ \ . 
    \end{split}{}
\end{equation}
Each different crossover pattern $\{\xi_i\}$ comes with a probability $C(\xi)$. Let $r$ be the tunable overall recombination parameter, dimensions $[t^{-1}]$. Under the simplifying assumption that any genome pair has the same recombination rate $r$, which is the case of a \emph{panmictic} population where any individual is equally likely to interact with anyone else, the recombination term in the master equation is written
\begin{equation}\label{eRT1}
    \frac{d}{dt}\Big|_{\substack{rec}} P(g,t) = r \sum_{\xi,g'}\  C(\xi)\Big[P_2(g^{(1)},g^{(2)},t) - P_2(g,g',t)\Big]\ ,
\end{equation}
where $g^{(1)}, g^{(2)}$ are found by inverting eq.(\ref{eRPC}). The sum runs over all possible recombination patterns and all possible sequences $g'$ and $P_2$ is the two-genome distribution (read two-particle distribution).

To close the equations we need a further assumption, which will also be part of the definition of the QLE phase in sec.(\ref{sINST}). This is that the two-genome distributions in eq.\eqref{eRT1} factorize:
\begin{equation}\label{eMCH} 
    P_2(g_{\alpha},g_{\beta})=P(g_{\alpha})P(g_{\beta})\ ,
\end{equation}

We postpone to sec.(\ref{sec:multi-genome factorization}) a detailed discussion on the validity of this assumption. As for now, it is worth stressing that as in physics so in biology eq.(\ref{eMCH}) is \emph{never} exactly true. In a realistic biological environment, several phenomena introduce correlations between different individuals \emph{e.g.} competition for limited resources, geographical separation, existence of classes of individuals, or phylogenetic effects. In the theoretical arguments, such correlations will be assumed to be weak enough for eq.(\ref{eMCH}) to hold approximately. Inserting these assumptions in eq.(\ref{eRT1}),
\begin{equation}\label{eRT2} 
\frac{d}{dt}\Big|_{\substack{rec}} P(g,t) = r \sum_{\xi,g'}\  C(\xi)\Big[P(g^{(1)},t)\ P(g^{(2)},t) - P(g,t)\ P(g',t)\Big]\ .
\end{equation}

The final expression for master equation can now be obtained by using eq.\eqref{eFT}, \eqref{eMT} and \eqref{eRT2} into eq.(\ref{eMEI}): 
\begin{equation}\label{eMEE}
    \begin{split}
        \frac{d}{dt} P(g,t) = & \ [F(g)-\av{F}_t] P(g,t) + \mu \sum_{i=1}^L [P(M_i g, t)- P(g,t)] +\\\
        &+ r \sum_{\xi,g'}\  C(\xi)\Big[P(g^{(1)},t)\ P(g^{(2)},t) - P(g,t)\ P(g',t)\Big]\ ,
    \end{split}
\end{equation}
In what follows, the time dependence $t$ will be dropped in order to lighten the notation.
%%%

\subsubsection{Dynamics of genotype distribution}\label{ssDoGD}
It is possible to parameterize the distribution $P(g,t)\ \forall g$ by its cumulants. The cumulants of first and second order $\chi_i=\av{s_i}$ and $\chi_{ij}=\av{s_is_j} - \av{s_i} \av{s_j}$ are of special interest. Using eq.(\ref{eMEE}), it is also possible to derive the equations that describe their dynamics. A step-by-step derivation can be found in an extended version of this paper \cite{dichio2021}. In both cases, the structure of the calculation is 
\begin{equation}\label{e-scalc}
    \frac{d}{dt}\av{O} = \frac{d}{dt}\sum_g O(g) P(g) = \sum_g O(g)\frac{d}{dt} P(g)
\end{equation}
where $O(g)$ is a combination of spin terms and $\frac{d}{dt} P(g)$ is evaluated thorough eq.(\ref{eMEE}).

In the case of the first order cumulants $\chi_i=\av{s_i}$ (mean allele values in the population), a simplification comes from the fact that recombinations have no effect of their dynamics. Indeed, the effect of recombination is to reshuffle alleles in the population without changing their overall frequency. Therefore, in evaluating eq.(\ref{e-scalc}) the recombination term of the master equation can be ignored. The result is:
\begin{equation} \label{eEFOC}
    \dot\chi_i = \av{s_i[F(g)-\av{F}]} - 2\mu \chi_i \ ,
\end{equation}
where no specific ansatz has yet been done for both the fitness function $F(g)$ and the probability distribution $P(g)$.

An analogous result can be derived for the second order cumulants $\chi_{ij} = \av{s_is_j}-\av{s_i}\av{s_j}$; this time recombinations matter, since they act on the pairwise statistics (correlations between loci) computed at the population level. For $i\ne j$:
\begin{equation} \label{eSOC}
    \dot\chi_{ij} = \av{(s_i-\chi_i)(s_j-\chi_j)[F(g)-\av{F}]} - (4\mu+rc_{ij})\chi_{ij}\ , 
\end{equation}
where we have defined 
\begin{equation}\label{eDCIJ} % Definition c_ij
    c_{ij} = \sum_{\xi} C(\xi) [\xi_i(1-\xi_j)+(1-\xi_i)\xi_j]\ .
\end{equation} 

This latter quantity can be easily interpreted as the probability that, in the offspring, the alleles at the two loci $i,j$ come from different parents.  When recombinations are completely random, we expect $c_{ij}=1/2\ \forall i,j$. Two possible models for $c_{ij}$ are the following:
    \begin{itemize}
        \item[$\circ$] \textbf{Crossover rate.} If there is recombination between two genomes, then each locus undergoes a crossover with fixed probability $\omega$, called crossover rate; as a consequence, $c_{ij}=2\omega(1-\omega)$ uniformly $\forall i,j$. 
        \item[$\circ$] \textbf{Neighbouring variability.} More realistically, if two loci are very far apart then they can be expected to be mostly uncorrelated. In \cite{ZA-1}, the authors assumed a fixed probability $\rho$ that a recombination causes a crossover between any pair of neighbouring loci. 
        After the recombination event, each two neighbouring loci will come from the same parent with probability $P(SP) = 1-\rho$, from different parents with probability $P(DP)=\rho$. As a first approximation, the number of such $SP/DP$ events for neighbors along the genomic chain can be assumed to be binomially distributed.
        As a result, one gets: 
        \begin{equation}
            c_{ij} = \sum_{\ k\ \text{odd}}^n {n \choose k}\rho^k(1-\rho)^{n-k} = \frac{1}{2} \Big[1-(1-2\rho)^{n} \Big]\ \label{eCIJZA}
        \end{equation}
        In the limit $|i-j|=n\rightarrow\infty$ for $\rho\le \frac{1}{2}$ one finds (consistently) $c_{ij}\sim 1/2$. 
\end{itemize}

The quantities $\chi_{ij}$, the evolution of which is described by eq.(\ref{eSOC}), are central to this review. Non-zero $\chi_{ij}$ is often taken to be an order parameter of linkage disequilibrium. In the literature LD can also refer to a non-zero norm (absolute value) of $\chi_{ij}$. Similarly, vanishing $\chi_{ij}$ is often taken to be a witness of linkage equilibrium.

In a population evolving under mutation and recombination without selection, $\chi_{ij}$ exponentially decays to zero. According to eq.(\ref{eSOC}), selection acts in the opposite direction, driving $\chi_{ij}$ away from zero. The tendency of natural selection is indeed to fix the most fit alleles in a population. If this process is run to completion all individuals are identical and variability is lost. In that limit, the population may be said to be in a (trivial) state of linkage equilibrium, as all quantities $\chi_{ij}$ vanish.

%%%%%%%%%%
\section{Direct coupling analysis (DCA)}\label{cISF}

We now make a break in the presentation of statistical genetics and turn to a set of technical tools called direct coupling analysis (DCA) which we will need later. The starting point is a general task of statistical inference: suppose we have $M$ independent draws from a Gibbs-Boltzmann distribution; the task of DCA is to find the parameters of the distribution from the samples. In statistics the corresponding problem would be separated into retrieving the interaction graph (model learning) and determining the parameters of the interactions of the energy function (parameter inference), together referred to \textit{learning and inference in an exponential family} \cite{Wainwright-2008a}. In statistical physics the same basic task has also been called an \textit{inverse Ising/Potts problem}~\cite{NZB}.

A distinguishing characteristics of DCA is that while maximum likelihood and analogous Bayesian point estimate methods are feasible for small enough instances, for larger instances these methods become computationally demanding. A number of alternative inference methods have therefore been proposed, of which the most widely used are mean-field or variational methods \cite{Kappen-1998a,Wainwright-2008a}, and pseudo-likelihood maximization \cite{Besag-1975a,Ravikumar-2010a}. More recent algorithms of the same general type as pseudo-likelihood were introduced in \cite{Vuffray-2016a,Berg-2017a,Lokhov-2018a} and are known to exhibit better performance on some model problems. Previous high-impact applications of DCA to biological data are somewhat out of the main emphasis of the current review. We therefore present them separately in sec.(\ref{secDCA-biology}).

\subsection{DCA for bi-allelic genome distributions}\label{sAIP}
In this section we discuss DCA when all variables take two values (Ising model). Generalization to Potts model is unavoidable in most of the applications surveyed in sec.(\ref{secDCA-biology}), but are not needed here. From the methodological point of view of different DCA flavours, Ising and Potts model are similar, see \emph{e.g.} \cite{CFFMW}.

Let us consider an Ising model with $L$ binary spin variables $s_i=\pm1$, with $i=1,\dots,L$. The Hamiltonian for an Ising system reads
\begin{equation}\label{eIH}
    \mathscr{H}_{\vec{J},\vec{h}}(\vec{s}) = -\sum_i h_is_i - \sum_{i<j}J_{ij}s_is_j,
\end{equation}
where $\vec{J}$ is the matrix of pairwise couplings between the spin variables ($J_{ii}=0\ \forall i$) and $\vec{h}$ is the vector of local magnetic fields. Collectively they are referred to as the \emph{parameters} of the Ising problem. The equilibrium distribution is the Boltzmann distribution
\begin{equation}\label{eIBD}
    p(\vec{s})=\frac{1}{\mathcal{Z}}e^{- \mathscr{H}_{\vec{J},\vec{h}}(\vec{s})}\ .
\end{equation}
In inference problems we can set the inverse temperature $\beta$ to $1$ without loss of generality. When the task is to infer parameters from data we cannot distinguish $\beta$ from an overall scale factor of the parameters. The normalization $\mathcal{Z}$ in eq.\eqref{eIBD} is the standard partition function
\begin{equation}\label{ePFIIP}
    \mathcal{Z}(\vec{J},\vec{h}) = \sum_{\vec{s}} e^{- \mathscr{H}_{\vec{J},\vec{h}}(\vec{s})}\ .
\end{equation}
The expected value of a function $Q(\vec{s})$ of the spin variables is defined as
\begin{equation}
    \av{Q} = \sum_{\vec{s}} p(\vec{s})Q(\vec{s})\ .
\end{equation}
The alternative term \textit{inverse Ising problem} is explained by the fact that in the \textit{forward Ising problem} the parameters $\vec{J},\vec{h}$ of the Boltzmann distribution eq.(\ref{eIBD}) are known and the task is to compute statistical observables \textit{e.g.} $\chi_i, \chi_{ij}$. In an \emph{Inverse Ising Problem} (IIP) the paradigm could be the opposite.
However, important DCA methods such as pseudo-likelihood do not start from statistical observables but directly from the samples. For this reason we prefer the less circumscribed term DCA.

\subsection{Numerical methods for DCA applied to Ising distributions}\label{sML}
Let $p(x_1,\dots x_M|\theta)$ be the probability to observe $M$ samples drawn from a probability distribution given by a set of parameters denoted $\theta$. We recall that according to the maximum likelihood criterion the best estimate $\theta^{ML}$ of the parameters from the samples is given by
\begin{equation}\label{eMLES}
    \theta^{ML}=\argmax_{\theta} p(x_1,\dots, x_M|\theta)\ .
\end{equation}
In a Bayesian context eq.\eqref{eMLES} is a point estimate (one predicted parameter value) assuming a flat (information-free) \emph{prior} information $p(\theta)$ on the parameter. To avoid dealing with small numbers, it is common practice to maximize the logarithm of the likelihood. The log-likelihood per sample is defined as
\begin{equation}\label{eLLF}
\mathscr{L}_D(\vec{J},\vec{h})=\frac{1}{M}\log p(D|\vec{J},\vec{h}) \ ;
\end{equation}
For $M$ independent samples $D=\{ \vec{s}^{m}\}$ drawn from a Boltzmann distribution of an Ising model, \textit{i.e.} eq.(\ref{eIBD}), the log-likelihood per sample reads
\begin{align}
    \mathscr{L}_D(\vec{J},\vec{h}) &= \sum_ih_i\frac{1}{M}\sum_m s_i^m + \sum_{i<j}J_{ij} \frac{1}{M}\sum_m s_i^ms_j^m - \log\pf(\vec{J},\vec{h})\notag\\
    &= \sum_ih_i\av{s_i}^D + \sum_{i<j}J_{ij}\av{s_is_j}^D - \log\pf(\vec{J},\vec{h})\ ,\label{eLLH}
\end{align}
where $\av{s_i}^D$ and $\av{s_is_j}^D$ are the corresponding sample averages. The maximum likelihood problem, the solution of which can formally be written
\begin{equation}\label{eMLEIM}
    \{\vec{J}^{ML}, \vec{h}^{ML} \} = \argmax_{\vec{J},\vec{h}} \mathscr{L}_D(\vec{J},\vec{h})\ ,
\end{equation}
is computationally costly. \emph{Boltzmann machine learning} is a gradient-descent algorithm with an adjustable learning rate $\eta$ so that at equilibrium (converged values) one has
\begin{equation} \label{e-emp-mle}
\begin{split}
    0 = \frac{\de \mathscr{L}_D}{\de h_i}(\vec{J}^n,\vec{h}^n) &=  \av{s_i}^D - \av{s_i}\ , \\
    0 = \frac{\de \mathscr{L}_D}{\de J_{ij}}(\vec{J}^n,\vec{h}^n) &=  \av{s_is_j}^D - \av{s_is_j}\ .
\end{split}
\end{equation}
The computational cost here appears both in that  one has to estimate ensemble averages which are computationally costly, and that convergence may be slow.

The simplest DCA method is \emph{mean-field} (MF) inference, for historical reasons also often called naive mean-field (nMF) inference. The original derivation was based on a mean-field approximation of the partition function and using a fluctuation-dissipation relation \cite{Kappen-1998a}, reviewed \textit{e.g.} \cite{Wainwright-2008a,NZB} and in \cite{dichio2021}. The most straight-forward derivation is on the other hand to take the Ising probability distribution eq.(\ref{eIBD}) as a Gaussian probability distribution over continuous variables, from which it immediately follows that:
\begin{equation}\label{eMFC}
    \vec{J}^{MF} = -\vec{\chi}^{-1}\ .
\end{equation}
At the cost of a quite strong approximation, the task of computing the couplings requires now a simple matrix inversion of the empirical covariance matrix $\vec{\chi}$. This can be done in a polynomial time $\sim L^3$, whereas the original maximum likelihood maximization requires an exponential time $\sim 2^L$. Several more refined methods of this sort exist, based on modifications of the thermodynamic potential, reviewed in \cite{NZB}.

The next most common DCA method is \emph{pseudo-likelihood maximization} (PLM). It is based on another basis than MF and starts from the probability of $s_i$ conditional on the observation of all the other variables $\vec{s}_{\backslash i}$
\begin{equation}\label{ePLME}
    p(s_i|\vec{s}_{\backslash i}) = \frac{1}{1 + e^{-2s_i(h_i + \sum_{j\ne i}J_{ij}s_j)}} =\frac{1}{2} \Bigg[ 1+s_i\tanh\Big( h_i + \sum_{j\ne i} J_{ij} s_j \Big)\Bigg]\ 
\end{equation}
By the form of the Gibbs-Boltzmann distribution (distributions in exponential families) the conditional probability depends only on the field $h_i$ and on the couplings $J_{i\bullet}$ between $i$ and every other spin. From the conditional probability one can form a log-likelihood per sample $\mathscr{L}^i_D$ for eq.(\ref{ePLME}), which reads
\begin{equation}\label{e-pseudol}
    \mathscr{L}^i_D(J_{i\bullet},h_i) = \frac{1}{M}\sum_m \log\frac{1}{2}\Bigg[ 1+s_i^m\tanh\Big( h_i + \sum_{j\ne i} J_{ij} s_j^m \Big)\Bigg]\ .
\end{equation}
This can be maximized by setting to zero the derivatives with respect to the parameters:
\begin{equation}
\begin{split}
    \av{s_i}^D &= \Big\langle \tanh\Big( h_i^{PL} + \sum_{j\ne i} J_{ij}^{PL} s_j \Big) \Big\rangle^D \ \\
    \av{s_is_j}^D &= \Big\langle s_j\tanh\Big( h_i^{PL} + \sum_{k\ne i} J_{ik}^{PL} s_k \Big) \Big\rangle^D\ ,\label{eSOCAS}
\end{split}
\end{equation}
the solution of which yields a set of $L$ estimated parameters $h_i^{PL}, J_{i\bullet}^{PL}$. There are $L$ functions like eq.(\ref{e-pseudol}). In the most common variant of PLM (asymmetric PLM) these functions are maximized independently. In this case the inferred couplings $J_{ij}^{PL}\ne J_{ji}^{PL}$, while in the underlying probabilistic model there is only one parameter $J_{ij}$. An output routine is then needed, the most commonly used is to take the average $\frac{1}{2}(J_{ij}^{PL}+J_{ji}^{PL})$ as final estimate. The computational cost of PLM is $L^2$  for the minimization of each $\mathscr{L}^i_D$; hence $L^3$ in total, the same scaling as MF but with a larger pre-factor. A theoretical advantage of PLM is that is statistically consistent \emph{i.e.} it yield the same parameter estimate as maximum-likelihood in the limit of infinite data.

A standard procedure to numerically test a version of DCA is to simulate data from a Gibbs-Boltzmann distribution with known fields and couplings, and then compare the results of the inference with the input values of the parameters. 

If $J_{ij}^0$ are the input parameters to the simulation and $J^*_{ij}$ the inferred couplings, the reconstruction error can be visualized and quantified in different ways. A qualitative measure is a scatter-plot where the values of $J^*_{ij}$ are regressed on $J_{ij}^0$. This usually gives a clear indication of when DCA does not work, by the appearance of a "cloud of points". In many applications of DCA it has turned out that the most relevant predictions are those $J^*_{ij}$ of largest value, see \cite{AurellBarbierDecelleMulet2022} for a recent discussion of theoretical and methodological implications. 

In this review we illustrate inference errors by scatter-plots and quantify them by a $L_2$-metric, which for the $J_{ij}$ parameters is defined as:
\begin{equation}\label{eREIIP}
    \gamma_J = \sqrt{\frac{\sum_{i<j}(J^*_{ij}-J_{ij}^0)^2}{\sum_{i<j}(J_{ij}^0)^2}}\ .
\end{equation}
In later sections the emphasis will be on prediction errors on epistatic fitness parameters. In the simplest version we then compare inferred fitness $f^*_{ij} = \left(\mu+r c_{ij}\right) J^*_{ij}$ to underlying fitness $f_{ij}^0$, where $\mu$, $r$ are other evolutionary parameters discussed above and the derived quantity $c_{ij}$ is given in eq.(\ref{eCIJZA}). The comparison is done by scatter-plots and by the use of metrics analogous to $\gamma_J$.

The phenomenology of the dependence of $\gamma_J$ on Ising/Potts parameters and number of samples has been investigated in many studies, reviewed in \cite{NZB}. All DCA methods fail for few enough samples, and for small enough Ising/Potts parameters at a given number of samples. The reason for the second is that there is then not enough information about the underlying probability distribution from the samples, which are overwhelmed by statistical noise. All DCA methods also face difficulties for large enough Ising/Potts parameters, as it is difficult to independently sample from such distributions (low-temperature phase in statistical physics).

\subsection{Biological applications of DCA, a brief survey} \label{secDCA-biology}

Biological sequence data analysis has been a prominent area of applications of DCA in the last decade. In those applications a Gibbs-Boltzmann distribution is taken as a given, and the emphasis has been on the methodological challenges, and on the biological interpretation of the results. That is hence different from the perspective of this review, where the focus is on a QLE phase as the mechanism behind Gibbs-Boltzmann distributions. Nevertheless, as much of the methodological developments have been directly motivated by these applications of DCA we here provide a brief survey. 

The flagship application of DCA has been to predict spatial contacts in protein structures from tables of homologous (similar) proteins. This is based on several lines of biological knowledge. The most basic is that proteins can be grouped into protein families with similar protein structures, but more variable protein sequences. The second is that epistasis within one protein-coding gene is mostly associated to changes around contacts in the structure; changing one amino acid close to another amino acid can change the stability of the whole structure. The third, mostly empirical, is that large DCA terms have turned out to be significantly better predictors of residue-residue spatial proximity than correlations in the allele distributions at two loci. While this does not prove that the distribution of protein sequences in a protein family is a Gibbs-Boltzmann distribution -- likely not exactly true -- it shows that this is a useful starting point for predictions. One feature of DCA applied to biological data analysis of this type is that the problems are usually under-sampled (there are more parameters to the model than data). A given DCA must therefore be regularized, which
adds another layer of methodological variants.

The first result in this direction which had wide resonance used mean-field inference regularized by pseudo-counts \cite{Morcos-2011a}; later contributions using mean-field inference with other regulation schemes are \textit{e.g.} \cite{Hopf-2012a,Jones-2012a,RLS2014}. DCA by pseudo-likelihood maximization with various regularizations was introduced in the field slightly later \cite{Ekeberg-2013a,Ekeberg-2014a}. These results were deemed sufficiently informative that DCA methods were incorporated into the protein structure prediction pipelines in the CASP tournament, and have been reviewed \textit{e.g.} in \cite{Stein-2015a} and \cite{CFFMW}. They were also later combined with other information sources in meta-algorithms achieving significantly better performance, see \textit{e.g.} \cite{Jones-2015a,Golkov-2016a,Michel-2017a,Hopf-2017a,Ovchinnikov-2017a}. As has been widely reported, in the last years DCA-based protein structure methods (and other methods) have been overtaken by AI/deep learning approaches~\cite{Senior-2020,Hiranuma-2021}. Although a full theory of the success of such methods is not at hand, a likely interpretation is that AI/deep learning is able to learn both the Gibbs-Boltzmann terms of DCA as well as deviations from a Gibbs-Boltzmann distribution based on the biophysics of protein structure. That this has been possible is ultimately due to the very large number of solved protein structures on which it has been possible to train AI/deep learning methods.  

For other biological inference tasks with less abundant number of training examples and/or where the goal is to uncover new biology DCA remains an important tool. Applications include predicting protein interaction partners \cite{Baldassi-2014a,Uguzzoni-2017a}, context-dependence of mutations in beta-lactamase TEM-1 \cite{Figliuzzi-2016a}, secondary and tertiary RNA structure prediction \cite{DeLeonardis-2016a} and inference of epistatic interactions from population-wide whole-genome sequencing of bacterial and viral pathogens \cite{Skwark-2017a,Schubert2018,Zeng-COVID19}. The approach has also been applied to HIV in a well-known series of papers \cite{FERGUSON2013,Chakraborty2013pre,Chakraborty2018pnas}. 
A particularly promising recent contribution in this direction is the prediction of how mutable are SARS-CoV-2 positions from the single SARS-CoV-2 reference sequence and sequences of other coronaviruses~\cite{Rodriguez-Rivas2022}. In that case predictions could be validated against mutagenesis data and the unprecedented large number of sequenced SARS-CoV-2 genomes, more than 10 million to date~\cite{GISAID}.

In applications of DCA to data, an empirical method
called \textit{sequence re-weighting} has been used from the 
beginning to separate correlations between loci (LE) due to inheritance from 
LE due to multi-loci fitness functions \cite{Morcos-2011a},
more systematic approaches were recently
introduced in \cite{horta, Horta2021}.
The \textit{profile- and phylogeny-aware sequence randomization}
method of \cite{horta}
is based on scrambling a multiple sequence alignment (MSA)
such that single-locus allele frequencies and inter-sequence
genomic distances are both preserved. Since shared inheritance is 
typically determined from genomic closeness, the 
phylogeny inferred from the
set of fictitious individuals represented by the rows in the
scrambled MSA will be the same as (or similar to) that in
the original population. On the other hand, due to scrambling
all effects of synergetic contributions to fitness (epistasis)
is lost. Therefore, if the same DCA terms appear using
both the scrambled and the original MSA they are likely due to
inheritance, and should not be retained. 
In \cite{horta} this approach was assessed 
using two sets of respectively nine proteins and their MSAs
(data set DS1), and 60 proteins and their MSAs (data set DS2),
and in \cite{Zeng-COVID19} the same approach was used on a data set of
50,000 SARS-CoV-2 genomes.
The randomization step was then a
significant computational overhead. In more recent investigations
of larger sets of SARS-CoV-2 genomes, consequently either 
a screening based on metadata (geographic position) \cite{CRESSWELLCLAY2021} or excluding known large clones
(SARS-CoV-2 Variants of Concern, or VoCs) \cite{ZengPRE2022}
were used.

On the theme of this review, the results of \cite{Skwark-2017a}, \cite{Schubert2018} and especially \cite{FERGUSON2013,Chakraborty2013pre,Chakraborty2018pnas} and \cite{Rodriguez-Rivas2022} are evidence that respectively populations of \textit{Streptococcus pneumoniae}, \textit{Neisseria gonorrhoeae}, HIV and coronaviruses are in or close to a QLE phase: all these pathogens are known to exhibit relatively strong recombination.

%%%%%%%%%%
\section{Quasi-linkage equilibrium (QLE)}\label{sINST}
As briefly stated in sec.(\ref{sec:introduction}), the QLE state was discovered by Kimura in 1965 in a bi-allelic 2-loci model~\cite{Kimura}, and developed extensively by Neher and Shraiman in a genome-wide setting (multi-loci models) \cite{NS-1}. As shown in the latter, and as will be discussed below, QLE appears when allele frequencies change slowly, and correlations are small and steady. This is the case when genetic interactions are weak effects compared to recombination. However, logically this may not be the only setting in which QLE phase can appear. We therefore start from the following formal definition:

\begin{definition-non}\label{def:QLE}
A population is said to be in a quasi-linkage equilibrium (QLE) phase if two conditions are met: (1) multi-genome distributions factorize \emph{i.e.} $P_k(g_1,g_2,\dots,g_k,t)=P(g_1,t)P(g_2,t)\cdots P(g_k,t)$; and (2) single-genome distributions lie in an exponential family with no higher terms than in the fitness function.
\end{definition-non}

\noindent 
The first part of the Definition will be discussed below.
In this review, we consider fitness functions on biallelic genomes with at most pairwise epistatic interactions. The second part of the above Definition then implies that one-genome distributions are Gibbs-Boltzmann distributions of an Ising model
\begin{equation}\label{eICE} 
    P(g,t)=\frac{1}{\mathcal{Z}(t)}\ \exp\Bigg(\sum_i\ h_i(t)\ s_i + \sum_{i<j}\ J_{ij}(t)\ s_i s_j\Bigg) \ ,
\end{equation}
As in many formal definitions in biology, it must be understood that in a real
population they hold only to a higher or lower degree. It may even be that there is no real
population where one-genome
distributions are exactly of the Gibbs-Boltzmann type, or where
multi-genome distributions exactly factorize.
Nevertheless, the formal statement emphasizes that if there are significant deviations from the definitions
in some population or some model, then we are not concerned with them
when we discuss QLE.

We stress that there is no explicit or implicit assumption of detailed balance of a stochastic process. While there are parallels to the statistical physics of non-ideal gases, which we will discuss, for the moment it is more useful to imagine that eq.(\ref{eICE}) emerges for reasons unrelated to thermodynamic equilibrium. The tasks of the theory of the QLE state are then to determine when the conditions hold and what is then the relation between the Ising model parameters $h_i(t)$ and $J_{ij}(t)$ (physically time-dependent external magnetic fields and interactions) and evolutionary model parameters.

The denominator $\pf(t)$ eq.(\ref{eICE}) is a normalization, physically a partition function. For simplicity, in the following the time-dependence of $h_i(t)$, $J_{ij}(t)$ and $\pf(t)$ will be suppressed. 

\subsection{QLE in the KNS theory}\label{sQLEKNST}
The authors of \cite{NS-1} investigated in detail a QLE regime in which selection is weak on the time scale of recombination $\sigma\ll r$. Selection-induced epistatic couplings between loci are then weak and steady and can be treated as perturbations. 

The relation to evolutionary parameters can be derived self-consistently by assuming that $\vec{J}$ in eq.(\ref{eICE}) are small in absolute value \emph{i.e.} $|J_{ij}|\ll1$. The empirical correlations can then be calculated by first evaluating the partition function perturbatively for small $|J_{ij}|\ll1$ and then taking the appropriate derivatives (for a derivation, see \cite{dichio2021}):
\begin{equation}
    \chi_{ij} = \frac{\de^2 \log \pf}{\de h_i\de h_j}\sim J_{ij}(1-\chi_i^2)(1-\chi_j^2)\ ,\label{eSOCQLEij}
\end{equation}
The magnetizations $\chi_i$ are not larger than one in absolute value. Hence, if $J_{ij}$ are small in absolute value, then the empirical correlations $\chi_{ij}$ are also small in absolute value.

The relation between evolutionary (dynamic) Ising parameters in general follows from comparing the master equation eq.(\ref{eMEE}) to the time derivative of the distribution $\dot{P}(g)$ using eq.(\ref{eICE}). To this we add the above discussed expansion for small couplings (for details, see \cite{dichio2021}).

Following \cite{NS-1}, we further assume that the mutation rate is small and can be set to zero. In fact, this is a non-trivial simplification since if the mutation rate is exactly zero QLE in an infinite population would only be a long-lived transient as the population drifts towards fixation, see eq.(\ref{eEFOC}). For the fitness function, the parametrization eq.(\ref{eFL}) is used.

The result is that if $P(g)$ is and remains a Gibbs-Boltzmann distribution for an Ising model, the dynamics of the parameters $\vec{h,J}$ must satisfy
\begin{align}
    \dot h_i &= f_i + r\sum_{j\ne i} c_{ij}J_{ij}\chi_j\ ,\label{eDF}\\
    \dot J_{ij} &=f_{ij} - rc_{ij}J_{ij}\ . \label{ePCQLE}
\end{align}
where $f_i$ and $f_{ij}$ are the fitness parameters, $r$ an overall rate of recombination and $c_{ij}$ a quantification of the amount of recombination between loci $i$ and $j$ per generation. In the case where the recombination rate is high $\sigma/r \ll 1$, eq.(\ref{ePCQLE}) is a relaxation which will rapidly reach a steady state:
\begin{equation}
    J_{ij} = \frac{f_{ij}}{r c_{ij}} \label{eCRQLE-1}
\end{equation}
This result is the simplest relation 
between evolutionary parameters ($f_{ij}$, $c_{ij}$ and $r$) and Ising parameters ($J_{ij}$). We note that it is not the case that $J_{ij}$ directly measure epistatic interactions of the fitness function. For $i$ and $j$ sufficiently closely located on the genome the factor $c_{ij}$ will be small, and the steady-state $J_{ij}$ will be large even if the $f_{ij}$ is only of moderate size. This is a special case of the more general fact that closely spaced loci can be in linkage disequilibrium in a recombining population (which holds also outside QLE). Nevertheless, for sufficiently distant loci and sufficiently high rate of recombination, $c_{ij}$ is approximately constant (taking in fact the value one half). Hence for such pairs of loci $J_{ij}$ is approximately proportional to $f_{ij}$, the proportionality being $\frac{2}{r}$. 
 
Substituting the steady-state eq.(\ref{eCRQLE-1}) in eq.(\ref{eDF}), one finds for the first-order Ising parameter
\begin{equation}\label{e-hi}
    \dot h_i = f_i + \sum_{j\ne i} f_{ij}\chi_j \equiv \hat{f_i} \ ,
\end{equation}
where $\hat{f_i}$ is the effective strength of selection on locus $i$. In contrast to eq.(\ref{eCRQLE-1}), the dynamics for $h_i$ is a drift. Unless $\hat{f_i}\approx 0$ and if no other effects set in, $h_i$ will increase towards $+\infty$ or decrease towards $-\infty$ which means that the distribution over alleles at locus $i$ will drift towards fixation. In a finite population this tendency is eventually countered by genetic drift.

As for the dynamics of the first and second order cumulants, they can be understood as follows. In view of eq.(\ref{eSOCQLEij}, \ref{eCRQLE-1}), the off-diagonal second order cumulants rapidly approach the steady state  
\begin{equation}\label{eASOCQLE}
    \bar\chi_{ij} = \frac{f_{ij}}{rc_{ij}}(1-\chi_i^2)(1-\chi_j^2)\ , \quad i\ne j\  .
\end{equation}
The first order cumulants instead evolve according to the following equations:
\begin{equation}\label{eEFOCG}
    \dot\chi_i \stackrel{(a)}{=} \av{s_iF}-\chi_i\av{F} \stackrel{}{=}\de_{h_i}\av{F} \stackrel{(b)}{\sim}\sum_j \de_{\phi_i}\chi_j\de_{\chi_j}\av{F}\stackrel{(c)}{=}\sum_j\chi_{ij}\de_{\chi_j}\av{F}\ ,
\end{equation}
where in $(a)$ we have used eq.(\ref{eEFOC}) with $\mu=0$; in $(b)$ the chain rule of differentiation; in $(c)$ the fact that $\chi_{ij}=\de\chi_i/\de h_j$. We see from the RHS of eq.(\ref{eEFOCG}) that the allele averages evolve so to maximize $\av{F}$, there are $L$ such equations and they are all coupled by the correlations $\chi_{ij}$.

We hence see that this type of QLE with small effective Ising parameters $J_{ij}$ really merits the designation "quasi-linkage equilibrium". The only relevant dynamic equations are the $L$ eq.(\ref{eEFOCG}) which define the $L$-dimensional QLE manifold, which is not identical to linkage equilibrium (LE), because the $J_{ij}$  are non-zero, but which can be put in one-to-one relation to a state in LE. As long as this type of QLE holds, the genotype distribution (hence the population average of any trait) is confined on such manifold and can be parametrized by the set of time-dependent first cumulants $\chi_i(t)$.

\subsection{Inference of epistasis}\label{ss-INF-QLE}
We have shown how eq.(\ref{eCRQLE-1}) opens up an interesting connection between theory and experiments, under the QLE assumption. Indeed, if experimental data on the evolution of a population are available, then leveraging DCA methods described in sec.(\ref{cISF}) it is possible to infer the couplings of the underlying Boltzmann distribution (which holds in QLE). This in turn, through eq.(\ref{eCRQLE-1}), allows to characterize the epistatic interactions and interpret them as resulting from biological genetic expression patterns and constraints induced by the environment.

We now proceed to \emph{in silico} testing of epistasis inference based on eq.(\ref{eCRQLE-1}). The testing strategy is based on the following steps: 

\begin{enumerate}
    \item \textbf{Simulating evolution.} The simulation tool \sml{FFPopSim} \cite{ZN} allows simulations of the evolution of a population of biallelic genomes, based on the master equation eq.(\ref{eMEE}). The output is a time series of snapshots of the evolving population \emph{i.e.} the information on the genomic sequences of all individuals present at each time. As our goal is testing the QLE regime, the evolutionary parameters are instantiated accordingly, see
    tab.(\ref{tNST}). 
    \begin{table}[h]
    \centering
    \begin{tabular}{cccc}\toprule
      &\sml{FFPopSim} & Value  & Description  \\\midrule
    \multirow{3}{*}{\rotatebox[origin=c]{90}{Structure}} 
                    &$N$ & $200$ & carrying capacity \\
                    &$L$ & $25$ & n. of loci \\
                    &$T$ & $2.5\times10^3$ & n. of generations \\
    \multirow{4}{*}{\rotatebox[origin=c]{90}{Drivers}} 
                    & $\omega$ & $0.5$ & crossover rate \\
                    & $r$ \cellcolor{Gray}& $[ 0.0,\ 1.0 ]$ \cellcolor{Gray} & recombination rate\cellcolor{Gray}\\
                    & $\mu$ \cellcolor{Gray}& $[ 0.005,\ 0.1] $ \cellcolor{Gray}& mutation rate\cellcolor{Gray} \\
                    & $\sigma_e$ \cellcolor{Gray}& $[ 0.001,\ 0.02] $ \cellcolor{Gray}&\cellcolor{Gray} $f_{ij}\sim\mathcal{N}(0,\sigma_e)$
        \\\bottomrule
    \end{tabular}
    \caption{Parameters for the QLE simulations in \texttt{FFPopSim}. $N$ is the average size of the population, $L$ is the fixed number of sites for each genome, $T$ is the simulation time and $\omega$ is the crossover rate. These parameters are held fixed. The other (gray) parameters are varied. Namely, the recombination rate $r$, the mutation rate $\mu$ and the standard deviation of the epistatic fitness components $\sigma_e$. A Sherrington-Kirkpatrick (SK) fitness function is postulated where $f_i=0$ and $f_{ij}\sim\mathcal{N}(0,\sigma_e)$ $\forall i,j$.}\label{tNST}
    \end{table}
    \item \textbf{Inferring couplings.} 
    In the simulation all genomes present at a single time are the starting point of the DCA methods discussed in sec.(\ref{cISF}), namely MF and PLM. The output are the inferred couplings $\vec{J^*}$. Because of random drift, unavoidable in finite-size simulations, averages over the population fluctuate in time; in order to cope with fluctuations, the empirical averages for DCA are optionally computed not only on the final state of the population but on the whole time series \textit{e.g.} $\av{s_i}=\frac{1}{NT}\sum_{t=1}^{NT}s_i(t)$. The MF and PLM inference on data obtained from the whole time series are referred to as \emph{alltime}-MF and \emph{alltime}-PLM; unless otherwise specified, this will be implied hereinafter.
    \item \textbf{Testing epistasis.} From the knowledge of $\vec{J^*}$ and of the evolutionary parameters, eq.(\ref{eCRQLE-1}) can be turned around so to become an inference formula
    \begin{equation}
    f_{ij}^* = J_{ij}^*\cdot r c_{ij}\ , \label{eCRQLE}
    \end{equation}
    $f_{ij}^*$ being the inferred epistatic interactions. $c_{ij}$ is computed as in eq.(\ref{eCIJZA}) with $\rho=\omega=0.5$.
    In analogy to eq.(\ref{eREIIP}),
    reconstruction error $\varepsilon$ is quantified as
    \begin{equation}\label{eRMSE}
        \varepsilon = \sqrt{\frac{\sum_{i<j}(f^*_{ij}-f_{ij})^2}{\sum_{i<j}f_{ij}^2}}\ .
    \end{equation}
\end{enumerate}
\begin{figure}
\centering
    \includegraphics[scale=1]{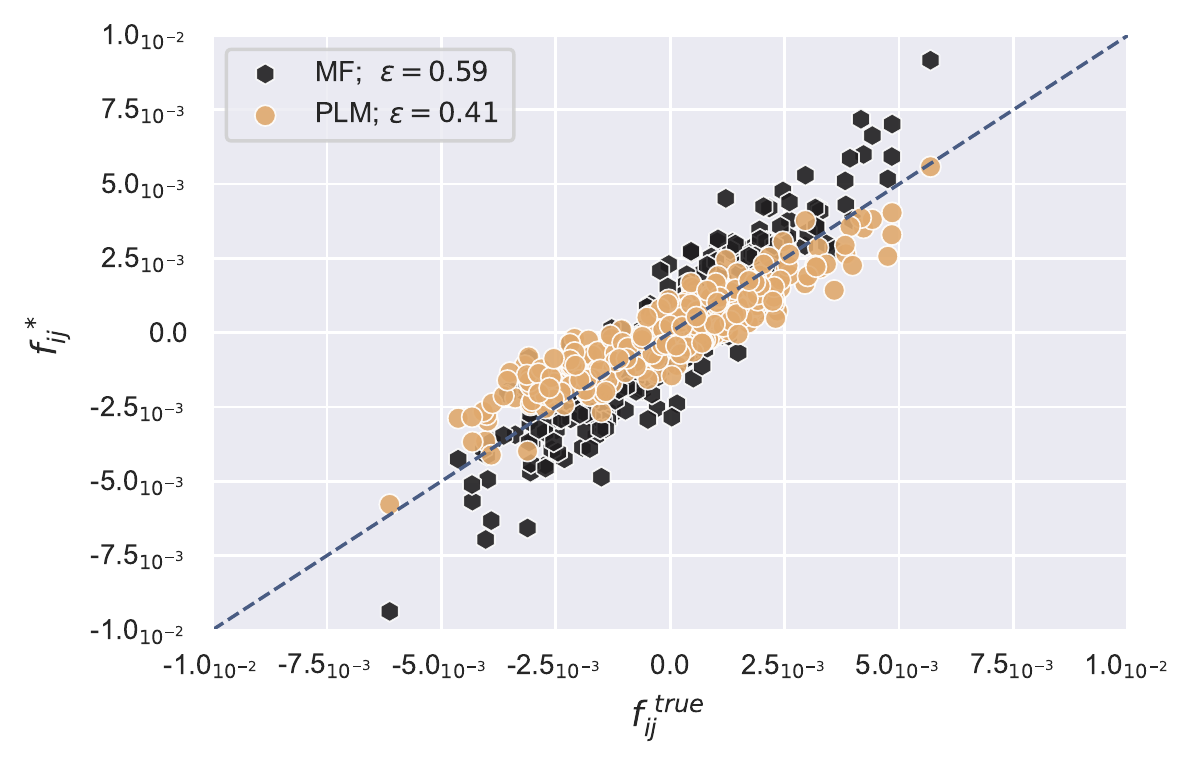}
    \caption{Reconstruction of epistatic parameters of the fitness in QLE. Example of a scatter plot for the reconstructed fitness components $f_{ij}^*$ (y-axis) versus the true parameters $f_{ij}$ (x-axis). Simulation as in Tab.(\ref{tNST}). Here $r=0.5, \mu=0.05, \sigma_{e}=0.002$. The root mean square error $\epsilon$ is the reconstruction error as in eq.(\ref{eRMSE}). Both MF (black) and PLM (brown) are used for the inference procedure, their performances are similar in this QLE regime.}\label{fSP}
\end{figure}

\noindent
Fig.(\ref{fSP}) shows a typical outcome when inference has been successful. The data points lie on a line not too far from the diagonal indicating that $f^*_{ij}$ is a good predictor of $f_{ij}$. The deviation from the diagonal, also visible in the plot, is a signature of systematic deviations indicating that a better theory should be possible. We will discuss one such improved (more general) theory below.

\begin{figure}[ht!]
    \centering
        \includegraphics[width=\textwidth]{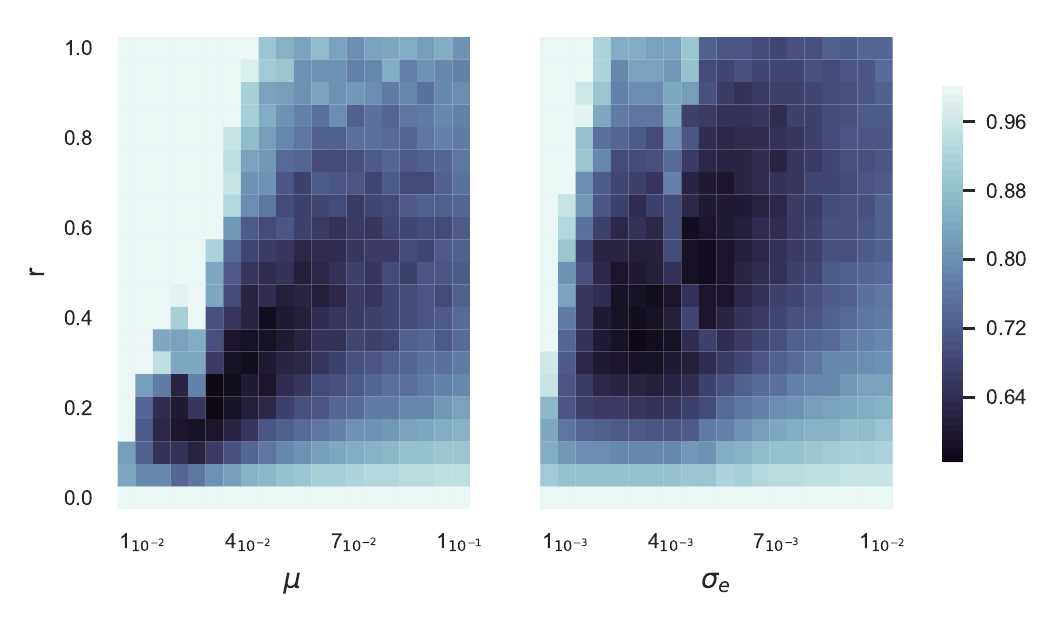}
    \label{sfSIGMAR}
    \caption{Phase diagrams $\mu - r$ and $\sigma_e - r$ for the reconstruction of the epistatic fitness components, from eq.(\ref{eCRQLE}). Simulations as in tab.(\ref{tNST}). Alltime-PLM is used for the inference of couplings $J_{ij}$ from data. Heat-maps based on the reconstruction error $\varepsilon$ as in eq.(\ref{eRMSE}). Similar results can are obtained by using alltime-MF inference. Re-plotted from \cite{ZA-1}.}\label{fNSTPS}
\end{figure}

By repeating the testing procedure for a range of evolutionary parameters, one can explore the performances and limits of validity of the inference strategy. In fig.(\ref{fNSTPS}), the parameter space is explored in the directions $\mu - r$, $\sigma_e - r$. 

A number of observations can be made. First, reconstruction fails for very low mutation rates $\mu$. This can be explained by the structure of the population being essentially frozen and any selection-induced correlation is not reflected in the data (for finite $N,T$). Secondly, inference is not possible for low enough recombination rate $r\sim0$, in which case $f_{ij}^*=J_{ij}^*\cdot rc_{ij}\sim 0$ which implies $\varepsilon\sim 1$ regardless of the inference method employed. On the other hand, reconstruction error increases for higher values of $\mu,r$ as well. In the case of mutation, this is coherent with the assumption of negligible $\mu$ underlying eq.(\ref{eCRQLE}). High recombination rate on the other hand results in weaker couplings $J_{ij}$ inferred from data, which become subject more and more to small-sample noise.

\subsection{Derivation of QLE from a Gaussian ansatz}
\label{sec:Gaussian}

We will now present another derivation of QLE which does not rely on the perturbative analysis of above~\cite{MAU}. It relies instead on a Gaussian ansatz, and a closure of the defining master equations. We will show that it leads to another prediction formula for the epistatic fitness terms which applies in a wider parameter range. 

Let us consider again an evolutionary process in which selection is weak (QLE), due to recombinations and/or mutations happening at a sufficiently fast  pace, \emph{i.e.} $\sigma/r\ll1$ or $\sigma/\mu \ll 1$ (or both). Note that this time we do not constrain $\mu$ to negligible values. Since genomes change by mutations or recombination (or both) no clones -- individuals with the exact same genotype -- will be present in the population at any time $t$.

Consider now the time derivative of the average moments $\av{s_i}$, $\av{s_is_j}$, $\av{s_is_js_k}$ etc. The dynamic equations for these observables are not closed. A \textit{Gaussian closure} ansatz means to evaluate all averages on the right hand side of these equations as if they were taken with respect to a multivariate Gaussian trial function over (fictitious) continuous variables given by
\begin{equation}\label{eGCI} 
    P(g,t) = \frac{1}{\pf}\exp\Bigg[-\frac{1}{2}\sum_{i,j} (s_i-\chi_i)(\chi^{-1})_{ij}(s_j-\chi_j)\Bigg]\ ,
\end{equation}
In above $\pf$ is a normalization and $\chi$ is the covariance matrix \textit{i.e.} $\chi_{ij} = \av{s_is_j}-\av{s_i}\av{s_j}$. 

The dynamic equations for the first and second moments are computed by evaluating the various terms on the right hand side of the master equation for $P(g)$ multiplied by $s_i$ and $s_is_j$ and averaged over the trial distribution eq.(\ref{eGCI}). From them, those for the first and second cumulants follow:

\begin{alignat}{2}
    \dot\chi_i &\stackrel{}{=}&& 
    \sum_j f_{j}\av{s_is_j} + \sum_{j<k} f_{jk}\av{s_is_js_k} - \sum_j f_j \chi_i\chi_j - \sum_{j<k} f_{jk} \chi_i\av{s_js_k}-2\mu\chi_i \label{eFOCP}\\ 
    \dot\chi_{ij} &\stackrel{}{=}&& 
    \sum_k f_{k}\av{s_is_js_k} + \sum_{k<l} f_{kl}\av{s_is_js_ks_l} - \sum_k f_{k}\chi_k\av{s_is_j} - \sum_{k<l} f_{kl}\av{s_is_j}\av{s_ks_l}\ + \notag\\ & && - \chi_i(\dot\chi_j+2\chi_j\mu)- \chi_j(\dot\chi_i+2\chi_i\mu) - (4\mu+rc_{ij})\chi_{ij}\ ; \label{eSOCP} 
\end{alignat}
in the latter, valid for $i\ne j$, we have used eq.(\ref{eEFOC}) and left implicit $\dot\chi_i$ . 

In the general case, evaluating the expectations $\av{s_is_js_k}$ and $\av{s_is_js_ks_l}$ requires the knowledge of higher order cumulants \emph{i.e.} $\chi_{ijk}, \chi_{ijkl}\dots$ However, under the Gaussian ansatz eq.(\ref{eGCI}), all $3$- and $4$-point moments can be expressed in terms of first and second order cumulants $\chi_i,\chi_{ij}$. In this sense, the Gaussian ansatz is a  \emph{closure} (GC) for the eq.(\ref{eFOCP}-\ref{eSOCP}). The resulting dynamics expressed only in terms of the first and second cumulants is \cite{dichio2021}

\begin{align}
    \dot\chi_i =& \sum_j\chi_{ij}\Big(f_j+\sum_kf_{jk}\chi_k-2f_{ij}\chi_i\Big) -2\mu\chi_i\ , \label{eEFGC} \\
    \dot\chi_{ij} =& -(4\mu + rc_{ij})\chi_{ij} - 2\chi_{ij}(f_i\chi_i + f_j\chi_j) + 2f_{ij}\chi_{ij}(\chi_{ij}+2\chi_i\chi_j)\ + \notag \\
    &-2\chi_{ij}\sum_k\Big[f_{ik}(\chi_{ik}+\chi_i\chi_k) +f_{jk}(\chi_{jk}+\chi_j\chi_k)\Big]+\sum_{k,l}f_{kl}\chi_{ik}\chi_{jl}\ .\label{eESOCVC}
\end{align}

Note that, given the Gaussian ansatz in eq.(\ref{eGCI}) and the evolutionary parameters, these are \emph{exact} equations and fully determine the dynamics of the probability distribution in a $L(L+1)/2$-dimensional space.

Similarly to sec.(\ref{sQLEKNST}), we note that at large values of recombination and/or mutation rates, the dynamics of the $\chi_{ij}$ in eq.(\ref{eESOCVC}) rapidly reach a steady state. In order to understand this quantitatively, a systematic expansion for small $\epsilon = 1/(4\mu+rc_{ij})\rightarrow 0^+$ can be carried out. This is done in \cite{ZM} by assuming: 
\begin{equation}\label{ePESOC}
    \chi_{ij} = \chi_{ij}^{(0)}+\epsilon\chi_{ij}^{(1)}+\epsilon^2\chi_{ij}^{(2)}+\epsilon^3\chi_{ij}^{(3)}+\mathcal{O}(\epsilon^4)
\end{equation}
and imposing $\dot\chi_{ij}=0$ in eq.(\ref{eESOCVC}), by pairing terms corresponding to the same order $\epsilon^n$. Up to the first order $\chi_{ij}^{(1)}$, one finds
\begin{equation}
    \chi_{ij} = \frac{f_{ij}}{4\mu+rc_{ij}}(1-\chi_i^2)(1-\chi_j^2)\ ,\label{eIFLGCA} 
\end{equation}
which is valid under the condition that $L\sigma(f) < 1$ \cite{dichio2021}. In \cite{ZM} the authors also show how it is possible to get the same result by generalizing eq.(\ref{eCRQLE-1}) to the case where mutations are not negligible and interpreting eq.(\ref{eSOCQLEij}) as a DCA inference method. 

Note however eq.(\ref{eEFGC}-\ref{eESOCVC}) have a number of advantages. In the first place, they allow for a direct characterization of the dynamics of the whole probability distribution, which is fully determined by those of the cumulants (known explicitly). Moreover, they can be formally studied well outside the expansion eq.(\ref{ePESOC}), under more specific conditions. 

In \cite{MAU} for instance the authors investigate a QLE phase for a model in which the fitness function has two competing maxima and correlations decay exponentially with the distance along the genome. The $r-\mu$ phase diagram can be studied analytically and a transition is found from a paramagnetic phase (low $\mu$) - in which the genomes in the population are broadly distributed and encompass the two fittest genomes - to a ferromagnetic regime (high $\mu$) - in which one of the two maximally fit genomes eventually takes over.

\subsection{Broader, easier inference of epistasis}\label{sGCUE}
Let us consider again the goal of inferring the epistatic interactions of the fitness function from the observation of an evolutionary process. Turning around eq.(\ref{eIFLGCA}) into an inference formula, we have:
\begin{equation}
    f_{ij}^* = \chi_{ij}\cdot\frac{4\mu+rc_{ij}}{(1-\chi_i^2)(1-\chi_j^2)}\ ,\label{eIFGC} 
\end{equation}
 
There are two major advantages of this latter formula with respect to eq.(\ref{eCRQLE}). In the first place, it accounts for an arbitrary non-zero mutation rate. In the second place, it relates the epistatic interactions directly to the cumulants $\chi_i, \chi_{ij}$, without any DCA-based intermediate step, making the computation much less demanding.

The test of eq.(\ref{eIFGC}) versus eq.(\ref{eCRQLE}) is performed in \cite{ZM} along a similar strategy to the one outlined in sec.(\ref{ss-INF-QLE}). In particular, the epistatic interactions $f_{ij}$ are inferred from the \emph{all-time} averages $\chi_i,\chi_{ij}$ in two different ways. On the one hand, directly from eq.(\ref{eIFGC}), under the Gaussian ansatz (GA). On the other hand, they are calculated according to the KNS formula eq.(\ref{eCRQLE}) by first reconstructing the couplings $\vec{J}$ through standard DCA methods (MF or PLM). In both cases, the reconstruction errors are defined as in eq.(\ref{eRMSE}).

\begin{figure}[htbp]
    \centering
    \includegraphics[width=\textwidth]{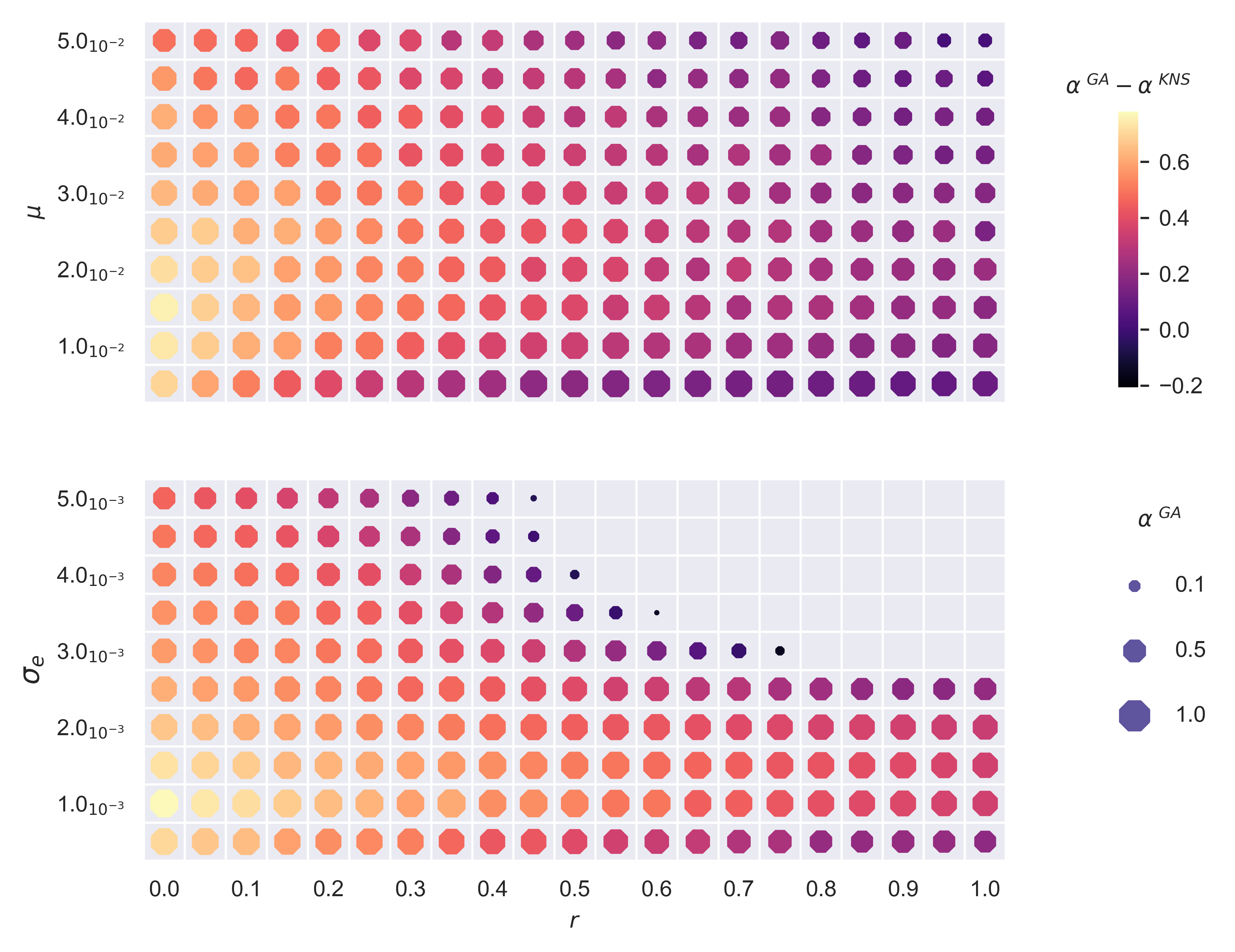}
    \caption{Reconstruction of the epistatic fitness components in the phase spaces $r-\mu$ (fixed $\sigma_e=0.004$) and $r-\sigma_e$ (fixed $\mu=0.2$).
    The size of each dot represents the accuracy $\alpha = 1-\varepsilon$ of the reconstruction based on the gaussian ansatz GA, eq.(\ref{eIFGC}); $\varepsilon$ is the reconstruction error. Bigger dots are then found where the GA reconstruction has greater accuracy (smaller error). The colormap is the difference of between the GA accuracy and the one based on the KNS reconstruction, eq.(\ref{eCRQLE}). Violet-yellow dots are found where the GA-reconstruction outperforms the KNS counterpart. Both equally fail ($\alpha\sim0, \varepsilon\sim1$) in the upright part of the plot below. Other parameters as in tab.(\ref{tSTP}). Re-plotted from \cite{ZM}.}
    \label{fKNSvsGA}
\end{figure}

The parameters for the evolutionary simulations are summarized in tab.(\ref{tSTP}). Note in particular that compared to tab.(\ref{tNST}), a broader range of mutation rates $\mu$ is tested. Moreover, a fitness landscape is chosen for which not only epistatic components $f_{ij}$ but also additive ones $f_i$ are non zero, implying locus-specific evolutionary (dis)advantages. Two sets of simulations are performed. In fig.(\ref{fKNSvsGA}) we plot the accuracy $\alpha = 1-\varepsilon$ for the reconstruction based on eq.(\ref{eIFGC}) ($\varepsilon$ is the reconstruction error) and compare it with the one corresponding to the KNS reconstruction.

   \begin{table}[h]
    \centering
    \begin{tabular}{cccc}\toprule
      &\sml{FFPopSim} & Values  & Description  \\\midrule
    \multirow{3}{*}{\rotatebox[origin=c]{90}{Structure}} 
                    &$N$ & $200$ & carrying capacity \\
                    &$L$ & $25$ & n. of loci \\
                    &$T$ & $10,000$ & n. of generations \\
    \multirow{5}{*}{\rotatebox[origin=c]{90}{Drivers}} 
                    & $\rho$ & $0.5$ & crossover rate \\
                    & $\sigma_a$ & $0.05$ & $f_{i}\sim\mathcal{N}(0,\sigma_a)$\\
                    & $r$ \cellcolor{Gray}& $[ 0.0,\ 1.0 ]$ \cellcolor{Gray} & outcrossing rate\cellcolor{Gray}\\
                    & $\mu$ \cellcolor{Gray}& $[ 0.05,\ 0.5] $ \cellcolor{Gray}& mutation rate\cellcolor{Gray} \\
                    & $\sigma_e$ \cellcolor{Gray}& $[ 0.004,\ 0.04] $ \cellcolor{Gray}&\cellcolor{Gray} $f_{ij}\sim\mathcal{N}(0,\sigma_e)$
        \\\bottomrule
    \end{tabular}
    \caption{\texttt{FFPopSim} evolutionary parameters for the test KNS vs GC under QLE conditions \cite{ZM}. $N$ is the average size of the population, $L$ is the fixed number of sites for each genome, $T$ is the simulation time and $\omega$ is the crossover rate, $\sigma_a$ is the standard deviation of the additive components of the fitness function. These parameters are held fixed. The other (gray) parameters are varied. Namely, the recombination rate $r$, the mutation rate $\mu$ and the standard deviation of the epistatic fitness components $\sigma_e$. The fitness function $F(g) = \sum_if_is_i+\sum_{i<j}f_{ij}s_is_j$ is used, where both the additive and epistatic components are gaussian distributed around zero: $f_i\sim\mathcal{N}(0,\sigma_a)$, $f_{ij}\sim\mathcal{N}(0,\sigma_e)$ $\forall i,j$.}   \label{tSTP}
    \end{table}

In the first set, fig.(\ref{fKNSvsGA}) top row, the parameter space is explored in the directions $r-\mu$, while holding fixed $\sigma_e=0.004$. Reconstruction based on eq.(\ref{eIFGC}) has higher accuracy (lower error), as expected for high mutation rates. Similarly to what has been discussed in sec.(\ref{ss-INF-QLE}), for too high recombination and mutation rates (top-right corner), the resulting noise worsens the accuracy of the (weak) empirical allele statistics for a finite-time simulation, and by consequence the performances of both the reconstruction strategies. 

In the second set, fig.(\ref{fKNSvsGA}) bottom row, the parameter space is explored in the directions $r-\sigma_e$, while holding fixed $\mu=0.2$. The reconstruction based on the gaussian ansatz, accounting for non-negligible mutation rates, outperforms almost everywhere the KNS reconstruction. The accuracy of the epistatic reconstruction however decreases for increasing $\sigma_e$, which can be explained in view of the assumption $L\sigma_e<1$ made in the derivation of eq.(\ref{eIFGC}). The failure of the GA inference for high $\sigma_e,r$ (top-right corner) is instead due to a breakdown of the QLE phase, as we will discuss in the next section.

As a final note, for the Gaussian ansatz to hold, the $\{f_i\}$ also have to be small. Increasing the overall magnitude of additive terms of the fitness function, the evolutionary process has a stronger tendency to drive some alleles to fixation. The minor allele at such loci will then be present in only a few copies in a finite population, and often not be present at all. At the genome scale, the chances of observing several copies of the same genotype (clones) will increase. The actual distribution then differs from the one assumed by the Gaussian ansatz, making this inaccurate.

\subsection{Multi-genome factorization}
\label{sec:multi-genome factorization}
We now turn to the assumption of multi-genome factorization,
exact or approximate. For two genomes, as used above,
this means 
$P_2(g_1,g_2,t)\approx P_1(g_1,t)\cdot P_1(g_2,t)$.
We note that this is the assumption of molecular chaos 
(Stosszahlansatz) in the theory of gases. 
We further note that in the detailed derivation 
in \cite{NS-1} and \cite{Gao2019}, also given in \cite{dichio2021},
there appears an equation quadratic in $P(g,t)$ on the right-hand
side, analogous to the collision term in the Boltzmann equation.
We can therefore discuss the limitations 
of the multi-genome factorization assumption 
borrowing from the language of gas dynamics.

First, direct test of factorization of probability distributions
from data is hard because the number of data required is very large.
One has to go for indirect tests.
Those are either of the type if predictions obtained
from a factorization assumption (and other assumptions) hold in the
data, or from general arguments about the evolutionary process.
As to the first, successful
inference of evolutionary parameters from snapshots of the population
distribution in some parameter regimes is evidence that the assumptions
behind these inference formulae hold, and those include
two-genome factorization.

Turning to general arguments factorization of the probability distributions
of two particles entering a collision holds if they arrive from
afar, and do not share a common history. In statistical
genetics this translates to two genomes recombining not having
a common ancestry. In the numerical models used in this review
pairs of genomes are picked uniformly random (random mating).
In real populations the two main obstacles to random
mating are physical isolation and sexual selection.
The first refers to that 
two individuals can only mate if they actually meet. 
Every individual the genome of which has a chance to survive
hence belongs to a group where mating happens frequently enough,
and that group is or is not in contact with other groups 
with which mating 
is less frequent. Evolutionary models of this kind
in population biology
are the \textit{island models} cf.\cite{BMK}.
and \textit{stepping-stone models} \cite{KimuraWeiss1964}.
With this type of obstacle
to random mating each group can be considered separately
and it may conceivable be that for the same species,
a QLE phase is found in some groups, but not in some others.
Sexual selection means that pairs of individuals 
mate more or less frequently depending on the two
genomes and especially on how similar they are.
In the numerical tests in this review no such 
effects are taken into account, but 
the consequences for the relation between 
population dynamics paraments and distributions
where discussed in \cite{Gao2019}, within the context
of assuming a QLE phase in the terms developed in \cite{NS-1}.
It will be an interesting task for the future to assess
whether multi-genome factorization is present in the presence
of sexual selection by using the formulae derived in \cite{Gao2019}.

Coming back to the main topic of this review, whether factorization of multi-genome 
distributions holds depends not only if mating is random,
but also on the overall relative strength of recombination 
compared to mutations and selection, as discussed throughout this review.

%%%%%%%%%%
\section{QLE breakdown}\label{cVI}
So in statistical physics as in its applications to population genetics, when stepping beyond the equilibrium approximation a rich variety of new behaviours emerges. Their theoretical understanding is, however, more challenging.  In a larger perspective, a number of mathematical models have been proposed to tackle different non-equilibrium problems of population genetics. In models without recombination there is no interchange of genetic material between individuals. The evolution of the distribution of genotypes in a population is then a stochastic process, and a rich set of tools can be brought to bear. Time reversal of stochastic processes  is the basis of Kingman's coalescent which allows to estimate properties of genealogies \cite{kingman1982a,kingman1982b}; a theory with many later developments, see \textit{e.g.} \cite{mohle1994,chang1999,carinci2015}. In the forward dynamics, in a phase quite far from QLE, the competition between clones and competitions between mutations in different clones have been studied extensively \cite{Park2007,Fogle2008}, as have effects of fast adaptation \cite{Brunet2008} and time-changing fitness \cite{Mustonen2009}. A comprehensive review of them is beyond the scope of this review, where instead we focus on when and how a QLE phase breaks down, how to characterize transient phases, and which other dynamical states can be reached.  The reader interested in different approaches can find further useful entries in  \cite{sella2005,manrubia2021,Lassig,neher2013a} and references therein.  

\subsection{The role of drift}\label{ssMR}
Genetic drift - \emph{i.e.} random fluctuations of the population statistics due to its finite size $N<\infty$ - plays a fundamental role in the loss of a QLE phase. In the KNS theory it can be taken into account by adding noise terms to the dynamics of the appropriate observables, for instance eq.(\ref{eEFOC},\ref{eSOC}) in the KNS theory of sec.(\ref{sMNST}) \cite{NS-1}.

Rather than the cumulants, a more transparent observable to understand the role of fluctuations is the empirical fitness distribution \emph{i.e.} the histogram of the values $F(g)$ for each genome $g$ in a population at time $t$. The greatest interest is in the fittest individuals in the population, as they affect the most the average fitness $\av{F}$. By consequence, rare events involving the latter can potentially change the fate of the whole population, under appropriate conditions. A formal understanding exists in simple scenarios, from which however some general lessons can be learnt.

\begin{figure}
    \begin{center}
    \includegraphics[width=.9\textwidth]{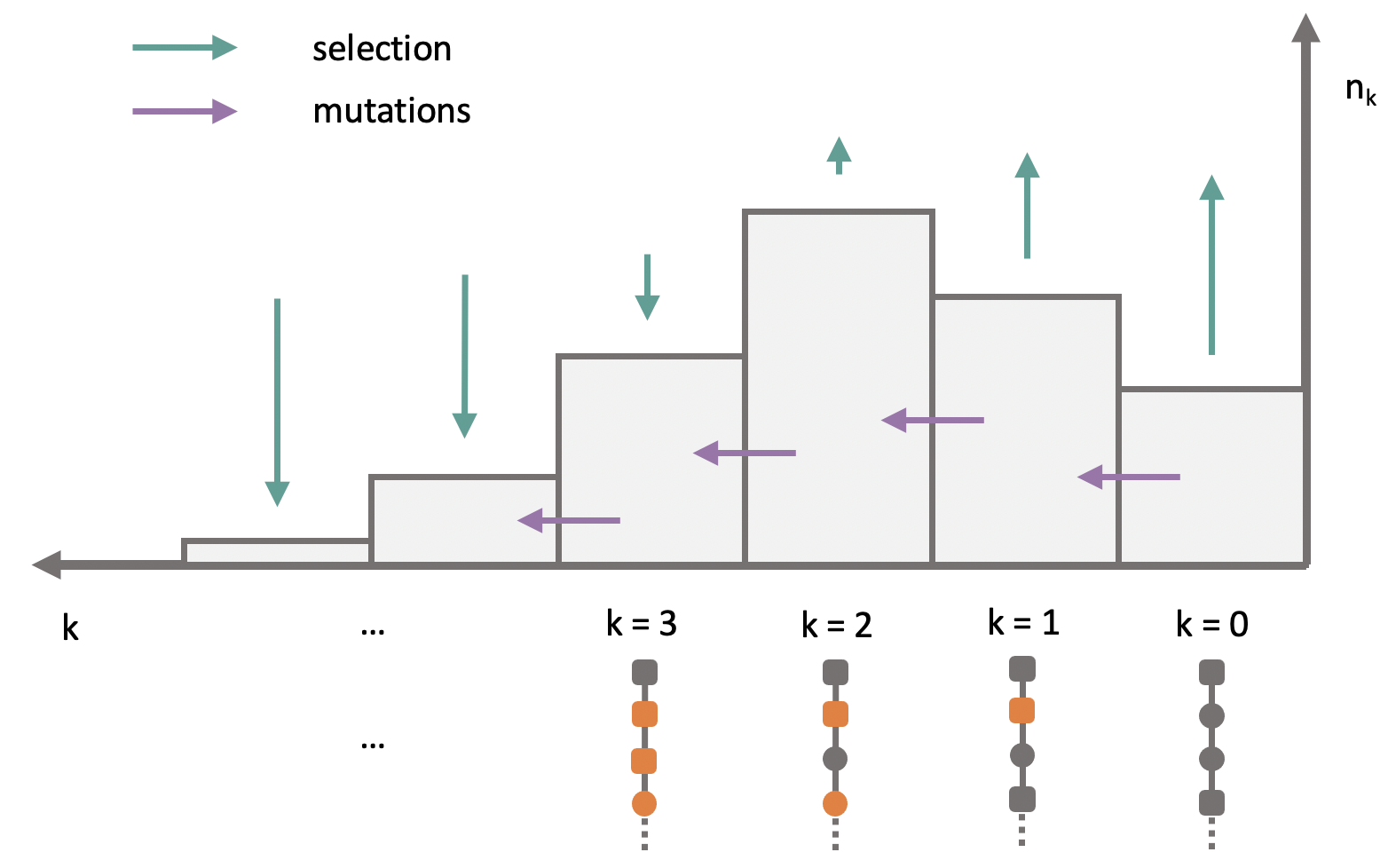}
    \end{center}
    \caption{Deleterious mutation–selection balance. The population is distributed among classes of individuals carrying $k$ deleterious mutations. Classes with few mutations grow due to selection (green arrows), but lose individuals through mutations (violet arrows), while classes with many mutations are selected against but replenished by mutations. For illustration purposes, a cartoon of a  genotype is also sketched, rounds and squares stand for the two possible states of each Ising locus, mutations with respect to the most fit class ($k=0$) are highlighted in orange ($k=1,2,3\dots$).}\label{f-selmut}
\end{figure}

In \cite{NS-4}, the authors consider the following model: $N$ individuals of an asexual population ($r=0$), grouped into discrete classes, each characterized by the number $k$ of deleterious mutations, happening at rate $u$ and causing a fixed fitness loss $s \ll 1$, fig.(\ref{f-selmut}).
The equation 
\begin{equation}\label{e-fit-class}
    \frac{d}{dt}n_k = s(\bar{k} - k)n_k-un_k+un_{k-1}+\sqrt{n_k}\eta_k
\end{equation}
drives the stochastic evolution of $n_k$ = number of individuals in the $k$-th class, $\eta_k$ is the noise term (drift). 
The fitness distribution travels toward higher fitness values because of selection promoting individuals fitter than the average $\bar{k}$, while at the same time being pushed back by random deleterious mutations.

As for the fate of the class $n_0$ (fittest class), the picture is the following: the dynamics of $n_{i}$ for $i\ge1$ is slaved to the stochastic trajectory of $n_0(\tau)$. The effect of the fluctuations of the fittest class on those of the mean, calculated as the cross correlation $\av{\delta n_0(0)\delta \bar{k}(\tau)}$, is  delayed by a time of order $\sim s^{-1}\log\lambda$, with $\lambda=u/s$. The latter, in turn, generate a delayed restoring force opposing the fluctuation of $n_0$. All fluctuations are controlled by the factor $(Ns)^{-1}$.

Understanding the dynamics of fluctuations is a general problem in population genetics; depending on the population size, it can have profound implications for its evolutionary behaviour. An important application of these insights is that to the problem of the Muller's ratchet \cite{Muller,felsenstein1974,charlesworth1997}: a \emph{click} of Muller's ratchet is the loss of the most fit class of individuals; the rate of the ratchet is given by the inverse of the mean time between successive clicks of the ratchet. 
In \cite{NS-4}, starting from eq.(\ref{e-fit-class}) and implementing the results outlined above, the time $t_{click}$ between two clicks of the ratchet is approximated as:
\begin{equation}\label{e-ratch}
    t_{click}=\frac{2.5\zeta(\lambda)}{\alpha(\lambda)s\sqrt{Nse^{-\lambda}}}e^{Ns\alpha(\lambda)e^{-\lambda}}\ ,
\end{equation}  
where $\alpha(\lambda)\sim\mathcal{O}(1)$ and $\zeta(\lambda)\sim \log{\lambda}$. Repeated clicks of the Muller's ratchet lead to the accumulation of deleterious mutations; if the selection is too weak, this inevitably leads to the degradation of each genotype, sometimes referred as decay paradox \cite{loewe2006}. In the absence of recombination or epistasis \cite{kondrashov1994,bell1988}, the only way to escape the mutational meltdown of a population is the appearance of beneficial mutations (adaptation) \cite{schultz1997,Desai}. %
An extension of the model above to the case where beneficial mutations are allowed can be found in \cite{goyal2012}. For any population size $N$ and total mutation rate $U$, a mutation–selection balance exists \emph{i.e.} an appropriate influx of beneficial mutations which is able to counterbalance deleterious ones. 

The study of the interplay between random beneficial/deleterious mutations, selection strength and population size is a general, fundamental problem \cite{desai2007}. Notably, the appearance of a particularly fit individual or group of individuals -- in this case, due to beneficial mutations -- can lead to the emergence of clones \emph{i.e.} a set of individuals with the same genotype. The rise of large clones changes dramatically the structure of the population, inducing strong correlations between different sites of the genome, hence causing the breakdown of a QLE state. This is an example of a transition to a clonal-competition (CC) regime, that we discuss in the next section.

\subsection{The emergence of clones}\label{ss-em-clo}
\begin{figure}[htbp]
\centering
    \includegraphics[width=\columnwidth]{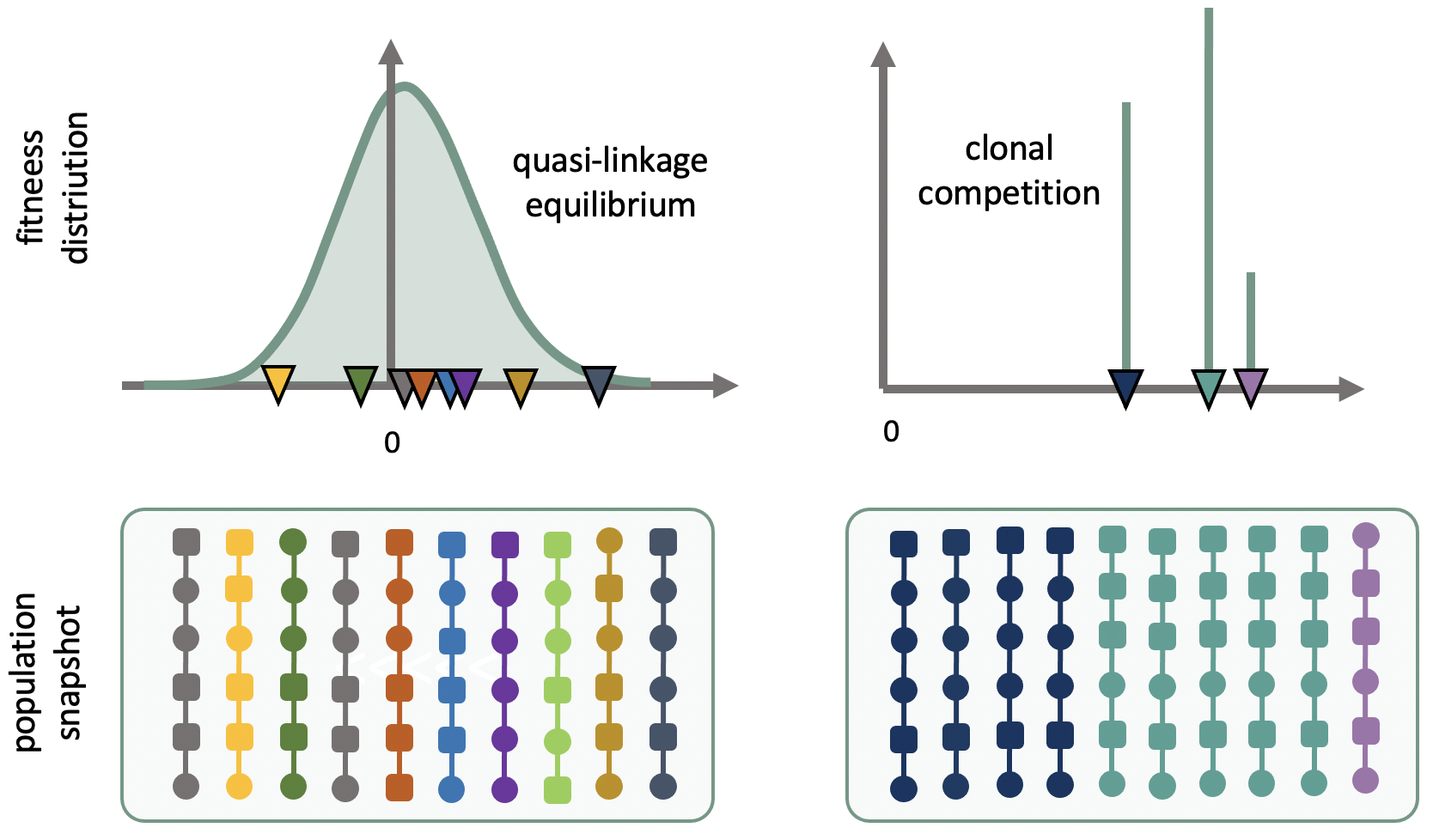}
    \caption{Quasi-linkage equilibrium  vs clonal competition (right). A population of $N=10$ individuals with genotype of length $L=6$ is represented, each chain is an individual, squares and circles represent the two possible site states. In a quasi-linkage equilibrium state (left), individuals with the same genotype are rare and the fitness distribution is broad. In a clonal competition regime (right), few different genotypes are present in the population, each of them characterising a number of individuals (clones).}
  \label{fETEG1}
\end{figure}

The CC transition (or clonal condensation) consists in the emergence of the large clones which compete between each other and spread into the population, see fig.(\ref{fETEG1}) for an illustration. Since in the limit of large mutation/recombination rates a QLE phase exists, this transition can be expected to happen at low enough $r,\mu$ with respect to selection strength.

A suitable order parameter to describe this transition is the probability $Y=\av{\delta(\ \norm{g-g'})}$ that two random individuals in the population have the same genotype. The quantity $Y$ plays an analogous role to the Parisi order parameter $\overline{P(q)}$ (average overlap) in spin glass theory \cite{MezardMontanari}. In the CC phase one expects $Y\sim\mathcal{O}(1)$, while in a QLE phase $Y\sim0$. 

A quantitative understanding of the CC transition was attained in \cite{NS-3} for the case in which the mutational influx is negligible, $\mu\sim0$, and recombination alone fuels variability in the population. This analysis is in turn built on the phenomenological considerations of \cite{NS-2}. This scenario is relevant \emph{e.g.} when two diverged strains of a population are merged together, the hybridization being driven solely by recombination events by individuals from the two strains. 

The model in \cite{NS-3} is the following. A generic fitness function is represented as $F=A+E$, where $A$ is an additive part, corresponding to the 1-spin terms, inherited in a recombination event, and an epistatic contribution $E$. The latter corresponds to 2-spin and higher order interactions, lost upon recombination. When two individuals recombine, the additive fitness value of the offspring is drawn from $\sim\mathcal{N}(\av{A},\sigma_A)$ while its epistatic fitness term is drawn from $\sim\mathcal{N}(0,\sigma_E)$. The governing master equation reads
\begin{equation}\label{e-mod-ns}
    \dot{P}(A,E) = (F-\av{F}-r)P(A,E)+\frac{r}{2\pi\sigma_E\sigma_A}e^{-\frac{(A-\av{A})^2}{\sigma_A^2}-\frac{E^2}{2\sigma_E^2}} \ .
\end{equation}
The parameters of interest are: the elapsed time $t$, which plays the role of the inverse temperature; the ratio $r/\sigma$, which quantifies the strength of the recombination with respect to the selection pressure $\sigma^2=\sigma_A^2 +\sigma_E^2$; the heritability $h$, defined by 
\begin{equation}
    h^2 = \frac{\sigma_A^2}{\sigma_A^2+\sigma_E^2} \ ,
\end{equation}
which encodes the structure of the fitness function, from  purely epistatic $(h=0)$ to purely additive $(h=1)$. 

When $r/\sigma\gg1$ is sufficiently large, eq.(\ref{e-mod-ns}) admits a factorized QLE solution $P(A,E)=\theta(A,t)\omega(E)$ where
\begin{equation}
    \theta(A,t)=\frac{e^{-\frac{(A-\sigma_At)^2}{2\sigma_A^2}}}{\sqrt{2\pi\sigma_A^2}}\ , \quad \omega(E) = \frac{r}{r+\av{E}-E}\frac{e^{-\frac{E^2}{2\sigma_E^2}}}{\sqrt{2\pi\sigma_E^2}}\ .
\end{equation}
In words, the solution has a steady epistatic fitness distribution while it travels towards higher additive fitness values with velocity $\sigma_A$. 
This kind of behaviour is also seen in a full description of QLE phase, if one considers the transient state after \emph{e.g.} the birth of a single beneficial mutation. Pairwise statistics are then stationary while single-site statistics drift towards fixation. For a discussion and possible observation in very large SARS-CoV-2 genomic data sets, see \cite{Zeng-COVID19-2}.

The travelling solution breaks down when there are individuals with epistatic fitness value $E>r+\av{E}$, beyond this threshold the population has not anymore a QLE-like distribution but is made of growing clones, the sizes of which are strongly fluctuating (random), and which depend on when and where they appeared in the population. Therefore, the breakdown of the QLE solution depends on the value of $r$, and a CC phase sets in for sufficiently low recombination rates. The details of this transition however depend of the structure of the fitness function, as quantified by $h$.

In the purely epistatic case $h=0$, the mean fitness increases until a balance is reached between the selection of the fittest genotype $E_{max}-\av{E}$ and the recombination rate $r$. 
The CC phase $Y_t(r,0)\sim\mathcal{O}(1)$ is found for $r<r_c\sim \sigma\sqrt{2\log N}$ and $$t>t_c(r)=\sigma^{-1}\sqrt{\frac{1}{2}\log N}\cdot\frac{r_c}{r-r_c}$$
Since there is no heritability, the dynamic of the population is the one of a "record process" \cite{Krug2007}: when a new fitter genotype appears in the population, it can grow until replacing the previous record holder; the more time goes on, the lower the chances of seeing such a record replacement.

In the case in which $h>0$, at low $r$ the population is dominated by few large clones whose fitness function is partly non-heritable. None of them overwhelms the population but rarely they are overtaken by new fit clones appeared by chance, the more frequently the higher the parameters $h$ and $r$. At the same time, since the epistatic contributions to fitness are lost upon recombination, the population has a tendency to partition in a fraction of fit clones and a cloud of other individuals with random epistatic fitness values. In this case, one finds $Y_t(r,h)\sim\mathcal{O}(1)$ when $r<r_c = \sigma(\sqrt{2}-\sqrt{\gamma})\sqrt{2\log N}$, where $\gamma=v/\sigma^2$ is the ratio between the velocity with which the population moves toward higher fitness values (in the purely additive case, $v=\sigma_A^2$, in general $v<\sigma_A^2$), and the selection pressure. 

\subsection{Transition to non-random coexistence (NRC)}
We will here describe another loss of stability of QLE, not towards a CC phase but towards a new phase of non-random coexistence (NRC). For this stability the population size ($N$) is an essential parameter. Qualitatively speaking, we consider a regime characterized by a high mutation and recombination rates $\mu,r$ yet at the same time strong selection $\sigma_e$. For concreteness and comparison to simulations, the fitness function is taken to be the Sherrington-Kirkpatrick form from spin glass theory:
\begin{equation}\label{e-SKFF}
    F(g)=\sum_{i<j}f_{ij}s_is_j\ ,
\end{equation}
where the $f_{ij}$ are independent draws from a Gaussian distribution with mean zero and standard deviation $\sigma_e$. For completeness, in eq.(\ref{e-SKFF}) the conventions $f_{ii}=0$ and $f_{ij}=f_{ji}$ hold for all diagonal and off-diagonal elements. 

\subsubsection{Hallmarks of the NRC}\label{sHNRCP}

\begin{figure}[ht!]
      \begin{tabular}[t]{c}
          \begin{subfigure}[h!]{\columnwidth}
            \centering
            \includegraphics[width=\textwidth]{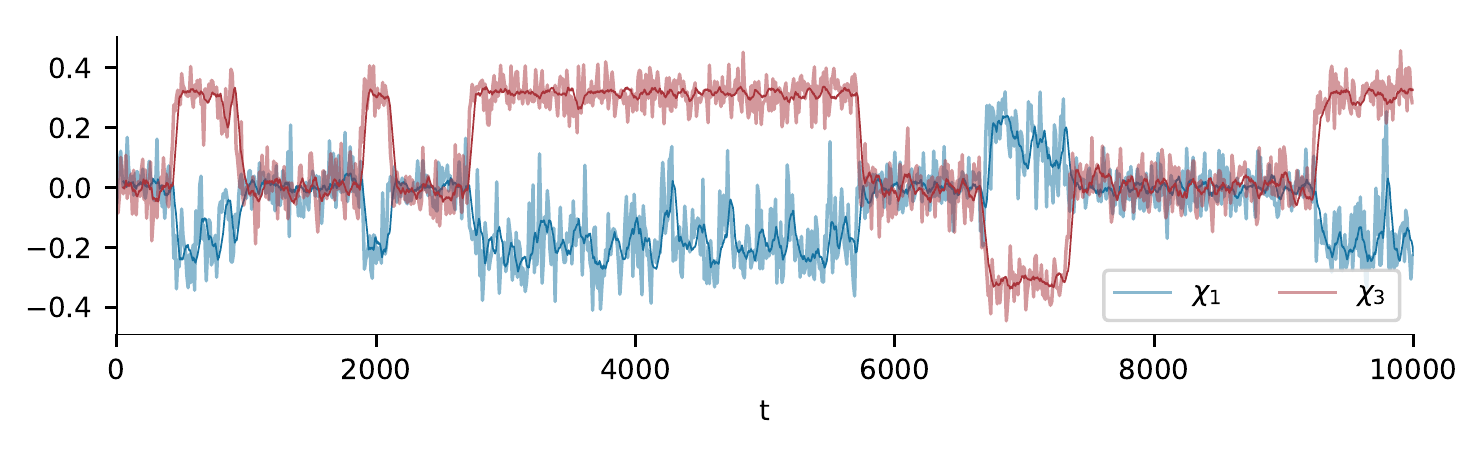}
            \caption{Dynamics of first order cumulants $\chi_i=\av{s_i}$ (average over the population) in the intermittent regime QLE-NRC. Due to the $g\rightarrow-g$ symmetry of the fitness function, two kind of trajectories are observed, mirrored with respect to zero - examples here are $\chi_1,\chi_3$. Solid blue/red lines represent the average over $[t-\Delta t,t]$, with $\Delta t = 50$. The NRC is characterized by $|\chi_i|\sim\alpha$ for some $0<\alpha<1$.}
            \label{fFOC-NRC}
          \end{subfigure}\\
            \begin{subfigure}[t]{\columnwidth}
            \centering
            \includegraphics[width=\textwidth]{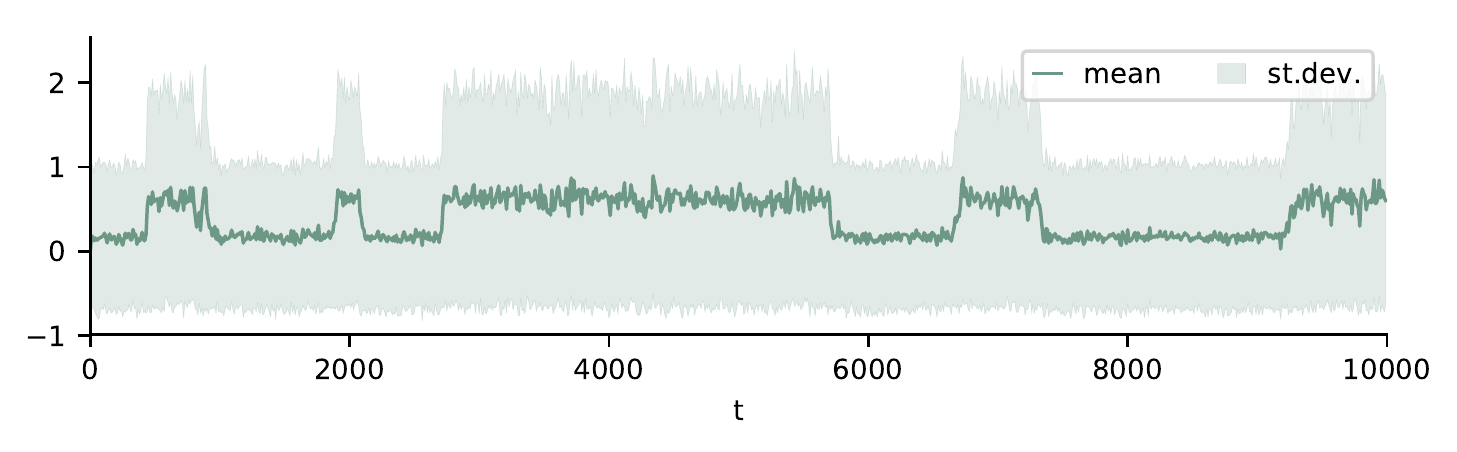}
            \caption{Dynamics of the total fitness mean and st.dev. over the population. An NRC regime is characterized by an increase of the fitness mean and st.dev. in the population with respect to the QLE values.}
            \label{fFS-NRC}
            \end{subfigure}\\
            \begin{subfigure}[t]{\columnwidth}
            \centering
            \includegraphics[scale=.75]{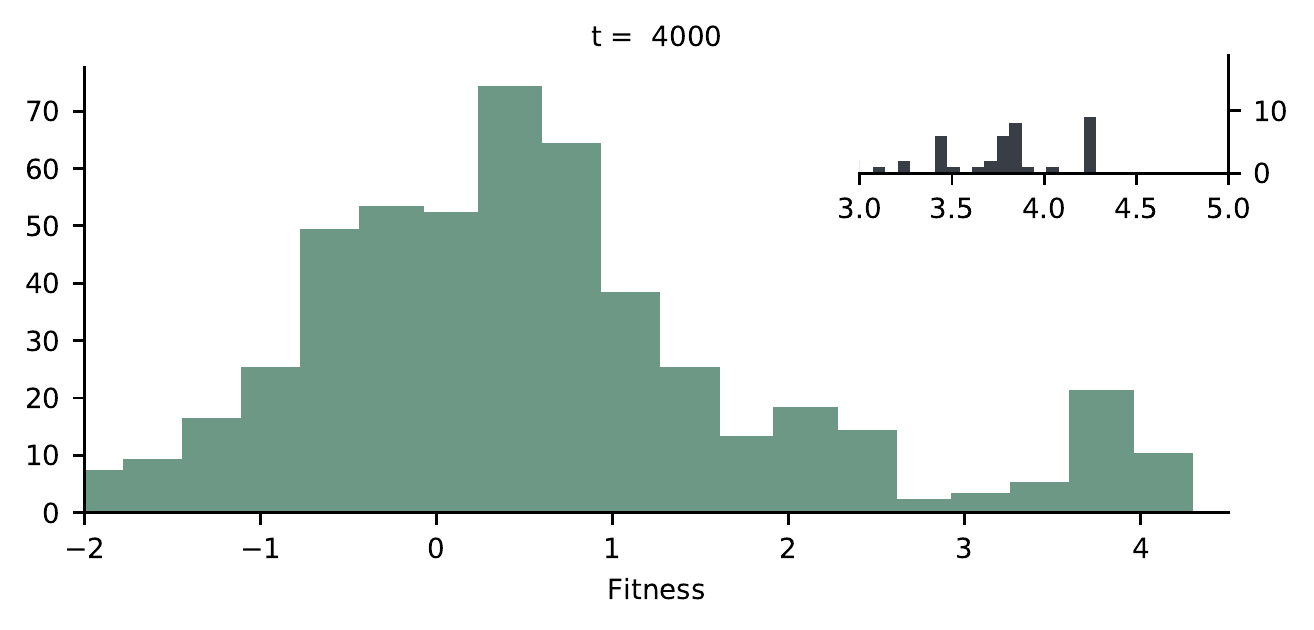}
            \caption{Snapshot of the fitness distribution at $t=4000$ (NRC-like); inset: zoom in the high fitness region. Differently from the QLE shape $F(g)\sim\mathcal{N}(0,L\sigma_e)$, in a NRC regime the distribution is bimodal, a group of individuals exists in the high fitness region.}
            \label{fFD-NRC}
          \end{subfigure}\\
    \end{tabular}
    \caption{QLE-NRC instability, SK fitness function. Simulation parameters: $N=500$, $L=25$, $T=1.0\times10^4$, $\mu=r=\omega=0.5$, $\sigma_e=0.024$, meanings as in tab.(\ref{tNST}). Code available in \cite{StudioDarwin}.}\label{f-interm}
\end{figure}

% QLE expectation
A QLE behaviour for a population evolving under a purely epistatic fitness function eq.(\ref{e-SKFF}) is characterized by a frequent reshuffling of the genomic sequences. In the absence of locus-specific (additive) fitness components, this simply results in a population for which at the steady state $\chi_i\sim0 \ \forall i$. This can be formally seen by imposing $\dot{h}_i=0$ in eq.(\ref{e-hi}) with $f_i=0\ \forall i$. Almost by definition, $|\chi_{ij}|\ll1$. Moreover, under the same QLE conditions, $F(g)$ is simply a sum of ${L \choose 2}$ Gaussian numbers $f_{ij}$, therefore $F(g)\sim\mathcal{N}(0,L\sigma_e)$ has again a Gaussian shape.

% NRC 
Holding $r,\mu$ fixed, for increasing $\sigma_e$ -- hence the magnitude of the fitness function, hence the selection strength -- the QLE is expected to break down eventually. As it turns out however, this transition, is not abrupt. In fig.(\ref{f-interm}), an illustrative simulation is shown where for high enough selection strength the population dynamics jumps stochastically between a QLE-like behaviour ($\chi_i\sim0$) and a novel bi-stable phase.

We have called the latter Non-Random Coexistence (NRC), since for each locus $i$ we observe $|\chi_i|\sim\alpha$ with $0<\alpha<1$, \emph{i.e.} both alleles are present in population without either fixating or disappearing. Here one of them is dominant, though not completely so, see fig.(\ref{fFOC-NRC}). As shown in \cite{dichio2021}, this also implies stronger correlations between loci with respect to the QLE values. At higher selection strengths, the transitions NRC $\rightarrow$ QLE become rarer and the former takes over. Any change in the population structure must be ultimately due to the synergy between chance (drift) and selection. Therefore, we can advance in the understanding of the NRC regime by looking at the fitness statistics. Fig.(\ref{fFS-NRC}) reveals a significant increase in the mean and st.dev. of the total population fitness in correspondence to the NRC regime with respect to the QLE value. A typical instantaneous snapshot of the fitness distribution in the NRC state is shown in fig.(\ref{fFD-NRC}). In contrast to QLE, the fitness distribution is bimodal. A large fraction of the population sits in the core, which has the shape of an asymmetrically placed Gaussian, biased towards higher fitness values by the effect of strong selection. At the same time, a group of fit individuals coexists, which are sustained by selection. They involve larger fractions of the population for higher selection strengths and disappear when the population transits back to the QLE state.

%clonal structure and consequences
The presence of a group of genotypes with high fitness values is compatible with a clonal condensation as described previously in sec.(\ref{ss-em-clo}), for the case $\mu=0, h=0$. However, here clones are pruned at a much higher rate because of frequent random mutations that constantly reshuffle genotypes, regardless of their fitness. By consequence, no individual clone is able to take over and in the right tail of the fitness distribution we rather see a group of similar genotypes, differing by a few mutations. In the language of the previous section, we still have $Y\sim0$ despite the strong selection and the prominent high fitness nose of the distribution, implying that the NRC regime is of different nature than the CC phase.
In \cite{dichio2021}, the results described so far are also tested in the slightly different scenarios in which the fitness function is not frustrated (\emph{e.g.} $f_{ij}>0 \ \forall i,j$) or $r\sim 0$.

\subsubsection{Heuristics and the role of $N$}
\label{heuristics-role-N}

\begin{figure}[h!]
        \centering
        \includegraphics[scale=1]{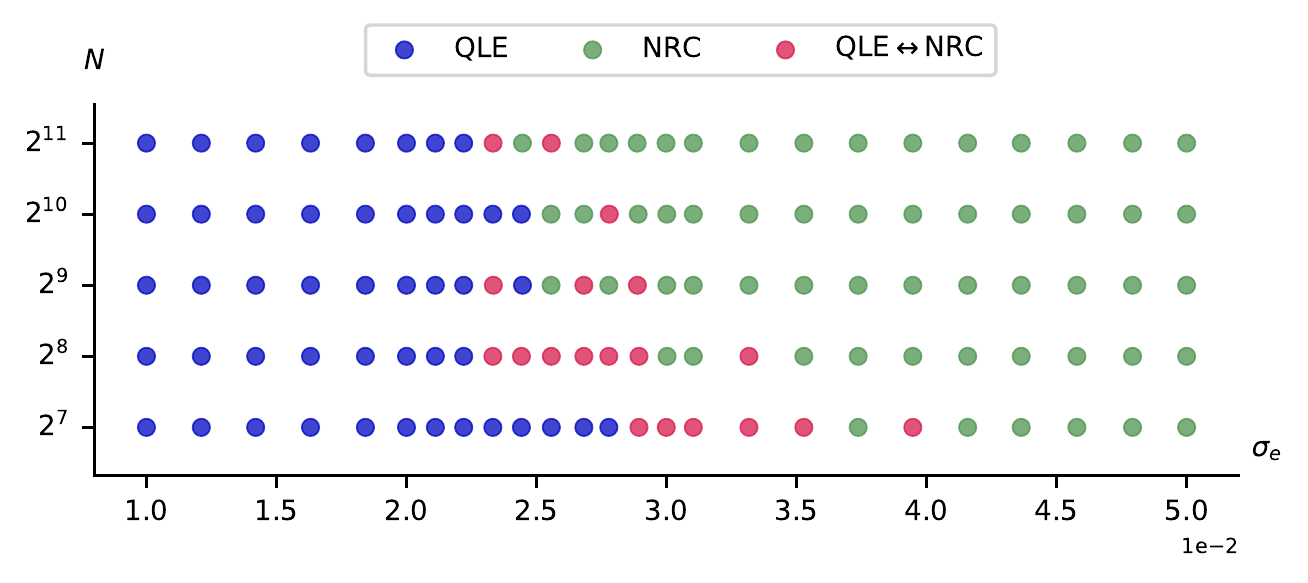}
        \caption{Qualitative phase space in the $\sigma_e$, $N$ directions. Fixed parameters of the simulations: $L=25$, $T=2.0\times10^4$, $\mu=r=\omega=0.5$, SK fitness function. Simulations are run for each pair ($\sigma_e$, $N$). If at any point a transition QLE$\leftrightarrow$NRC is observed, the corresponding point is marked as red, instability. If the population dynamics is in the QLE/NRC throughout all the simulation, the point is marked as blue / green, respectively. The two regimes are classified by setting a threshold on the fitness average. Note that in the intermediate region, the system dynamics depends on the details of the fitness landscape (quenched disorder). Because of this and of the finite simulation time, the instability region is likely to be broader.}
    \label{fIS-NRC}
\end{figure}

  \begin{figure}
        \centering
        \includegraphics[width=\textwidth]{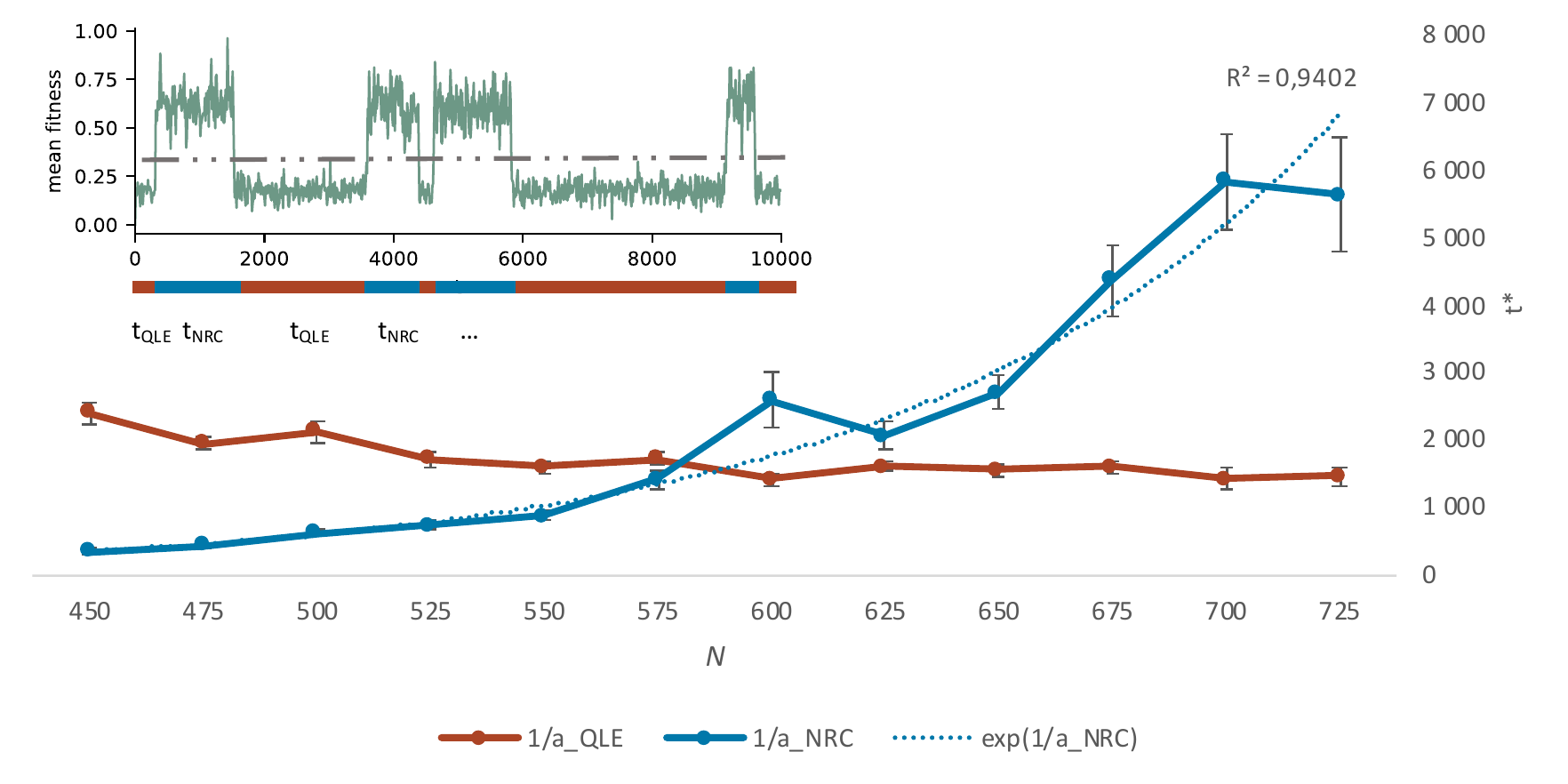}
        \caption{Estimated escape times $t_{QLE}, t_{NRC}$ from the QLE and NRC phase. For each $N$, a simulation is run with the following parameters: $L=25$, $T=1.5\times10^{6}$, $\mu=r=\omega=0.5$, $\sigma_e=0.029$ with SK fitness function.
        Inset: The dynamics undergoes multiple transitions QLE $\rightarrow$ NRC and back - an example is shown for $T=1.0\times10^{4}$ generations. The escape times from the QLE ($t_{QLE}$) and from the NRC ($t_{NRC}$) are computed by setting a threshold on the fitness mean. Main: 
        the distributions for the two kind of escape times can be constructed for each simulation. As it turns out, they are both exponentially distributed and therefore fitted with $y(T)=\gamma e^{-a T}$ \cite{dichio2021}. The inverses $t^*_{\textsc{QLE}}\sim1/a_{\textsc{QLE}},\ t^*_{\textsc{NRC}}\sim1/a_{\textsc{NRC}}$ are taken as estimations of the escape times from the respective phases and are here shown as a function of the population size $N$. While $t_{\textsc{QLE}}$ is almost insensitive to $N$, $t_{\textsc{NRC}}$ is compatible with a behaviour $\sim \exp{N}$, as confirmed by the coefficient of determination $R^2\sim 1$.}
  \label{fETA-NRC}
    \end{figure}
    
Not fully rigorous arguments (heuristics) can be used to clarify the QLE $\leftrightarrow$ NRC transition. At the boundary between the two regimes, the two major parameters are the population size $N$ and selection strength $\sigma_e$. Their interplay is hence pivotal.

% evidences
Fig.(\ref{fIS-NRC}) shows the phase diagram for the population dynamics in the direction $N-\sigma_e$. For fixed $\sigma_e$, larger population undergo more easily a transition to a NRC regime. For fixed $N$ a QLE regime is found for low selection strength. Increasing $\sigma_e$, one finds an intermittent behaviour like the one discussed in fig.(\ref{f-interm}) and finally a NRC regime takes over.
The intermittent behaviour at intermediate $\sigma_e$ is further explored in fig.(\ref{fETA-NRC}). It shows that the estimated escape times from the QLE and NRC phases -- \emph{i.e.} the number of generations after which the system has a transition out of the current phase -- depend on $N$, details on the estimation procedure can be found in \cite{dichio2021}. In particular, the average time to wait in order to observe a transition NRC$\rightarrow$QLE depends exponentially on the population size $t_{NRC}\propto e^N$, while $t_{QLE}$ is almost insensitive to $N$.

Generally speaking, the role of $N$ is twofold. On one hand, a population can be pictured as an ensemble of $N$ walkers in the genotype space. A larger population explores this space more in depth, increasing the chances of finding configurations with particularly high fitness. On the other hand, larger populations are also more resilient to rare events \emph{e.g.} the disappearance by chance of all copies of an existing fit genotype.

% Emerging picture
Two separate processes are therefore relevant for the QLE$\rightarrow$NRC transition: the appearance of a genotype with high fitness and the establishment around it of a group of similarly fit individuals. The former is enhanced by the population size $N$, mutations and recombinations. The latter is promoted by selection $\sigma_e$ and hampered by mutations, recombination and drift-induced fluctuations. 
For low enough $\sigma_e$, there exists no genotype that can be efficiently promoted by selection against mutations and recombinations, and the population dynamics is QLE-like. For increasing the selection strength, there are more and more of them in the genotype space and they are found faster in larger populations. A genotype manages to establish itself if its fitness is high enough that selection -- by increasing the number of its copies in the population -- is able to preserve its existence (at least one copy) against mutations, recombination and drift.

The extinction of fit genotypes and the consequent transition NRC $\rightarrow$ QLE is driven by a similar mechanism to Muller's ratchet. Consistently with eq.(\ref{e-ratch}), we find the same exponential dependence $t_{NRC}\propto e^N$. In larger populations there are more individuals in the high-fitness mode of the NRC fitness distribution, which is therefore more resilient against random (drift) or entropic forces (mutations, recombinations).

\subsubsection{Possible observation of NRC in experiments}
\label{possible-observation-NRC}

We now change gears. The long term evolution experiment (LTEE) has followed $12$ initially identical \emph{E.coli} populations for more than $70,000$ generations (since 24 February 1988) \cite{lenski1991}. These high-profile experiments have revealed both random differences between the populations and common evolutionary changes applying to all of them. The almost 35 years of evolutionary history of bacteria correspond (very approximately) to some millions of year of human history, \textit{i.e.} to before the emergence of modern man.

\begin{sidewaysfigure}
\includegraphics[width=\columnwidth]{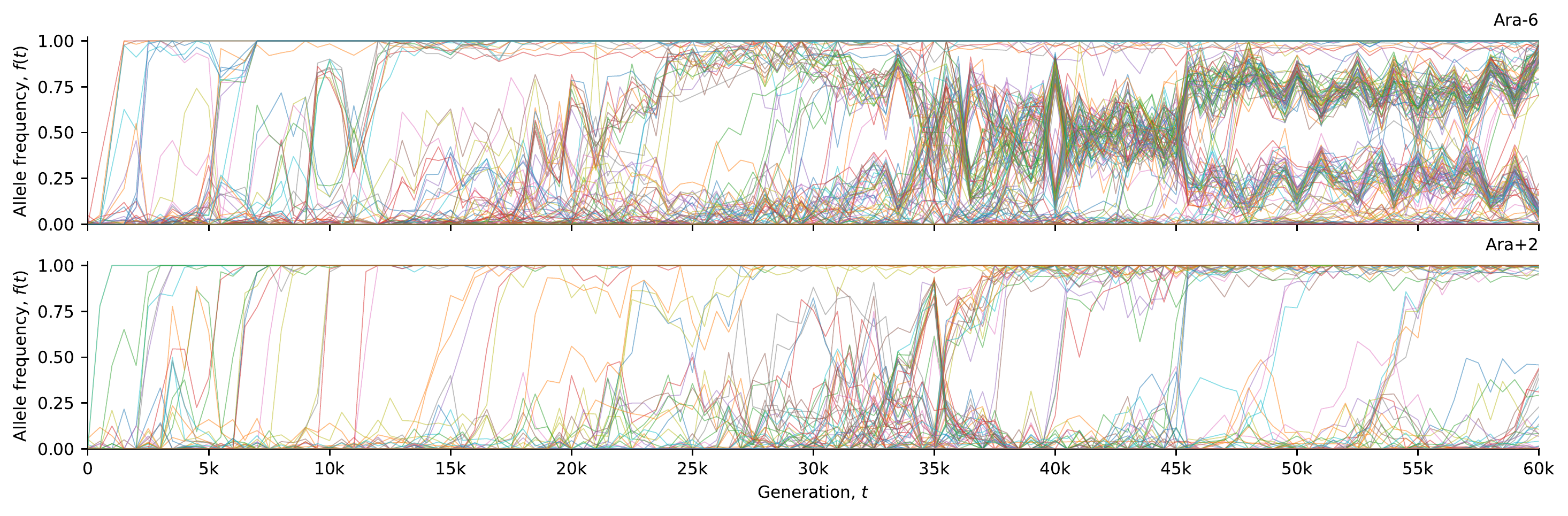}
\caption{Allele frequency trajectories $\nu_i=(1-\chi_i)/2\in[0,1]$ of all \emph{de novo} mutations detected in $2$ of the $12$ LTEE populations, labeled respectively \texttt{Ara-6} and \texttt{Ara+2}. Population \texttt{Ara-6} (top row) shows an example of quasi-stable coexistence of clades. Population \texttt{Ara+2} (bottom row) does not show such coexistence, but instead mutations fix rapidly. The analysis in \cite{Good} reported quasi-stable coexistence in $9$ out of $12$ LTEE populations. Figure previously unpublished, private communication from B.H. Good and M.M. Desai, reproduced here with permission. Data obtained from same set of experiments described in \cite{Good}.}
\label{fGood}
\end{sidewaysfigure}

A recent analysis \cite{Good} of this dataset has observed several mutations segregating into (at least) two intermediate-frequency clades that coexist for long periods, a phenomenon the authors denoted \emph{quasi-stable coexistence}. The authors of \cite{Good} have generously made available previously unpublished data which show the effect in a somewhat more pronounced manner, which we include here as fig.(\ref{fGood}) [with permission].

The mechanism behind quasi-stable coexistence has been somewhat of a mystery. The authors of \cite{Good} conjecture negative frequency-dependent selection (removal of deleterious alleles at a rate which depends on the current fraction of such alleles), or interaction at the phenotype layer between individuals and the environment. Neither of these mechanisms has however been independently demonstrated. On the other hand, it is clear that the phenomenology of quasi-stable coexistence is quite similar to NRC. In the LTEE the number of bacteria in each population fluctuates daily in a cycle of dilution and seeding into fresh medium, transition of the bacteria from stationary to growth phase, and then growth to saturation in that medium. The number remains quite large throughout the cycle, hence $N$ can be taken to be essentially infinite. In the perspective of the above discussion we hence expect the transition to quasi-stable coexistence to be essentially irreversible (as the backward rate NRC $\rightarrow$ QLE depends exponentially on $N$). This is in accordance with observations. While a strict Sherrington-Kirkpatrick epistatic fitness function for all loci is unrealistic, it is enough for the argument that it is a reasonable approximation for the set of loci of substantial variability in the population. The proposed mechanism is conjectural, but a spin-glass-like fitness function is in essence not far from well-known conceptual models in population biology such as the Kauffman model \cite{Drossel2008,Kauffman1969}. We hence advance an NRC phase as another possible explanation for the observed quasi-stable coexistence in LTTE.

%%%%%%%%%%
\section{Summary and discussion}\label{sec:discussion}
In this work we have provided a self-contained review of the dynamics of a population evolving under selection (Darwinian evolution, survival of the fitness), mutation and recombination. We have shown how high rates of mutation or recombination (or both) relative to the strength of selection naturally lead to the quasi-linkage equilibrium (QLE) phase first described by Kimura~\cite{Kimura}. The QLE phase is characterized by weak correlations between different loci and multi-genome distributions approximately factorizing into products of one-genome distributions. Under these conditions, the distributions of individuals in the population over genotypes are well described by distributions in the exponential family, whose parameters are related to evolutionary parameters, including fitness.

The last property is a kind of distributional phenotype-genotype relation (dPGR), because it is phenotype which determines fitness, and it is the distribution law over genotypes which is determined by fitness. We have derived a dPGR under different albeit quite similar assumptions for the one-genome distribution in the QLE state, namely from a Gibbs-Boltzmann distribution with Ising Hamiltonian and from a multivariate Gaussian distribution. When it holds, evolutionary parameters can be inferred from Gibbs-Boltzmann model parameters which in turn can be inferred from data, using dPGR to make the translation. Alternatively, under the Gaussian ansatz, a dPGR relates directly statistics of the data to epistatic fitness parameters. We have demonstrated that the above scheme works in computational experiments, and we have also demonstrated how to derive higher-order inference schemes. We suggest that QLE is a fertile field for future applications of statistical physics concepts to population genetics on the whole-genome level.

On a different track we have considered the dynamics at a relatively stronger selection, which induces a breakdown of the QLE dynamics. We have illustrated the importance for this phenomenon of a finite population size in simple cases, as the study of the Muller's ratchet. We have also discussed recent theoretical results investigating the clonal condensation transition in the absence of mutations. Finally, we have described a new behaviour which breaks time translation symmetry. For sufficiently strong selection the state of the population switches randomly between a behavior qualitatively similar to QLE, and a phase of non-random coexistence (NRC) where the population is dominated by one or more fit genotype and their neighbours. We have determined how the stability boundaries of the QLE phase going towards the NRC phase and vice versa change with model parameters, including population size $N$. We surmise that also the NRC phase constitutes an interesting application area of statistical physics, albeit to non-equilibrium processes not in detailed balance.

On a general level the conclusion of this work is that the statistical physics analogy is both useful and of limited use in population genetics. It is a fact that reasonable models of evolution in certain parameter ranges settle down to stationary distributions of the same form as in equilibrium statistical mechanics, which is the basis for theories such as those developed in~\cite{Ao2007,Ao2008,Wang2015}. A whole collection of methods, collectively known as  direct coupling analysis (DCA) can then be used to infer evolutionary parameters from the distribution of sequences in a population. Naturally, if and when one has access to time series data, other inference methods can be used, and they will often be more powerful. Large-scale sequencing data are however rarely of this type: more typically one knows the distribution of sequences (genotypes) at one or at most a few time points, and one does not have information on which sequences gave rise to which other sequences except from their similarity. In this setting the knowledge that the distribution is of the Gibbs-Boltzmann type can be leveraged to extract parameters describing the dynamics from what is essentially static information.

On the other hand, the underlying dynamics is not in thermal equilibrium \textit{i.e.} does not obey detailed balance, and the range of possibilities is hence wider~\cite{Waddington1957,Zhou2012}. In the class of models considered in this work the distribution does not have to approach that of a stationary Gibbs-Boltzmann distribution with energy-like terms. Dynamics can go on indefinitely, and the distribution of genotypes in a population can fluctuate without ever reaching a stationary state. Models of life, even with all the simplifying assumptions made here, hence allow for rich repertoires not easily captured by models too closely patterned after equilibrium statistical mechanics.

\newpage

\appendix
\addtocontents{toc}{\protect\setcounter{tocdepth}{0}}
\section{Statistical Genetics and Direct Coupling Analysis in and out of Quasi-Linkage Equilibrium / Supplementary information}

\subsection{Derivation of eq.(\ref{eEFOC})} 
Using eq.(\ref{e-scalc}) and the definition $\chi_i=\av{s_i}$:

\begin{equation}
\begin{split}
    \dot\chi_i &= \frac{d}{dt}\Big(\sum_g s_iP(g)\Big) \\
    &= \sum_g s_i \frac{d}{dt}P(g) \\
    &= \sum_g \Big(s_i[F(g)-\av{F}] P(g) + \mu \ s_i \sum_{j=1}^L [P(M_j g)- P(g)]\Big) \\
    &\stackrel{}{=} \av{s_i[F(g)-\av{F}]} - 2\mu \av{s_i} \ .\notag
\end{split}
\end{equation}
where the last line follows from $\sum_g s_i P(M_j g) = (-1)^{\delta_{ij}}\langle s_i\rangle$\ .
\qed

\subsection{Derivation of eq.(\ref{eSOC})}
As a preliminary result, let us evaluate the time derivative of $\av{s_is_j}$ under the recombination term alone: 
\begin{equation}
\begin{alignedat}{2}
    \frac{d}{dt}\Big|_{\substack{rec}} \av{s_is_j} &= &&\sum_{\xi,g,g'} C(\xi) s_i s_j \Big[P(g^{(1)})P(g^{(2)}) - P(g)P(g')\Big]\\
    &\stackrel{(a)}{=}&&\sum_{\xi,g^{(1)},g^{(2)}} \Big[C(\xi)(\xi_i s_i^{(1)} + (1-\xi_i) s_i^{(2)})\ \times\\
    & &&\times(\xi_j s_j^{(1)} + (1-\xi_j) s_j^{(2)})P(g^{(1)})P(g^{(2)})\Big]\ + \\
    & && - \sum_{\xi,g,g'} C(\xi) s_i s_j P(g)P(g')  \\
    &\stackrel{}{=}&&\sum_{\xi}C(\xi)\Big[\xi_i\xi_j\av{s_is_j} + \xi_i(1-\xi_j)\langle s_i\rangle\av{s_j} \ + \\
    & && + (1-\xi_i)\xi_j\langle s_i\rangle\av{s_j} + (1-\xi_i)(1-\xi_j)\av{s_is_j} - \av{s_is_j} \Big]\\
    &\stackrel{}{=} && \av{s_is_j} \sum_{\xi}C(\xi) (2\xi_i\xi_j-\xi_i-\xi_j)\ + \\
    & && + \av{s_i}\av{s_j} \sum_{\xi}C(\xi) (\xi_i(1-\xi_j) + (1-\xi_i)\xi_j)\\
    &\stackrel{(b)}{=}&&\ -c_{ij} \ \chi_{ij}\ . \label{e-recomb}
\end{alignedat}
\end{equation}
In $(a)$ we have used eq.(\ref{eRPC}) and changed the first sum over $g,g'$ in a sum over $g^{(1)}$, $g^{(2)}$; in $(b)$ we have used the definition of $\chi_{ij}$ and $c_{ij}$, eq.(\ref{eDCIJ}). 
Now, for $i\ne j$,
\begin{equation}
\begin{alignedat}{2}
    \dot\chi_{ij} &= &&\frac{d}{dt} (\av{s_is_j}-\chi_i\chi_j) \\\
    &= && \frac{d}{dt} \av{s_is_j} - \dot\chi_i\chi_j - \chi_i\dot\chi_j \\\
    &\stackrel{(a)}{=}&& \av{s_is_j [F(g)-\av{F}]} + \mu \sum_gs_is_j\sum_{k=1}^L [P(M_k g)- P(g)]   + r\frac{d}{dt}\Big|_{\substack{rec}} \av{s_is_j} \ \\
    & && + \ 4\mu \chi_i\chi_j - \av{s_i[F(g)-\av{F}]}\chi_j - \chi_i\av{s_j[F(g)-\av{F}]}\\\
    &\stackrel{(b)}{=}&&\ \av{(s_i-\chi_i)(s_j-\chi_j)[F(g)-\av{F}]}-4\mu\av{s_is_j}+ r\frac{d}{dt}\Big|_{\substack{rec}} \av{s_is_j}+ 4\mu \chi_i\chi_j\\\
    &\stackrel{(c)}{=}&& \av{(s_i-\chi_i)(s_j-\chi_j)[F(g)-\av{F}]} - 4\mu\chi_{ij} - rc_{ij}\chi_{ij}\ . \notag
\end{alignedat}
\end{equation}
In $(a)$ we have used eq.(\ref{eMEE},\ref{eEFOC}); in $(b)$ we exploited $\sum_g s_is_jP(M_k g)=(-1)^{\delta_{ik}+\delta_{jk}}\av{s_is_j}$ and added $\chi_i\chi_j\av{F(g)-\av{F}}=0$; $(c)$ comes again from the definition of $\chi_{ij}$ and from eq.(\ref{e-recomb}). 
\qed

\subsection{Derivation of eq.(\ref{eMFC})}
A convenient thermodynamic potential for the IIP is he Gibbs free energy
\begin{equation}\label{eGFELT}
    \mathcal{G}(\vec{J},\vec{\chi}) = \max_{\vec{h}} \Big[ \sum_ih_i\chi_i + \mathcal{F}(\vec{J},\vec{h}) \Big]
\end{equation}
which is found by operating a Legendre transform of the Helmholtz free energy 
\begin{equation}\label{HfreeE}
    \mathcal{F}(\vec{J},\vec{h}) = -\log\pf(\vec{J},\vec{h})
\end{equation} with respect to the fields $\vec{h}$. In many non-trivial systems, one does not have access to the explicit expression for $\mathcal{F}$, hence $\mathcal{G}$, and must resort variational principles. 

For the MF approximation, a useful variational principle for the Gibbs free energy is the following: 
\begin{equation}\label{eGFEVP1}
    \mathcal{G}(\vec{J},\vec{\chi}) = \min_{q\in\mathscr{G}}\Bigg\{ -\sum_{i<j}J_{ij}\av{\sigma_i\sigma_j}_q - S[q] \Bigg\}\ .
\end{equation}
where $q$ is any probability distribution in the family $\mathscr{G}$ of those for which $\av{\sigma_i}_q=\chi_i$, and $\mathcal{S}[q]=-\av{\log q}_q$ is the entropy of the distribution $q$. See  \cite{NZB} for more details.

Mean-field (MF) approximation makes the following family of joint probability distribution 
\begin{equation}\label{eMFD}
    p^{MF}(\vec{s})=\prod_i\frac{1+\Tilde{\chi}_is_i}{2}\ ,
\end{equation}
implying that spins are independent variables and the effective magnetization $\Tilde{\chi}_i$ results from both the local field $h_i$ and from the couplings $J_{ij}$ with all other spins. There is only one distribution of the form eq.(\ref{eMFD}) for which $p^{MF}\in\mathscr{G}$, \emph{i.e.} the one for which $\Tilde{\chi}_i=\chi_i$. The Gibbs free energy is therefore:
\begin{equation}
    \mathcal{G}^{MF}(\vec{J},\vec{\chi})= -\sum_{i<j}J_{ij}\chi_i\chi_j + \sum_i\Big[ \frac{1+\chi_i}{2}\log\frac{1+\chi_i}{2} + \frac{1-\chi_i}{2}\log\frac{1-\chi_i}{2} \Big]\ .
\end{equation}
By eq.(\ref{eGFELT}), we can evaluate the fields $h_i$ under the MF approximation by taking derivatives of $\mathcal{G}^{MF}$ with respect to $\chi_i$, yielding
\begin{equation}
    h_i^{MF} = -\sum_{j\ne i} J_{ij}\chi_j + \frac{1}{2}\frac{1+\chi_i}{1-\chi_i} = -\sum_{j\ne i} J_{ij}\chi_j + \arctanh \chi_i\ .\label{eMFF}
\end{equation}
By taking derivatives of the latter expression with respect to $\chi_j$ for $i\ne j$ we find 
\begin{equation}
    J_{ij}^{MF} = \frac{\partial h_i^{MF}}{\partial\chi_j\ \ } \ .
\end{equation}
Finally, from linear response theory we can write $\chi_{ij} = \partial\chi_i(\vec{J},\vec{h})/\partial{ h_j}$, where $\chi_{ij}$ are the second order cumulants (connected correlations). Using the inverse function theorem $[\partial\vec{h}/\partial\vec{\chi}]_{ij} = [(\partial\vec{\chi}/\partial\vec{h})^{-1}]_{ij}$ and inserting in eq.(ix), we get:
\begin{equation*}
    J_{ij}^{MF} = -(\chi^{-1})_{ij}\ .
\end{equation*}
\qed

\subsection{Derivation of eq.(\ref{eSOCAS}) / alternative}
Consider a particular spin variable $s_i$ and distinguish the part $\mathscr{H}_i$ of the Hamiltonian that depends on $s_i$ from the rest, that we collectively indicate with $\mathscr{H}_{\backslash i}$:
\begin{equation}
    \mathscr{H}(\vec{s}) = \mathscr{H}_i + \mathscr{H}_{\backslash i} 
    = -h_is_i -\sum_{j\ne i} J_{ij}s_is_j + \mathscr{H}_{\backslash i}(\vec{s}_{\backslash i})\ .
\end{equation}
Summing up explicitly the terms related to $\sigma_i$ in the partition function,
\begin{equation}
    \pf(\vec{J},\vec{h}) = \sum_{\vec{s}_{\backslash i}} 2 \cosh\Big( h_i + \sum_j J_{ij} s_j \Big) e^{-\mathscr{H}_{\backslash i}(\vec{s}_{\backslash i})}\ .
\end{equation}
From eq.(\ref{HfreeE}) one gets the first and second moments (one and two spin expectations) involving $s_i$ by simply deriving with respect to the parameters:
\begin{equation}
    \av{s_i} = - \frac{\de \mathcal{F}}{\de h_i}(\vec{J},\vec{h})\ ;\quad \av{s_is_j} = - \frac{\de \mathcal{F}}{\de J_{ij}}(\vec{J},\vec{h})
\end{equation}
Using the definition of $\mathcal{F}$, one finds
\begin{equation}
\begin{split}
    \av{s_i} &= \Big\langle \tanh\Big( h_i + \sum_{j\ne i} J_{ij} s_j \Big) \Big\rangle \\
    \av{s_is_j} &= \Big\langle s_j\tanh\Big( h_i + \sum_{k\ne i} J_{ik} s_k \Big) \Big\rangle\ 
\end{split}
\end{equation}

The average in the RHS of the last equations is over the entire Boltzmann distribution eq.(\ref{eIBD})
and these are still exact equations. The PLM approximation is implemented when substituting the (computationally prohibitive) averages above with sample averages, labelled with the superscript $D$:
\begin{align*}
    \av{\sigma_i}^D &= \Big\langle \tanh\Big( h_i^{PL} + \sum_{j\ne i} J_{ij}^{PL} \sigma_j \Big) \Big\rangle^D \ ,\\
    \av{\sigma_i\sigma_j}^D &= \Big\langle \sigma_j\tanh\Big( h_i^{PL} + \sum_{k\ne i} J_{ik}^{PL} \sigma_k \Big) \Big\rangle^D\ .
\end{align*}
\qed

%\subsection{Derivation of eq.(\ref{eIBD}) / Max.Ent.}
%Consider the constrained maximization problem defined by eq.(\ref{eSHIBD}-\ref{econstr}). For each constraint we introduce a different Lagrange multiplier: $\eta$ for the normalization, $h_i$ for each of the $L$ first moments, $J_{ij}$ for each of the $L(L-1)/2$ second moments. The maximization is operated by setting to zero the derivatives of the following expression
%\begin{equation}
%\begin{split}
%    -\sum_{\vec{s}} & p(\vec{s})\log p(\vec{s}) + \eta\bigg[\sum_{\vec{s}}p(\vec{s})-1\bigg] +  \\
%    &+ \sum_ih_i\bigg[\sum_{\vec{s}}p(\vec{s})s_i-\av{s_i}^D\bigg] + \sum_{i<j}J_{ij}\bigg[\sum_{\vec{s}}p(\vec{s})s_is_j-\av{s_is_j}^D\bigg]\ .
%\end{split}
%\end{equation}
%We get the distribution
%\begin{equation*}
%    p(\vec{s})=e^{-1+\eta}e^{\ \sum_ih_is_i +\sum_{i<j}J_{ij}s_is_j}=\frac{1}{\pf}e^{\ \sum_ih_is_i +\sum_{i<j}J_{ij}s_is_j}\ ,
%\end{equation*}
%where $\eta = 1-\log\pf$ is fixed by the normalization and $\vec{h},\vec{J}$ are chosen so to reproduce the observed moments $\av{s_i}^D,\av{s_is_j}^D$. 
%We recognize in the previous equation the Boltzmann distribution eq.(\ref{eIBD}) with the Ising Hamiltonian eq.(\ref{eIH}). As an aside, the Shannon entropy of this distribution is readily computed to be
%\begin{equation}
%    \mathcal{S} = -\sum_ih_i\av{s_i}^D - \sum_{i<j}J_{ij}\av{s_is_j}^D +\log\pf\ .
%\end{equation}
%\qed

\subsection{Derivation of eq.(\ref{eSOCQLEij})}
Let us start by evaluating the partition function of eq.(\ref{eICE}) perturbatively for small $|J_{ij}|\ll1$
\begin{equation}\label{e-Zexp}
\begin{alignedat}{2}
    \pf   &=&& \sum_g e^{\sum_i h_i s_i + \sum_{i<j} J_{ij} s_i s_j}\\
        &\stackrel{(a)}{\sim}&& \sum_g e^{\sum_i h_i s_i} \Big(1+ \sum_{k<j} J_{kj} s_k s_j \Big)\\
        &\stackrel{}{=}&& \sum_g e^{\sum_i h_i s_i} + \sum_{k<j}J_{kj} \sum_g e^{\sum_i h_i s_i}  s_k s_j\\
        &\stackrel{}{=}&& \prod_{i} 2\cosh h_i + \sum_{k<j}J_{kj} \Bigg( \prod_{i\ne j\ne k} 2\cosh h_i\Bigg) (2\sinh{h_k})(2\sinh{h_j})\\
        &\stackrel{}{=}&&\ 2^L\Bigg( 1 + \sum_{k<j}J_{kj}\tanh{h_k}\tanh{h_j} \Bigg) \prod_i \cosh{h_i}\ ,
\end{alignedat}
\end{equation}
where in $(a)$ we have expanded to the first order in $|J_{ij}|$. 
Taking the derivative of $\log\pf$ with respect to $h_i$ gives the first moment $\chi_i = \de\log\pf/\de h_i$. Using the approximate expression eq.(\ref{e-Zexp}):
\begin{alignat}{2}
    \chi_i  &\stackrel{}{\sim}&&  \frac{2^L}{\pf} \Bigg[\Bigg( 1 + \sum_{k<j}J_{kj}\tanh{h_k}\tanh{h_j} \Bigg) \prod_{l\ne i}\cosh{h_l}\sinh{h_i} +\notag\\
    & && + \Bigg( 1 + \sum_{i\ne j}J_{ij}\frac{\tanh{h_j}}{\cosh^2{h_i}}\Bigg)\prod_i \cosh{h_i}\Bigg]\notag\\
     &\stackrel{}{\sim}&& \tanh{h_i} + \frac{1}{1 + \sum_{k\ne j}J_{kj}\tanh^2{h_k}\tanh{h_j}}\sum_{i\ne j}J_{ij}(1-\tanh^2{h_i})\tanh{h_j}\notag\\
    &\sim&& \tanh{h_i} + \sum_{i\ne j}J_{ij}(1-\tanh^2{h_i})\tanh{h_j}\label{eFOCQLE}
\end{alignat}
From the latter, we evaluate the second order cumulants as
\begin{alignat}{2}
    \chi_{ii} = \frac{\de^2\log\pf}{\de h_i^2} &\stackrel{(a)}{\sim}&&  \ 1- \tanh^2{h_i} = \stackrel{(b)}{\sim} 1-\chi_i^2\label{eSOCQLEii}\\
    \chi_{ij} = \frac{\de^2\log\pf}{\de h_i\de h_j} &\stackrel{(a)}{\sim}&& \ J_{ij} (1-\tanh^2h_i)(1-\tanh^2h_j) \stackrel{(b)}{\sim} J_{ij}(1-\chi_i^2)(1-\chi_j^2)\notag
\end{alignat}
where in $(a)$ we have derived eq.(\ref{eFOCQLE}) w.r.t. $h_i$ or $h_j$, while in $(b)$ we have used again eq.(\ref{eFOCQLE}), upon moving the sum to the LHS. All of the eq.(\ref{eFOCQLE} - \ref{eSOCQLEij}) are correct to the first order in $|J_{ij}|$.
\qed

\subsection{Derivation of eq.(\ref{eDF}-\ref{ePCQLE})}
A convenient starting point is the time evolution of the quantity $\log P(g)$. Using on the LHS the ansatz eq.(\ref{eICE}) and on the RHS the master equation eq.(\ref{eMEE}) one gets:
\begin{equation}
\begin{split}
    - \frac{\dot\pf}{\pf} + \sum_i\dot h_i s_i &+ \sum_{i<j}\dot J_{ij} s_is_j =\\
    &= F(g)-\av{F} + r \sum_{\xi,g'}\  C(\xi)P(g')\Bigg[\frac{P(g^{(1)})\ P(g^{(2)})}{P(g)P(g')} - 1\Bigg]\label{eQLEI}
\end{split}
\end{equation}
Let us analyze separately the last term in the square brackets. Setting $\bar\xi_i = 1-\xi_i$:
\begin{alignat}{2}
    \sum_{\xi,g'}\ & C(\xi)P(g')\Big[\frac{P(g^{(1)})\ P(g^{(2)})}{P(g)P(g')} - 1\Big] = \notag\\
    &\stackrel{(a)}{=} \sum_{\xi,g'}\  C(\xi)P(g') \Bigg(e^{\ \sum_{i<j} J_{ij} \big[(\xi_is_i+\bar\xi_is_i')(\xi_js_j+\bar\xi_js_j') + (\bar\xi_is_i+\xi_is_i')(\bar\xi_js_j+\xi_js_j') -s_is_j-s_i's_j'\big]}-1\Bigg) \notag \\
    %&\stackrel{}{=} \sum_{\xi,g'}\  C(\xi)P(g') \Bigg(e^{\ \sum_{i<j} J_{ij} \big[(\xi_i\xi_j+\bar\xi_i\bar\xi_j-1)(s_is_j+s_i's_j') + (\xi_i\bar\xi_j+\bar\xi_i\xi_j-1)(s_is_j'+s_i's_j) \big]}-1\Bigg) \notag \\
    &\stackrel{(b)}{\sim} \sum_{\xi,g'}\  C(\xi)P(g') \sum_{i<j} J_{ij} \big[(\xi_i\xi_j+\bar\xi_i\bar\xi_j-1)(s_is_j+s_i's_j') + (\xi_i\bar\xi_j+\bar\xi_i\xi_j)(s_is_j'+s_i's_j)\big] \notag \\
    &\stackrel{(c)}{=} \sum_{\xi}\  C(\xi) \sum_{i<j} J_{ij} \big[(\xi_i\xi_j+\bar\xi_i\bar\xi_j-1)(s_is_j+\av{s_is_j}) + (\xi_i\bar\xi_j+\bar\xi_i\xi_j)(s_i\av{s_j}+\av{s_i}s_j)\big] \notag\\
    &\stackrel{(d)}{=}\sum_{i<j} c_{ij} J_{ij} \big[(s_i\av{s_j}+\av{s_i}s_j) - (s_is_j+\av{s_is_j}) \big]\label{eQLEf}
\end{alignat}
In $(a)$ we have used eq.(\ref{eICE}), inverted the relations eq.(\ref{eRPC}) to express $s^{(1)}_i = \xi_is_i+\bar\xi_is_i'$, $s^{(2)}_i = \bar\xi_is_i+\xi_is_i'$, clearly $s_i^{(1)}+s_i^{(2)}-s_i-s_i'=0$ cancel for each field $h_i$; in $(b)$ we have expanded to the first order in $|J_{ij}|$; in $(c)$ we have averaged over $P(g')$; in $(d)$, finally, we have used $c_{ij} = \sum_{\xi} C(\xi) (\xi_i\bar\xi_j+\bar\xi_i\xi_j) = \sum_{\xi} C(\xi) (1-\xi_i\xi_j - \bar\xi_i\bar\xi_j)$.

Substituting eq.(\ref{eQLEf}) into eq.(\ref{eQLEI}) and using eq.(\ref{eFL}):

\begin{equation}
\begin{alignedat}{2}
    - \frac{\dot\pf}{\pf} + \sum_i\dot h_i s_i & && + \sum_{i<j}\dot J_{ij} s_is_j = \\
    =& && \ \bar F-\av{F} +\sum_if_is_i +\sum_{i<j}f_{ij}s_is_j\ +\\
    & && + r \sum_{i<j} c_{ij} J_{ij} \big[(s_i\av{s_j}+\av{s_i}s_j) - (s_is_j+\av{s_is_j}) \big]
\end{alignedat}
\end{equation}
Dynamical equations for $\vec{h,J}$ emerge when collecting together terms with the same monomials in $s_i$:
\begin{align*}
    \dot h_i &= f_i + r\sum_{j\ne i} c_{ij}J_{ij}\chi_j\\
    \dot J_{ij} &=f_{ij} - rc_{ij}J_{ij} 
\end{align*}
\qed

\subsection{Focus: evolutionary simulations based on FFPopSim}
\sml{FFPopSim} is implemented in \sml{C++} with a \sml{Python2.7} wrapper. It allows population genetics simulations for a population of haploid individuals, identified by their genomes $g = (s_1,\dots,s_L)$ with biallelic loci $s_i=\pm1$. 

Simulations are instantiated by specifying the structure of the evolving population, the rates of the evolutionary mechanisms and initial conditions. In an individual-based model, the fundamental object undergoing evolution is not the genotype, but the \emph{clone} $c_i = (g_i, n_i)$ \textit{i.e.} the pair of a genotype $g_i$ and the number $n_i(t)$ of individuals in the population that have that genotype at time $t$. The population $\mathcal{P}$ is hence a set of clones.

A discrete generation scheme is employed, in which every individual at every generation undergoes each of the processes that drive evolution with tunable probabilities. In particular: 

\begin{itemize}
    \item \textbf{Mutations}. Mutations are bit-flip operations in a genotype. Each individual mutates with probability $1-e^{L\mu}$. Every individual that has been selected for mutations, suffers at least one of them, the number $K$ being drawn from a Poisson distribution $\mathscr{P}_{L\mu}(K)$ with mean $L\mu$.\footnote{These probabilities are consequences of the discreteness of the computer simulation. The rate $\mu$ as introduced in sec.(\ref{ssM}) is referred to a continuous-time formulation of the evolution. Let $\mathcal{E}$ be the event that a mutation appears in an individual; suppose such events are independent and that the probability of two of them happening at the same time is negligible. If their average rate is $\mu$ then the number $k$ of events $\mathcal{E}$ in the time interval $\Delta t$ is $\sim \mathscr{P}_{\mu\Delta t}$ where $\mathscr{P}_{\lambda}(k) = \lambda^ke^{-\lambda}/k!$ is the Poisson distribution. The number of such mutations in a genome of length $L$ in the interval $\Delta t$ is the random variable $K = \sum_{i=1}^L k_i$ that, being the sum of $L$ i.i.d. Poisson random variables, is again Poisson distributed, with mean $L\mu\Delta t$ \textit{i.e.} $K\sim\mathscr{P}_{L\mu\Delta t}$. Finally, the probability that there is \emph{at least} one mutation is $1-\mathscr{P}_{L\mu\Delta t}(0) = 1 - e^{L\mu\Delta t}$.\label{footmu}} Target loci are chosen randomly.
    \item \textbf{Selection}. Let $n_i(t)$ be the size of the clone $i$ at time $t$. We enforce selection by updating  $n_i(t) \rightarrow n_i(t+1) \sim \mathscr{P}_{\lambda}$ where
    \begin{equation}\label{eLAMBDA}
        \lambda = \frac{1}{\av{e^F}}e^{F(g_i)+1-\frac{1}{N}\sum_jn_j(t)} \ .
    \end{equation}
    In words, we draw the size $n_i(t+1)$ of the $i$-th clone at time $t+1$ from a Poisson distribution with mean $\lambda$ as in eq.(\ref{eLAMBDA}), where $F(g)$ is the fitness function eq.(\ref{eFL}) and the average $\av{e^F}$ is over the entire population. We note that for $F(g)\ll 1$, $e^{F(g)}/\av{e^F} - 1\sim F(g)-\av{F}$, so that we retrieve eq.(\ref{eFT}). The growth rate adjustment $\exp(1-\sum_jn_j(t)/N)$ is implemented to constrain the population close to the carrying capacity $N$.
    \item \textbf{Recombination}. A fraction $r^*$ of the offspring at the previous fitness-step are designated for sexual reproduction.\footnote{Note that $r^*\ne r$, the latter described in sec.(\ref{ssR}). The reason is that in general the outcrossing rate $r^*$ is not the recombination rate $r$. In fact, $r^*$ is treated as a probability while $r$ is a rate that can take any positive value. Considering the discreteness of the computer simulation as done for mutations, we should have $r^*=1-e^{-r}$. However, as long as $r\ll1$ they approximately coincide $r^*\sim 1-(1-r) = r$.}     They are shuffled and randomly paired. For each pair a crossover pattern $\vec{\xi}$ is created and the recombination is implemented by discarding parents and replacing them with two new individuals accordingly. Crossovers happen independently between any two loci with rate $\omega$.
\end{itemize}
For the details of the technical implementation we refer to the documentation in the GitHub repository \cite{StudioDarwin}.

\subsection{Derivation of n-points expectations for gaussian distributions}
Let $\vec{X}=(X_1,\dots,X_n)$ be a vector of random variables with pdf $p(\vec{x})$, then the \emph{characteristic function} is 
\begin{equation}
\phi(\vec{q})=\int p(\vec{x})e^{i\vec{q}\vec{x}}\ .
\end{equation}
with $\phi(\vec{0})=1$, $|\phi(\vec{q})|\le 1$. The function $\phi(\vec{q})$ is a characterization of the probability distribution $p(\vec{q})$ \textit{i.e.} it completely determines its behaviour and properties. If the \emph{raw moments} $\ \av{\prod_i X_i}$ exist, then
\begin{equation}\label{eM} %Moments
    \big\langle\prod_i X_i\big\rangle = \Big[ \prod_i \Big( -i \frac{\de}{\de q_i} \Big)\phi(\vec{q}) \Big]_{\vec{q}=\vec{0}}\ .
\end{equation} 

Let us consider the specific case of a  \emph{multivariate Gaussian probability distribution}:
\begin{equation}
    p(\vec{s}) = \frac{1}{\pf} e^{-\frac{1}{2}(\vec{s}-\vec{\chi})^T\chi^{-1}(\vec{s}-\vec{\chi})}\ ,
\end{equation}
where $\vec{\chi}$ (bold) are the mean values $\chi_i=\av{s_i}$, $\chi$ is the covariance matrix \emph{i.e.} $\chi_{ij}=\av{s_is_j}-\av{s_i}\av{s_j}$, $\pf=[(2\pi)^n \det(\chi)]^{-\frac{1}{2}}$ is the normalization. The characteristic function of such distribution is\footnote{The cumulant generating function is defined as $\psi(\vec{q})=\log \phi(\vec{q}).$ In the case of a multivariate gaussian, $\psi(\vec{q})=-\frac{1}{2}(\vec{s}-\vec{\chi})^T\chi^{-1}(\vec{s}-\vec{\chi}).$ We see that all cumulants of order $\ge 2$ ($\sim\de^3\psi(\vec{q})/\de q^3$) vanish. Since moments and cumulants are alternative characterizations of a probability distribution, all moments of a gaussian p.d.f. can be expressed in terms only of first and second order cumulants.} 
\begin{equation}
    \phi(\vec{q})= e^{i\vec{q}^T\vec{\chi}-\frac{1}{2}\vec{q}^T\chi\vec{q}}
\end{equation}
The first four moments of this distribution can be easily computed using eq.(\ref{eM}) (the subscript $\ne$ indicates that all indices are different indices). 

\begin{alignat}{2}
    \av{s_i} &=&& -i \frac{\de}{\de q_i} \phi(\vec{q})\Big|_{\vec{q}=\vec{0}} \label{e-fm} \\
    %& = && -i \Big(i \chi_i - \sum_n \chi_{in} q_n \Big) e^{i\vec{q}^T\vec{\chi}-\frac{1}{2}\vec{q}^T\chi\vec{q}}\Big|_{\vec{q}=\vec{0}} \notag\\
    & = &&\ \chi_i \notag\\
    \av{s_is_j}_{\ne } & = &&\ (-i)^2 \frac{\de}{\de q_j}\frac{\de}{\de q_i} \phi(\vec{q})\Big|_{\vec{q}=\vec{0}}\label{e-sm}\\
    %& = && \ -\Big[- \chi_{ij} + \Big(i \chi_j - \sum_n \chi_{jn} q_n \Big)\Big(i \chi_i - \sum_n \chi_{in} q_n \Big)\Big]e^{i\vec{q}^T\vec{\chi}-\frac{1}{2}\vec{q}^T\chi\vec{q}}\Big|_{\vec{q}=\vec{0}}\notag\\
    &= && \ \chi_{ij} + \chi_i\chi_j \notag \\
    \av{s_is_js_k}_{\ne} & = &&(-i)^3 \frac{\de}{\de q_k}\frac{\de}{\de q_j}\frac{\de}{\de q_i}\phi(\vec{q})\Big|_{\vec{q}=\vec{0}} \label{e-tm} \\
    %& = && \ i \Big[ - \chi_{jk} \Big(i \chi_i - \sum_n \chi_{in} q_n \Big) -\chi_{ik}\Big(i \chi_j - \sum_n \chi_{jn} q_n \Big) \notag \\
    %& && - \chi_{ij} \Big(i \chi_k - \sum_n \chi_{kn} q_n \Big) + \Big(i \chi_k - \sum_n \chi_{kn} q_n \Big) \times \notag \\
    %& && \times \Big(i \chi_j - \sum_n \chi_{jn} q_n \Big)\Big(i \chi_i - \sum_n \chi_{in} q_n \Big)\Big]e^{i\vec{q}^T\vec{\chi}-\frac{1}{2}\vec{q}^T\chi\vec{q}}\Big|_{\vec{q}=\vec{0}}\notag\\
    & = && \ \chi_i\chi_{jk} + \chi_j\chi_{ik} + \chi_k \chi_{ij} + \chi_i\chi_j\chi_k\notag\\
    \av{s_is_js_ks_l}_{\ne}
    & = &&(-i)^4 \frac{\de}{\de q_l}\frac{\de}{\de q_k}\frac{\de}{\de q_j}\frac{\de}{\de q_i}\phi(\vec{q})\Big|_{\vec{q}=\vec{0}}\label{e-ffm} \\
    %& = &&\Big\{ \chi_{ij}\chi_{kl} + \chi_{ik}\chi_{jl} + \chi_{jk}\chi_{il} +  \chi_{kl}\Big(i \chi_j - \sum_n \chi_{jn} q_n \Big)\Big(i \chi_i - \sum_n \chi_{in} q_n \Big)\ +\notag \\
    %& && \ + \chi_{jl}\Big(i \chi_k - \sum_n \chi_{kn} q_n \Big)\Big(i \chi_i - \sum_n \chi_{in} q_n \Big) + \chi_{il}\Big(i \chi_j - \sum_n \chi_{jn} q_n \Big)\ \times \notag \\
    %& && \times \Big(i \chi_k - \sum_n \chi_{kn} q_n \Big) + \Big[ - \chi_{jk} \Big(i \chi_i - \sum_n \chi_{in} q_n \Big) -\chi_{ik}\Big(i \chi_j - \sum_n \chi_{jn} q_n \Big) \notag\\ 
    %& && - \chi_{ij} \Big(i \chi_k - \sum_n \chi_{kn} q_n \Big) + \Big(i \chi_k - \sum_n \chi_{kn} q_n \Big)\Big(i \chi_j - \sum_n \chi_{jn} q_n \Big)\times \notag \\
    %& &&\ \times \Big(i \chi_i - \sum_n \chi_{in} q_n \Big)\Big]\Big(i \chi_l - \sum_n \chi_{ln} q_n \Big)\Big\}e^{i\vec{q}^T\vec{\chi}-\frac{1}{2}\vec{q}^T\chi\vec{q}}\Big|_{\vec{q}=\vec{0}}\notag \\
    & = &&\  \chi_{ij}\chi_{kl} + \chi_{ik}\chi_{jl} + \chi_{jk}\chi_{il} +  \chi_{ij}\chi_k\chi_l + \chi_{ik}\chi_j\chi_l + \chi_{il}\chi_j\chi_k  + \notag \\
    & && \ + \chi_{jk}\chi_i\chi_l + \chi_{jl}\chi_i\chi_k + \chi_{kl}\chi_i\chi_j + \chi_i\chi_j\chi_k\chi_l \ .\notag
\end{alignat}

\qed
\subsection{Derivation of eq.(\ref{eEFGC})}
Consider the dynamics of the first cumulants, eq.(\ref{eFOCP}). $\forall i\in1,\dots,L$
\begin{alignat}{2}
    \dot\chi_i %&\stackrel{}{=}&& \sum_j f_{j}\av{s_is_j} + \sum_{j<k} f_{jk}\av{s_is_js_k} - \sum_j f_j \chi_i\chi_j - \sum_{j<k} f_{jk} \chi_i\av{s_js_k}-2\mu\chi_i \notag\\
    &\stackrel{(a)}{=}&& \sum_j f_j\chi_{ij} + \sum_{j\ne i} f_{ij}\chi_j + \sum_{\substack{j<k\\j , k\ne i}} f_{jk} [\chi_i(\chi_j\chi_k + \chi_{jk}) + \chi_j\chi_{ik} + \chi_k\chi_{ij}]\ + \notag \\
    & &&- \sum_{j<k} f_{jk} \chi_i (\chi_{jk} + \chi_j\chi_k) - 2\mu\chi_i \notag\\
    &\stackrel{}{=}&& \sum_j f_j\chi_{ij} + \sum_{j\ne i} f_{ij}\chi_j - \sum_{j\ne i} f_{ij} \chi_i (\chi_{ij} + \chi_i\chi_j) + \sum_{\substack{j<k\\j , k\ne i}} f_{jk} (\chi_j\chi_{ik} + \chi_k\chi_{ij}) -2\mu\chi_i \notag \\
    &\stackrel{}{=}&& \sum_{j} f_j\chi_{ij} - \sum_{j\ne i} f_{ij} \chi_i \chi_{ij} + \sum_{j\ne i} f_{ij}\chi_j(1-\chi_i^2) + \sum_{\substack{j\ne k \\ j,k\ne i}} f_{jk} \chi_j\chi_{ik} - 2\mu\chi_i\notag\\
    &\stackrel{(b)}{=}&& \sum_j f_j\chi_{ij} - \sum_{j\ne i} f_{ij} \chi_i \chi_{ij} + \sum_{j\ne i} f_{ij}\chi_j\chi_{ii} + \sum_{j\ne i}\sum_{\substack{k\ne i\\k\ne j}} f_{jk} \chi_j\chi_{ik} -2\mu\chi_i\notag\\
    &\stackrel{(c)}{=}&& \sum_j f_j\chi_{ij} - \sum_{j\ne i} f_{ij} \chi_i \chi_{ij} + \sum_{j\ne i}\sum_{k\ne j} f_{jk} \chi_j\chi_{ik} \pm \sum_{k\ne i} f_{ik}\chi_i\chi_{ik} -2\mu\chi_i\notag\\
    &\stackrel{}{=}&& \sum_j f_j\chi_{ij} - 2 \sum_{j\ne i} f_{ij} \chi_i \chi_{ij} + \sum_{j}\sum_{k\ne j} f_{jk} \chi_k\chi_{ij} -2\mu\chi_i \notag\\
    &\stackrel{(d)}{=}&& \sum_j\chi_{ij}(f_j + \sum_k f_{jk}\chi_k-2f_{ij}\chi_i) -2\mu\chi_i \ . \notag
\end{alignat}
In $(a)$ we expanded $\av{s_js_k}$ and, after distinguishing the case where $i\ne j\ne k$, we exploited eq.(\ref{e-tm}); in $(b)$ we used $\chi_{ii} = \av{s_i^2}-\av{s_i}^2 = 1-\chi_i^2$; in $(c)$ we added and subtracted a sum; in $(d)$ we used $f_{ii}=0\ \forall i$.
\qed

\subsection{Derivation of eq.(\ref{eESOCVC})}
We start by substituting in eq.(\ref{eSOCP}) the result we have just derived for $\dot{\chi}_i$, eq.(\ref{eEFGC}). $\forall i,j\in 1,\dots L $ with ($i\ne j$):
\begin{align}
    \dot\chi_{ij}&\stackrel{}{=} {\color{blue}\sum_k f_{k}\av{s_is_js_k}} + {\color{red}\sum_{k<l} f_{kl}\av{s_is_js_ks_l}} - {\color{violet} \av{s_is_j} \Big(\sum_k f_{k}\chi_k + \sum_{k<l} f_{kl}\av{s_ks_l}\Big)}\ + \notag \\ 
    & -{\color{cyan}\chi_i \sum_k\chi_{jk}\big(\hat{f_k}-2f_{jk}\chi_j\big)} -\chi_j \sum_k\chi_{ik}\big(\hat{f_k}-2f_{ik}\chi_i\big)- (4\mu+rc_{ij})\chi_{ij}\label{eSOCI}
\end{align}
where we have defined $\hat{f_k}=f_k+\sum_jf_{jk}\chi_j$. We will now analyze separately the terms highlighted in blue (\textcolor{blue}{B}), red (\textcolor{red}{R}) and violet (\textcolor{violet}{V}) and cyan (\textcolor{cyan}{C}). In order to substitute eq.(\ref{e-tm}-\ref{e-ffm}) we again decompose the sums distinguishing cases where some of the indices are equal.

\begin{alignat}{2}
     {\color{violet} V} & = && \ (\chi_{ij}+\chi_i\chi_j)\Big( \sum_k f_k\chi_k + \sum_{k<l}f_{kl}(\chi_{kl}-\chi_k\chi_l) \Big)\notag \\
     & = &&\ (\chi_{ij}+\chi_i\chi_j) \Big( \sum_{k\ne i,j}f_k\chi_k + f_i\chi_i + f_j\chi_j + \sum_{\substack{k<l\\k,l\ne i,j}}f_{kl}(\chi_{kl}+\chi_k\chi_l) + \notag\\
     & && + \sum_{k\ne i,j} \big[f_{ik} (\chi_{ik} + \chi_i\chi_k) + f_{jk} (\chi_{jk} + \chi_j\chi_k)\big]  + f_{ij}(\chi_{ij}+\chi_i\chi_j)\Big) \notag \\
     {\color{blue} B} & = &&\  \sum_{k\ne i,j} f_k\av{s_is_js_k} + f_i\chi_j +  f_j\chi_i  \notag\\
     & = && \sum_{k\ne i,j} f_k (\chi_i\chi_j\chi_k + \chi_i\chi_{jk} + \chi_j\chi_{ik} + \chi_k\chi_{ij}) + f_i\chi_j +  f_j\chi_i  \notag\\
     {\color{red} R} & = && \sum_{\substack{k<l\\k,l \ne i,j}} f_{kl} \av{s_is_js_ks_l} + \sum_{k\ne i,j} \big[f_{ik} \av{s_js_k} + f_{jk} \av{s_is_k}\big] + f_{ij} \notag\\
     & = && \sum_{\substack{k<l\\k,l \ne i,j}} f_{kl} \big(\chi_i\chi_j\chi_k\chi_l + \chi_i\chi_j\chi_{kl} + \chi_i\chi_k\chi_{jl} + \chi_i\chi_l\chi_{jk} + \chi_j\chi_k\chi_{il} + \chi_j\chi_l\chi_{ik} + \notag \\
     & && + \chi_k\chi_l\chi_{ij} +\chi_{ij}\chi_{kl} + \chi_{ik}\chi_{jl} + \chi_{il}\chi_{jk}\big) + \sum_{k\ne i,j} \big[f_{ik} (\chi_{jk}+\chi_j\chi_k)\ + \notag \\
     & && + f_{jk} (\chi_{ik}-\chi_i\chi_k)\big] + f_{ij}\notag \\
     {\color{cyan} C} & = &&\ \chi_i \sum_k\chi_{jk}\big(f_k+\sum_l f_{kl}\chi_l-2f_{jk}\chi_j\big)\notag \\
     & = &&\ \chi_i \sum_{k\ne i,j}\chi_{jk}\big(f_k+\sum_l f_{kl}\chi_l-2f_{jk}\chi_j\big) + \chi_i\chi_{ij}\big(f_i + \sum_l\chi_{il}\chi_l-2f_{ij}\chi_j \big) \ +\notag\\
     & && + \chi_i(1-\chi_j^2)(f_j+\sum_lf_{jl}\chi_l)\notag
\end{alignat}
In the last line we have used $\chi_{ii}=1-\chi_i^2$, $f_{ii}=0\ \forall i$ and the definition of $\hat{f_i}$. In addition, note that there is a term in eq.(\ref{eSOCI}) which is nothing but (\textcolor{cyan}{C}) after exchanging $i\leftrightarrow j$.
    
Summing the all the terms in eq.(\ref{eSOCI}) and simplifying:
\begin{alignat}{2}
    \dot\chi_{ij} & = && -(4\mu+r c_{ij})\chi_{ij} - 2f_i\chi_i\chi_{ij} - 2f_j\chi_j\chi_{ij} +f_{ij} (1-\chi_{ij}^2-\chi_i^2\chi_j^2+2\chi_i\chi_j\chi_{ij})\ + \notag \\
    & && + \sum_{k\ne i,j} f_{ik}(\chi_{jk}+\chi_j\chi_k-\chi_{ij}\chi_{ik}+\chi_{ik}\chi_i\chi_j-\chi_{ij}\chi_i\chi_k-\chi_i^2\chi_j\chi_k)\ + \notag \\
    & && + \sum_{k\ne i,j} f_{jk}(\chi_{ik}+\chi_i\chi_k-\chi_{ij}\chi_{jk}+\chi_{jk}\chi_i\chi_j-\chi_{ij}\chi_j\chi_k-\chi_i\chi_j^2\chi_k)\ + \notag \\
    & && + (\chi_i^2\chi_j-\chi_j-\chi_i\chi_{ij}) \sum_{l} f_{il}\chi_{l} + (\chi_i\chi_j^2-\chi_i-\chi_j\chi_{ij})\sum_l f_{jl}\chi_{l} \ + \notag\\
    & && +\sum_{\substack{k<l\\k,l\ne i,j}} f_{kl} (\chi_{ik}\chi_j\chi_l + \chi_{jk}\chi_i\chi_l + \chi_{il}\chi_j\chi_k+\chi_{jl}\chi_i\chi_k+\chi_{ik}\chi_{jl}+\chi_{il}\chi_{jk})\ +\notag \\
    & && - \sum_{k\ne i,j} \chi_i\chi_{jk}\sum_l{\chi_lf_{kl}}- \sum_{k\ne i,j} \chi_j\chi_{ik}\sum_l{\chi_lf_{kl}}\ . \label{e-SOC2} %Evolution Second Order Cumulant Explicit
\end{alignat}
This already is the final result for the dynamics of the second order cumulants, where all sums have no equal indices. For the sake of elegance, it is possible to rewind this "decomposition" and the result is precisely eq.(\ref{eESOCVC}).
\qed

\subsection{Derivation of eq.(\ref{eIFLGCA})}
Recalling the definition $\epsilon = 1/(4\mu+rc_{ij})$, we substitute the expansion eq.(\ref{ePESOC}) into eq.(\ref{e-SOC2}) and impose $\dot{\chi}^{(n)}_{ij}=0$ for each order $\epsilon^n$. As a result, we have: 
\begin{alignat}{2}
&\mathcal{O}(\epsilon^{-1}) && : \chi_{ij}^{(0)} = 0 \notag\\
&\mathcal{O}(1) && : \chi_{ij}^{(1)} = f_{ij}(1-\chi_i^2)(1-\chi_j^2)\label{eFO}\\
&\mathcal{O}(\epsilon) && :     \chi_{ij}^{(2)} = \sum_k f_{ik}\big(\chi_{jk}^{(1)}+\chi_{ik}^{(1)}\chi_i\chi_j-\chi_i\chi_k\chi_{ij}^{(1)}\big)\ +\notag\\
    & &&- \sum_{k,l}f_{kl}\chi_i\chi_l\chi_{jk}^{(1)}-\sum_lf_{il}\chi_l\chi_i\chi_{ij}^{(1)}\ + \notag\\
    & &&+ \sum_{k<l} f_{kl} \Big(\chi_{ik}^{(1)}\chi_j\chi_l + \chi_{il}^{(1)}\chi_j\chi_k-2f_i\chi_i\chi_{ij}^{(1)}\notag\ + \\
    & &&+ f_{ij}\chi_i\chi_j\chi_{ij}^{(1)}\Big) + \{i\leftrightarrow j\} 
\end{alignat}
where, in the last equation the terms $\chi_{ij}^{(1)}$ as specified in eq. (\ref{eFO}) are left implicit. To the first order in $\epsilon$, eq.(\ref{eIFLGCA}) is found.
\qed

\subsection{Supplementary plots: NRC}
\begin{figure}[ht!]
      \begin{tabular}[t]{c}
            \begin{subfigure}[t]{\columnwidth}
            \centering
            \includegraphics[width=\textwidth]{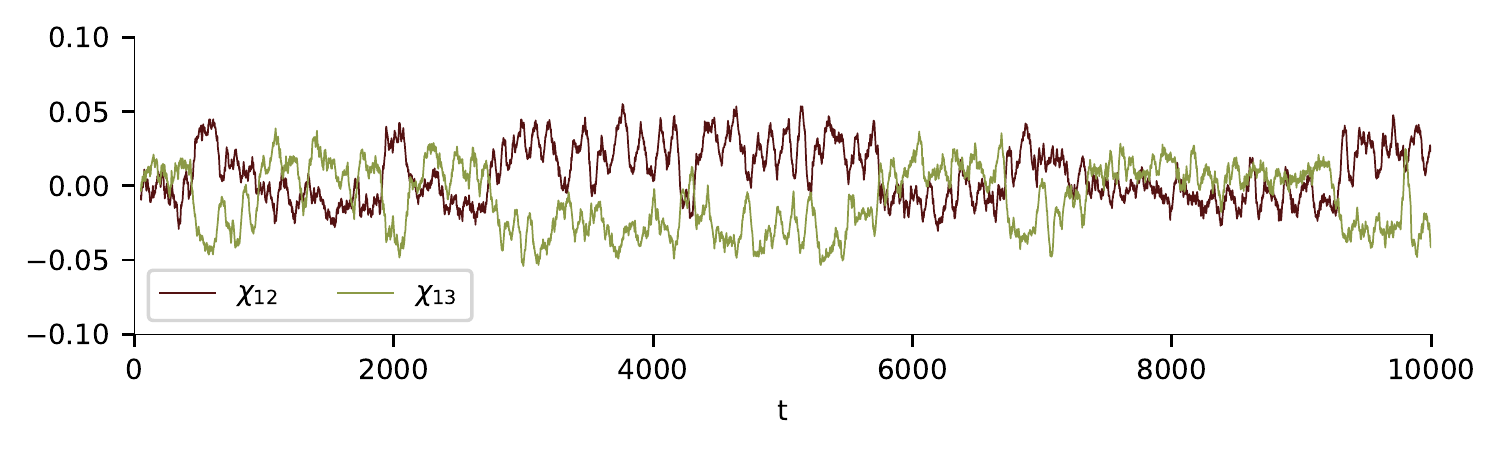}
            \caption{Dynamics of second order cumulants $\chi_{ij}=\av{s_is_j}-\av{s_i}\av{s_j}$. Two kind of behaviours are possible, since in a NRC phase trajectories at different loci are either correlated or anticorrelated - examples here are $\chi_{12}$ and $\chi_{13}$, average over $[t-\Delta t,t]$, with $\Delta t = 50$. Slightly stronger correlations appear in a NRC phase with respect to the QLE values $\chi_{ij}\sim0, \chi_{ii}\sim1$.}
            \end{subfigure}\\
            \begin{subfigure}[t]{\columnwidth}
            \centering
            \includegraphics[width=\textwidth]{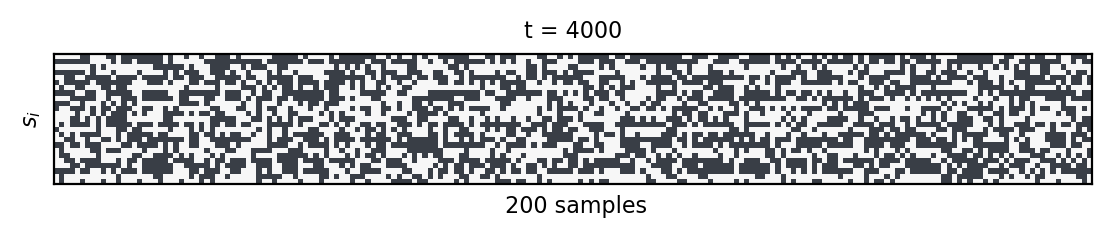}
            \caption{Snapshot of the population at $T=4600$ (NRC-like), $200$ out of $500$ individuals shown. Columns are individual genotypes of length $L$ (dark/light color for $s_i=\pm1$). Note that population averages correspond to averages along rows in this plot.}
          \end{subfigure}\\
          \begin{subfigure}[t]{\columnwidth}
            \centering
            \includegraphics[width=.8\textwidth]{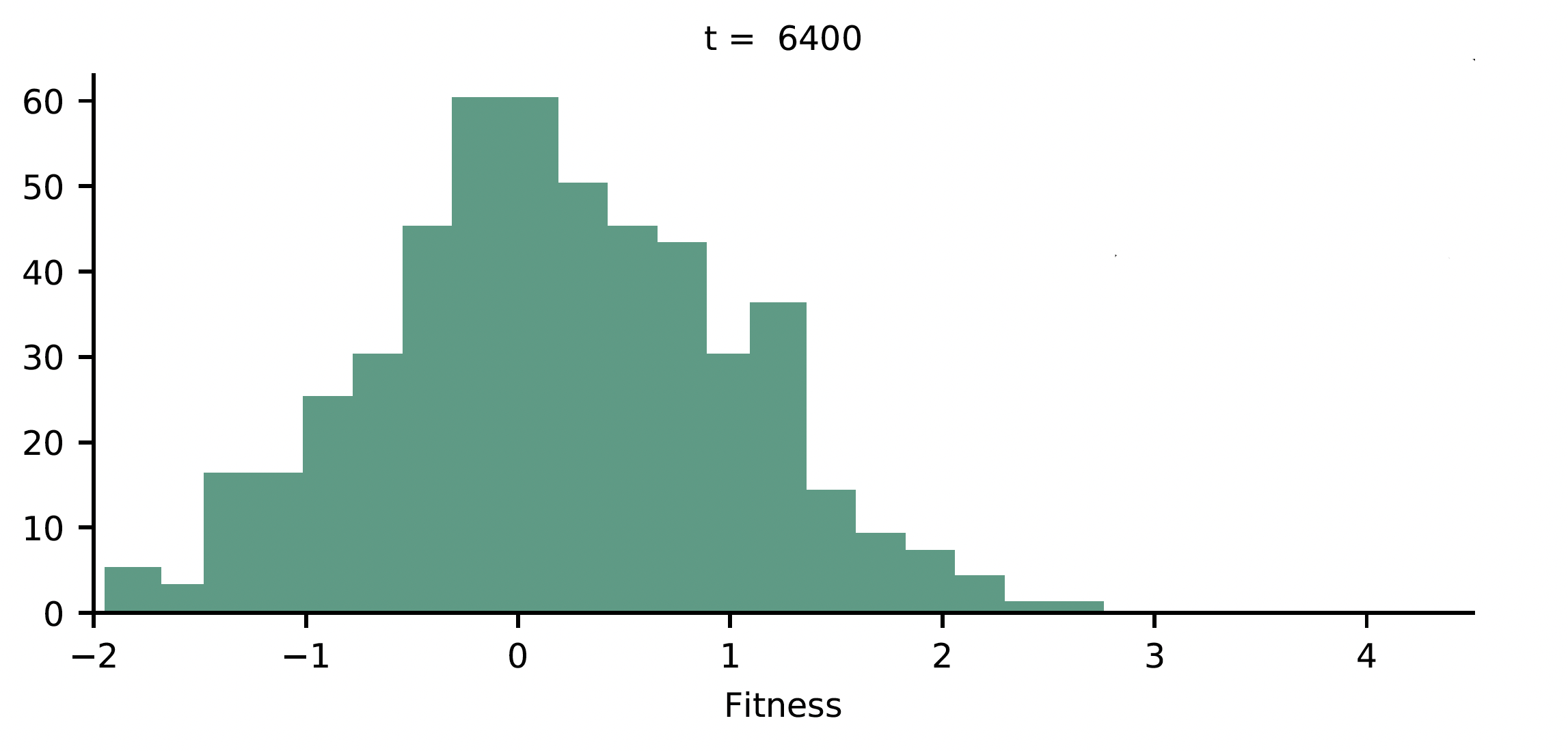}
            \caption{Snapshot of the fitness distribution at $T=6400$ (QLE-like)}
          \end{subfigure}
    \end{tabular}
    \caption{Supplementary plots for QLE-NRC instability, SK fitness function eq.(\ref{e-SKFF}). See fig.(\ref{f-interm}) for the main results. Parameters of the simulation: $N=500$, $L=25$, $T=1.0\times10^4$, $\mu=r=\omega=0.5$, $\sigma_e=0.024$, same meanings as in tab.(\ref{tNST}).}\label{f-interm-bis}
\end{figure}

\begin{figure}[ht!]
      \begin{tabular}[t]{c}
          \begin{subfigure}[h!]{\columnwidth}
            \centering
            \includegraphics[width=.8\textwidth]{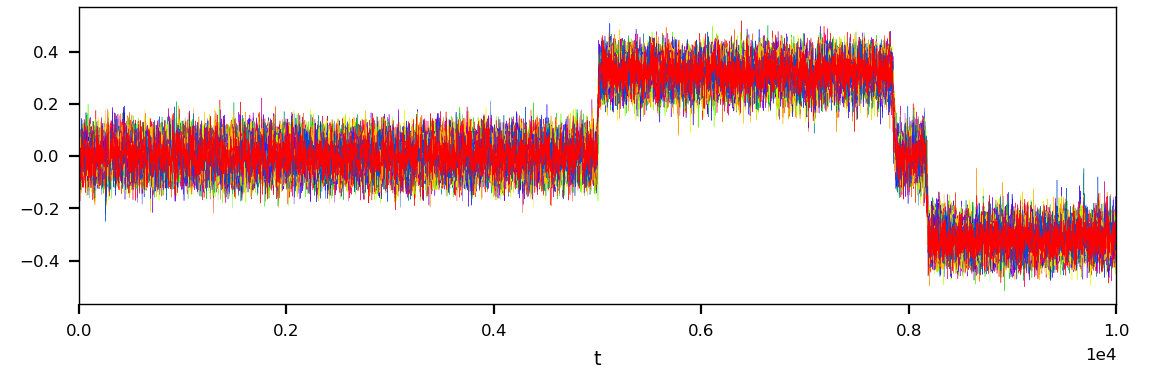}
            \caption{Evolution under an unfrustrated fitness function $F(g)=(\pm1)\sum_{i<j}|f_{ij}|s_is_j$ with $f_{ij}\sim\mathcal{N}(0,\sigma_e)$.
             The fitness landscape has here only two maxima, corresponding to $\bar g: \{s_i = +1 \ \forall i\}$ or $-\bar g$. In the case of a transition to a NRC-like behaviour, all $\chi_i$ move towards one of the two maxima all together, the coexistence of mirrored trajectories for different loci is hence removed. Parameters of the simulation: $N=500$, $L=25$, $T=1.0\times10^4$, $\mu=r=\omega=0.5$, $\sigma_e=0.0087$.}
          \end{subfigure}\\
            \begin{subfigure}[t]{\columnwidth}
            \centering
            \includegraphics[width=.8\textwidth]{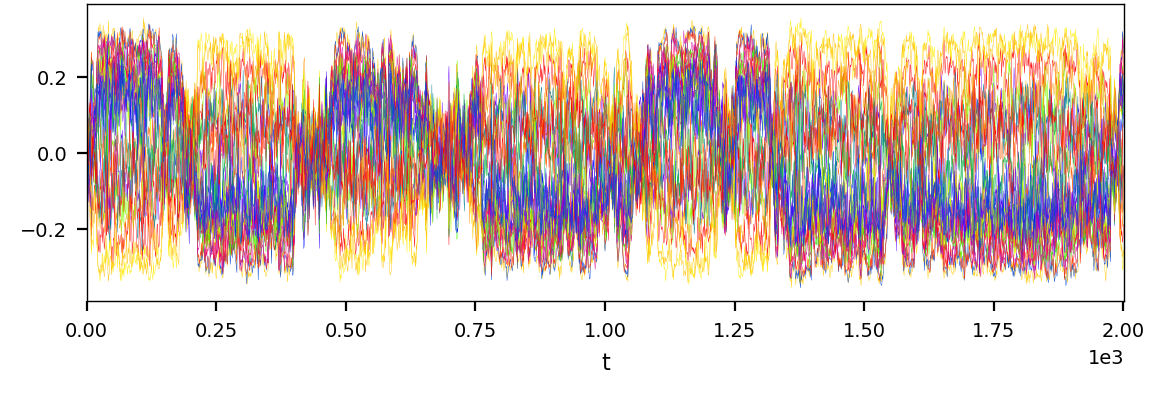}
            \caption{Evolution under strong selection and frequent mutations for asexual populations ($r=0$). In the absence of recombinations, the exploration of the genotype space is less efficient; moreover, fluctuations on the $\chi_i$ turn out to be much stronger and the resulting dynamics are less stable. Consequently, higher values of $\sigma_e$ and $N$ are needed to observe the oscillation in this figure, the first one enhancing selection, the second reducing noise. Interestingly, no such clear behaviour as in fig.(\ref{fFS-NRC}-\ref{fFD-NRC}) is observed in this case, therefore the heuristic arguments given in the main text might not apply to this case. Parameters of the simulation: $N=10000$, $L=25$, $T=2.0\times10^3$, $\mu=0.5$ $r=0.0$, $\sigma_e=0.09$.}
            \end{subfigure}
    \end{tabular}
    \caption{Intermittency QLE-NRC under different evolutionary conditions than the ones discussed in sec.(\ref{sHNRCP}). The dynamics of the first order cumulants $\chi_i \ \forall i$ are displayed, \emph{cf} fig.(\ref{fFOC-NRC}). }\label{fADDPLOTS}
\end{figure}

\begin{figure}[htbp]
    \centering
    \begin{tabular}[b]{cc}
    \begin{subfigure}[b]{0.45\textwidth}
    \includegraphics[width=\textwidth]{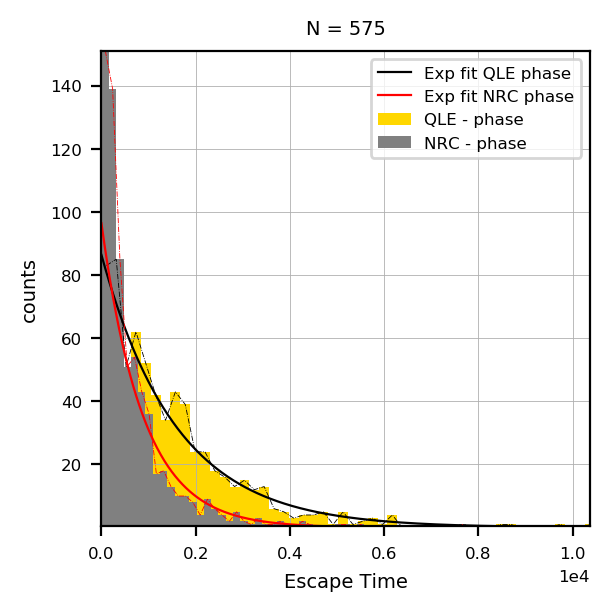}
    \end{subfigure}
    &
    \begin{subfigure}[b]{0.45\textwidth}
    \includegraphics[width=\textwidth]{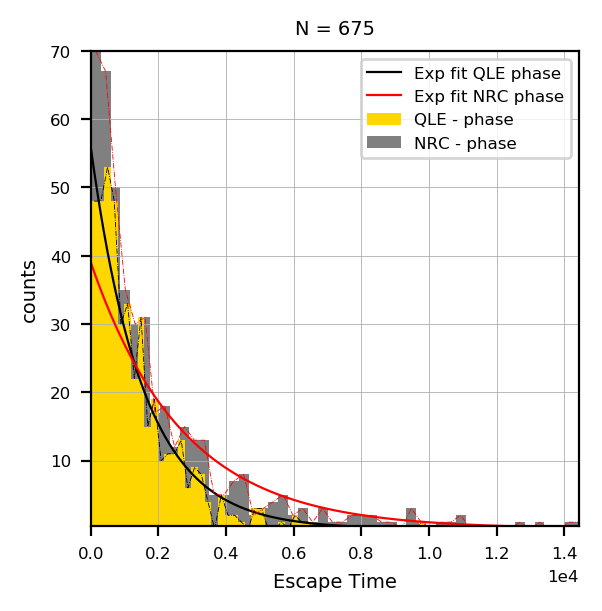}
    \end{subfigure}
    \end{tabular}
\caption{Distribution of escape times from QLE and NRC phases. Simulations are run in a region of the parameter space where the systems dynamics jumps back and forth between QLE and NRC behaviours. Among the values of $N$ indicated in fig.(\ref{fETA-NRC}, we here show $N=575,675$. By setting a threshold to the mean fitness statistics fig.(\ref{fFS-NRC}) we are able to classify QLE, NRC phase. Each time a transition QLE $\rightarrow$ NRC is observed, we record an instance of the empirical escape time $t_{QLE}$, analogously we record a instance of $t_{NRC}$ each time we observe a transition NRC $\rightarrow$ QLE. Here the histograms for $t_{QLE}$ and $t_{NRC}$ are shown in yellow, gray, respectively. Both are fitted with $y(T) = \gamma_{\pi} \ e^{- a_{\pi} T} ,\ {\pi} = \{\text{QLE, NRC}\}$, here shown as black and red curves, respectively. In fig.(\ref{fETA-NRC}, $t_{\pi}\sim1/a_{\pi}$ is taken as an estimate of the average escape time from the phase $\pi$.
Parameters of the simulations: $L=25$, $T=1.5\times10^6$, $\mu=0.5=r=\omega=0.5$, $\sigma_e=0.029$. }
\label{fMRHM}
\end{figure}

We show in this section additional illustration for the QLE interface with an NRC phase. In fig.(\ref{f-interm-bis}) additional illustration of the simulation showcased in fig.(\ref{f-interm}) are presented. In fig.(\ref{fADDPLOTS}) the QLE-NRC interface is investigated under slightly different conditions. Finally, in fig.(\ref{fMRHM}) details of the simulation and fitting procedure that lead to fig.(\ref{fETA-NRC}) are presented.

\newcommand{\newblock}{}
\bibliography{bibliography}

\section*{Acknowledgements}
We thank Profs Simona Cocco and R\'emi Monasson and Dott. Eugenio Mauri for numerous discussions and for a pleasant collaboration forming part of the background of the material presented in sec.(\ref{sINST}). We also thank Prof Joachim Krug for constructive remarks on the MS. VD warmly thanks Nordita (Stockholm, Sweden) and KTH (Stockholm, Sweden) for hospitality. The work of HLZ was sponsored by National Natural Science Foundation of China (11705097), Natural Science Foundation of Nanjing University of Posts and Telecommunications (Grant No. 221101, 222134). EA acknowledges support of the Swedish Research Council through grant 2020-04980. We finally thank Profs. B.H. Good and M.M. Desai for kindly allowing us to use previously unpublished Fig.\ref{fGood}.
\end{document}